\documentclass[12pt,oneside,reqno,titlepage, hyperfootnotes=false]{amsart}
\usepackage{lmodern}
\usepackage[T1]{fontenc}
\usepackage[latin9]{inputenc}
\usepackage{color}
\usepackage{array}
\usepackage{float}
\usepackage{bm}
\usepackage{amstext}
\usepackage{amsthm}
\usepackage{amssymb}
\usepackage{graphicx}
\usepackage[a4paper]{geometry}
\usepackage{setspace}
\usepackage{wasysym}
\usepackage[authoryear]{natbib}
\usepackage[pdfusetitle,
 bookmarks=true,bookmarksnumbered=false,bookmarksopen=false,
 breaklinks=false,pdfborder={0 0 0},pdfborderstyle={},backref=false,colorlinks=true]
 {hyperref}

\makeatletter

\providecommand{\tabularnewline}{\\}
\floatstyle{ruled}
\newfloat{algorithm}{tbp}{loa}
\providecommand{\algorithmname}{Algorithm}
\floatname{algorithm}{\protect\algorithmname}

\numberwithin{equation}{section}
\numberwithin{figure}{section}

\usepackage{amsfonts}
\usepackage{amstext}
\usepackage{amsthm}

\usepackage{hyperref}
\hypersetup{
    colorlinks = true,
    citecolor = black
}

\usepackage{threeparttable}
\usepackage{graphicx}
\usepackage{enumerate}
\usepackage{listings}
\usepackage{caption}
\usepackage{subcaption}
\usepackage{algorithm,algpseudocode}
\usepackage[hang,flushmargin]{footmisc}
\newcommand{\algorithmfootnote}[2][\footnotesize]{%
  \let\old@algocf@finish\@algocf@finish
  \def\@algocf@finish{\old@algocf@finish
    \leavevmode\rlap{\begin{minipage}{\linewidth}
    #1#2
    \end{minipage}}%
  }%
}

\DeclareMathOperator*{\argmin}{arg\,min}

\DeclareMathOperator*{\argmax}{arg\,max}

\usepackage{float}

\newtheorem{thm}{Theorem}

\newtheorem{lem}{Lemma}
\newtheorem*{lem*}{Lemma}
\newtheorem*{thm*}{Theorem}

\newtheorem*{asm1}{Assumption 1}
\newtheorem*{asm2}{Assumption 2}
\newtheorem*{asm3}{Assumption 3}

\theoremstyle{definition}

\makeatother

\begin{document}
\title{Risk and optimal policies in bandit experiments}
\author{Karun Adusumilli$^\dagger$ }
\begin{abstract}
We provide a decision theoretic analysis of bandit experiments under
local asymptotics. Working within the framework of diffusion processes,
we define suitable notions of asymptotic Bayes and minimax risk for
these experiments. For normally distributed rewards, the minimal Bayes
risk can be characterized as the solution to a second-order partial
differential equation (PDE). Using a limit of experiments approach,
we show that this PDE characterization also holds asymptotically under
both parametric and non-parametric distributions of the rewards. The
approach further describes the state variables it is asymptotically
sufficient to restrict attention to, and thereby suggests a practical
strategy for dimension reduction. The PDEs characterizing minimal
Bayes risk can be solved efficiently using sparse matrix routines
or Monte-Carlo methods. We derive the optimal Bayes and minimax policies
from their numerical solutions. These optimal policies substantially
dominate existing methods such as Thompson sampling; the risk of the
latter is often twice as high. 
\end{abstract}

\thanks{\textit{This version}: \today{}\\
\thispagestyle{empty}I would like to thank two anonymous referees
for valuable suggestions that substantially improved the paper. Thanks
also to Xiaohong Chen, David Childers, Keisuke Hirano, Hiroaki Kaido,
Jonas Lieber, Ulrich M{\"u}ller, Frank Schorfheide, Stefan Wager
and seminar participants at multiple universities and conferences
for helpful comments.\\
$^\dagger$Department of Economics, University of Pennsylvania}
\maketitle

\section{Introduction \label{sec:Introduction}}

The multi-armed bandit problem describes an agent who seeks to maximize
the welfare, i.e., the cumulative returns (aka rewards), generated
by sequentially selecting among various actions (aka arms), the effects
of which are initially unknown.\textcolor{blue}{{} }Compared to static
experiments, adaptive experiments such as bandit algorithms enable
fast learning and implementation of optimal actions, while minimizing
welfare-lowering experimentation. Due to this promise of large welfare
gains, they have been extensively studied in recent years and applied
in areas such as online advertising \citep{russo2017tutorial}, dynamic
pricing \citep{ferreira2018online}, public health \citep{athey2021shared}
and economics \citep{kasy2019adaptive,caria2020adaptive}.

The bandit problem can be formulated as a dynamic programming one,
but solving this exactly is typically infeasible. Instead, heuristic
solutions are commonly used, such as Thompson sampling (TS; see \citealp{russo2017tutorial}
and references therein) and Upper Confidence Bound (UCB; \citealp{lai1985asymptotically})
algorithms. There is by now a large theoretical literature on the
regret properties of stochastic bandit algorithms.\footnote{There is also an important, and parallel, literature on adversarial
bandits that this paper does not contribute to, see, e.g., \citealp{hazan2016introduction}.} Here, regret is the difference in welfare from pulling the best arm
and the agent's actual welfare. Existing results on lower bounds for
regret come in two forms. The first set of results, `instance dependent
bounds' (\citealp{lai1985asymptotically}), provide lower bounds on
rates of regret for `consistent' algorithms under a given set of reward
distributions for each arm. These results are of a large deviations
flavor. The second set of results specify the minimax rates of regret,
when nature is allowed to adversarially change the reward distributions
depending on $n$, the number of periods of experimentation allowed.
This rate is of the order $n^{-1/2}$ \citep[Ch.\ 9]{lattimore2020bandit}. 

Despite these advances, a number of questions still remain. Many algorithms,
including TS and UCB, attain the rate bounds described above, but
existing results are silent on selecting between them. Decision theory
under ambiguity suggests two common measures, Bayes and minimax risk,
for ranking algorithms. The importance of these measures is well recognized
in the literature, see \citet[Chs.\ 13, 35]{lattimore2020bandit},
but their characterization, and the subsequent derivation of optimal
algorithms, remain open questions (in fact, a common, but incorrect,
view is that these are intractable). We seek to answer these questions.

The first contribution of this paper is to define notions of asymptotic
Bayes and minimax risk for bandit experiments under diffusion asymptotics
(\citealp{wager2021diffusion,fan2021diffusion}). These asymptotics
consider the regime where the difference in expected rewards between
the arms scales at the minimax, $n^{-1/2}$, rate. This defines the
hardest instance of the bandit problem: if the reward gap scales at
a faster rate, identifying the optimal arm is straightforward, whereas
if it scales at a slower rate, there is too little difference between
the arms, so the asymptotic risk is trivially 0 in either case. The
$n^{-1/2}$ scaling thus provides a good approximation to the finite
sample properties of bandit algorithms. The same scaling occurs in
the analysis of treatment assignment rules by \citet{hirano2009asymptotics}. 

\citet{wager2021diffusion} and \citet{fan2021diffusion} study the
properties of TS under diffusion asymptotics, but do not address the
question of optimal policies under Bayes and minimax risk, as we do
here. We define Bayes risk using `non-negligible' priors, i.e., priors
applied on the mean rewards \textit{after} scaling them by the minimax,
$n^{-1/2}$, rate (see Section \ref{subsec:Bayes-risk}). This a major
departure from the existing literature that, starting from \citet{lai1987adaptive},
employs a fixed prior, but which leads to a trivial Bayes risk of
$0$ under the $n^{-1/2}$ scaling. This literature instead analyzes
Bayes risk using large-deviation methods without scaling the rewards,
but as it is based on analysis of tail probabilities and not distributional
approximations, the results are not sharp enough to select between
various policies, e.g., both TS and UCB attain the large-deviation
lower bound. By contrast, we characterize the minimal Bayes risk under
the $n^{-1/2}$ scaling as the solution to a $2^{\textrm{nd}}$-order
partial differential equation (PDE). 

We first demonstrate this characterization for Gaussian rewards, using
the theory of viscosity solutions to PDEs \citep{crandall1992user}.
The PDE machinery is indispensable because existing results (\citealp{wager2021diffusion,fan2021diffusion})
only apply to continuous policies, whereas the optimal Bayes policy
is generically deterministic, and hence discontinuous. Next, using
a limit of experiments approach, we show that the same PDE characterization
also holds asymptotically under both parametric and non-parametric
distributions of the rewards. Thus, any bandit problem can be asymptotically
reduced to one with Gaussian rewards. As part of this reduction, we
find that it is sufficient to restrict attention to just two state
variables per arm, apart from time: these are the number of times
the arm has been pulled in the past, and either the score process
(for parametric models) or the cumulative rewards from pulling the
arm (for nonparametric models). This reduction in dimension is perhaps
the main practical insight of this paper since the state space otherwise
grows linearly with $n$ (see Section \ref{subsec:Heuristics}).

We demonstrate the equivalence of experiments by extending the posterior
approximation method of \citet[Section 6.4]{le2000asymptotics} to
sequential experiments. The proof makes use of novel arguments involving
uniform approximation of log-likelihoods and posteriors in sequential
settings. It also differs from the standard approach based on asymptotic
representations; the latter is difficult to implement under diffusion
asymptotics as it requires the construction of couplings between continuous
time processes. The techniques introduced here are thus of independent
interest for analyzing other types of sequential experiments.

The PDE characterizing minimal Bayes risk is essentially a limit case
of the dynamic programming problem (DP) associated with the bandit
experiment. While it is infeasible to solve the DP problem directly,
we present ways to efficiently solve the PDE using finite-difference
and Monte-Carlo methods. This enables us to identify the Bayes optimal
policies. Compared to the latter, we find TS to be provably sub-optimal
as it over-explores; empirical illustrations drawn from real-world
examples find its Bayes risk to be twice as high in some cases. Conversely,
under independent Gaussian priors, the form of the optimal policy
is broadly similar to UCB (for one-armed bandits, this even holds
under any prior). In such cases, we show that MOSS (Minimax Optimal
policy in Stochastic Setting), a minimax-rate optimal version of UCB,
can effectively mimic the optimal policy after optimally tuning it
to the given prior. This is borne out by our empirical illustrations
and we thus recommend it over TS. Incidentally, such a tuned version
of MOSS, while natural, does not appear to have been considered before;
in fact, the standard implementation of MOSS performs even worse than
TS. It should be noted, however, that the similarity between the optimal
policy and UCB/MOSS fails for correlated and non-Gaussian priors. 

As an alternative to Bayes risk, we can use minimax risk. This is
simply Bayes risk under a least-favorable prior, and we numerically
compute both this prior and the minimax optimal policy. Intriguingly,
we find that optimally tuned MOSS (as proposed here) is close to minimax
optimal for one-armed bandits, even as an optimally tuned TS performs
much worse. This highlights the usefulness of our theory since it
would not have been possible to know the above without computing the
minimax lower bound; existing results give no reason to favor MOSS
over TS. 

Our framework easily accommodates various generalizations and modifications
to the bandit problem such as time discounting and best arm identification
(\citealp{russo2016simple}; \citealp{kasy2019adaptive}). The discounted
bandit problem has a rich history in economic applications, ranging
from market pricing \citep{rothschild1974two} to decision making
in labor markets \citep{mortensen1986job}. For discounted problems,
the optimal Bayes policy can be characterized using Gittins indices
(\citealp{gittins1979bandit}). However, except in simple instances,
e.g., discrete state spaces, computing the Gittins index is difficult
(see, \citealp[Section 35.5]{lattimore2020bandit}). Also, it does
not apply beyond the discounted setting; the optimal Bayes policy
in finite horizon settings is not an index policy (\citealp[Chapter 6]{berry1985bandit}).
Here, we take a different route and characterize the optimal Bayes
policy using PDEs.

\section{Diffusion asymptotics and statistical risk\label{sec:Diffusion-asymptotics-and}}

In this section, we provide a heuristic derivation of the PDE characterizing
minimal Bayes risk in the Multi-Armed Bandit (MAB) problem.

In the MAB problem, there are $K$ arms, and at each period $j$,
a decision maker (DM) chooses which arm $k\in\{0,\dots,K-1\}$ to
pull. Each pull generates a reward with an unknown mean $\mu_{k}$
that is specific to the arm. Suppose the experiment concludes after
$n$ periods, where $n$ is pre-specified. Knowledge of $n$ is reasonable
if it is the population size; indeed, the bandit setting blurs any
distinction between sample and population. In other cases, it might
be more reasonable to assume the DM employs discounting and allows
the experiment to continue indefinitely. The decision theoretic analysis
employed here requires modeling all aspects of decision making including
when to stop or how to discount, but our results are otherwise very
broadly applicable. We focus on the known $n$ case to avoid duplication
of effort, but see Appendix \ref{subsec:Discounting} for discounted
bandits. When $n$ is known, the number of periods that have elapsed
is a state variable, and after dividing by $n$ will be termed `time'.
Thus, time $t$ proceeds from 0 and 1, and is incremented by $1/n$
between successive periods. 

Let $A_{j}$ denote the action in period $j$, where $A_{j}=k$ if
arm $k$ is pulled. Suppose each time an arm $k$ is pulled, a reward,
$Y^{(k)}$, is drawn from the normal distribution $\mathcal{N}(\mu_{k,n},\sigma_{k}^{2})$,
where $\mu_{k,n}:=\mu_{k}/\sqrt{n}$. The scaling of the mean reward
by $\sqrt{n}$ follows \citet{wager2021diffusion} and \citet{fan2021diffusion}
and ensures the signal decays with sample size. The variances, $\sigma_{k}^{2}$,
are assumed to be known. Taking variances to be known is common practice
when working under local asymptotics as replacing unknown variances
with consistent estimates has no effect on asymptotic risk (see Section
\ref{sec:Conclusion}). In this section and the next, we provide a
detailed description of the MAB problem under such normally distributed
rewards. The utility of this analysis stems from the fact that more
general models - that assume either a parametric or non-parametric
distribution of rewards - reduce asymptotically to the normal setting
under the limit of experiments approach, see Sections \ref{sec:General-parametric-models}
and \ref{sec:The-nonparametric-setting}.

In what follows, we represent rewards using the so-called `stack-of-rewards
model' (\citealp[Section 4.6]{lattimore2020bandit}). This entails
the following: We exclusively use $j$ to refer to the periods of
experimentation, and $i$ to refer to the number of pulls of an arm.
$Y_{i}^{(k)}$ denotes the reward at the $i$-th pull of arm $k$,
and ${\bf y}_{i}^{(k)}:=\{Y_{i^{\prime}}^{(k)}\}_{i^{\prime}=1}^{i}$
denotes the sequence of rewards after $i$ pulls of that arm. We can
imagine that prior to the experiment, nature draws a stack of outcomes,
$\{Y_{i}^{(k)}\}_{i=1}^{n}$, corresponding to each arm $k$, and
at each period $j$, if $A_{j}=k$, the agent observes the outcome
at the top of the stack (this outcome is then removed from the stack).
Note that $\{Y_{i}^{(k)}\}_{i=1}^{n}$ are iid conditional on the
unknown parameters $\mu_{k}$.

Due to normality of the rewards, the only relevant state variables
are the number of times the arm was pulled, $q_{k}(t):=n^{-1}\sum_{j=1}^{\left\lfloor nt\right\rfloor }\mathbb{I}(A_{j}=k)$,
the cumulative rewards, $x_{k}(t):=n^{-1/2}\sum_{i=1}^{\left\lfloor nq_{k}(t)\right\rfloor }Y_{i}^{(k)}$,
and time $t$ (see Section \ref{sec:General-parametric-models} for
a formal argument about the sufficiency of these variables). The scaling
on $x_{k}(t)$ follows \citet{wager2021diffusion} and is equivalent
to rescaling the rewards $Y_{i}^{(k)}$ by the factor $1/\sqrt{n}$.
The DM chooses a policy rule $\pi(\cdot)\equiv\{\pi_{k}(\cdot)\}_{k}:\mathcal{S}\rightarrow[0,1]^{K+1}$
that determines the probability of pulling each arm $k$ given the
current state $s:=\{\{x_{k},q_{k}\}_{k},t\}$. 

For Lipschitz continuous $\pi$, \citet{wager2021diffusion} show
that $\{x_{k}(\cdot),q_{k}(\cdot)\}_{k}$ evolve in the large $n$
limit according to the stochastic differential equations (SDEs)
\begin{align}
dq_{k}(t)=\pi_{k}(s_{t})dt;\quad dx_{k}(t) & =\pi_{k}(s_{t})\mu_{k}dt+\sigma_{k}\sqrt{\pi_{k}(s_{t})}dW_{k}(t),\label{eq:diffusion limit}
\end{align}
where $\{W_{k}(\cdot)\}_{k}$ are independent one-dimensional Brownian
motions, and $\pi_{k}(s_{t}):=\pi_{k}(s(t))$. While (\ref{eq:diffusion limit})
is convenient for heuristics, there is in fact no guarantee that the
optimal policy possesses the requisite regularity properties for (\ref{eq:diffusion limit})
to formally hold. As it turns out, our formal results, in Section
3, do not rely on (\ref{eq:diffusion limit}). 

\subsection{Payoff and loss functions}

We take the loss function to be cumulative payoffs, where the payoff
is $0$ when the experiment concludes at $t=1$. We focus on the regret
payoff
\begin{equation}
R(A,\mu)=n^{-1/2}\left\{ Y^{(k^{*})}-\sum_{k}Y^{(k)}\mathbb{I}(A=k)\right\} ,\label{eq:regret payoff defn}
\end{equation}
where $k^{*}=\arg\max_{k}\mu_{k}$. It is the difference in rewards
between the optimal action, $A^{*}=k^{*}$, and the action $A$. Clearly,
regret is just a rescaling of the welfare payoff $W(A,\mu)=-\sum_{k}Y^{(k)}\mathbb{I}(A=k)/\sqrt{n}$.
While these payoffs are equivalent under Bayes risk, their behavior
under minimax risk is very different. Under the welfare payoff, the
minimax policy is trivial and excessively pessimistic: the DM should
never pull the arm. By contrast, the minimax risk under regret payoff
is non trivial. For this reason, we focus exclusively on regret (as
does most of the bandit literature). 

Our theory easily extends to other loss criteria, e.g., best arm identification
(\citealp{kasy2019adaptive}). The latter is discussed in Appendix
\ref{subsec:Best-arm-identification}.

\subsection{Bayes risk\label{subsec:Bayes-risk}}

Here we introduce asymptotic Bayes risk for bandit experiments.

\subsubsection{Priors and posteriors\label{subsec:Priors-and-posteriors}}

Suppose the DM places a prior, $m_{0}$, over $\bm{\mu}:=(\mu_{0},\dots,\mu_{K-1})$.
When the current state is $s\equiv\{\{x_{k},q_{k}\}_{k},t\}$, the
posterior density of $\bm{\mu}$ is\footnote{Here, and in the sequel, $\propto$ denotes `proportional to', i.e.,
equality up to a normalizing constant.} 
\begin{equation}
p(\bm{\mu}\vert s)\propto\prod_{k}p_{q_{k}}(x_{k}\vert\mu_{k};\sigma_{k}^{2})\cdot m_{0}(\bm{\mu});\quad p_{q}(\cdot\vert\mu;\sigma^{2})\equiv\mathcal{N}(\cdot\vert q\mu,q\sigma^{2}),\label{eq:posterior}
\end{equation}
where $\mathcal{N}(\cdot\vert\mu,\sigma^{2})$ is the normal density
with mean $\mu$ and variance $\sigma^{2}$. Importantly, the posterior
depends only on the $\left\lfloor nq_{k}\right\rfloor $ realizations
of the rewards, $\{{\bf y}_{nq_{k}}^{(k)}\}_{k}$, from each arm $k$
and is not affected by the past values of the actions (nor by past
values of $q_{k}$). Lemma \ref{Lemma 1} in Appendix \ref{subsec:Supporting-lemmas-for}
shows that this property holds generally, and is not limited to Gaussian
rewards. 

Since the prior is placed on the local parameter $\bm{\mu}$, it is
asymptotically `non-negligible'. In this regard, our approach differs
fundamentally from the previous literature (e.g., \citealp{lai1987adaptive})
on Bayesian bandits which employs a fixed prior. The rationale for
non-negligible priors is two-fold: First, it provides a better approximation
to finite sample properties. Indeed, any prior applied on the actual
mean, $\bm{\mu}/\sqrt{n}$, would be flat asymptotically, and its
Bayes risk simply $0$ under the $\sqrt{n}$ scaling of mean rewards.
Second, it enables us to characterize minimax risk as Bayes risk under
a least favorable prior (see Section \ref{subsec:Minimax-risk}).
The least favorable prior is non-negligible.

In practice, we are typically provided with a prior, $\rho_{0}$,
on the unscaled mean $\bm{\mu}_{n}=\bm{\mu}/\sqrt{n}$. To apply the
methods here, one needs to convert this to a prior, $m_{0}(\cdot)=\rho_{0}(\cdot/\sqrt{n})$,
on $\bm{\mu}$. To illustrate, suppose the DM places a Gaussian prior
$\mu_{k,n}\sim\mathcal{N}\left(\bar{\mu}_{k,0},\bar{\nu}_{k}^{2}\right)$
that is independent across $k$. We calibrate the scaled prior mean
and variance as $\mu_{k,0}=\sqrt{n}\bar{\mu}_{k,0}$ and $\nu_{k}^{2}=n\bar{\nu}_{k}^{2}$,
so $\mu_{k}:=\mu_{k,n}/\sqrt{n}\sim\mathcal{N}(\mu_{k,0},\nu_{k}^{2})$.
Then, if the current state is $s\equiv\{\{x_{k},q_{k}\}_{k},t\}$,
the posterior distribution of $\mu_{k}$ is 
\begin{equation}
\mu_{k}\vert s\sim\mathcal{N}\left(\frac{\sigma_{k}^{-2}x_{k}+\nu_{k}^{-2}\mu_{k,0}}{\sigma_{k}^{-2}q_{k}+\nu_{k}^{-2}},\frac{1}{\sigma_{k}^{-2}q_{k}+\nu_{k}^{-2}}\right).\label{eq:posterior distribution normal}
\end{equation}

\subsubsection{PDE characterization of Bayes and minimal Bayes risk\label{subsec:PDE-characterization-of}}

For a policy $\pi$, we define asymptotic Bayes risk, $V_{\pi}(s)$,
as the expected cumulative regret in the diffusion regime, where the
expectation is taken conditional on all information until state $s$.
We now informally derive a PDE characterization of $V_{\pi}(s)$. 

Consider the evolution of cumulative regret in a short time period,
$\Delta t$, following state $s$. The expected regret accrued within
this time period is approximately 
\[
\mathbb{E}_{\bm{\mu}\vert s}\left[\mu_{k^{*}}-\sum_{k}\mu_{k}\pi_{k}\right]\cdot\Delta t=\left(\mu^{\max}(s)-\sum_{k}\mu_{k}(s)\pi_{k}\right)\cdot\Delta t,
\]
where $\mu^{\max}(s):=\mathbb{E}_{\bm{\mu}\vert s}[\max_{k}\mu_{k}]$
and $\mu_{k}(s):=\mathbb{E}_{\mu_{k}\vert s}[\mu_{k}]$ are the posterior
means of $\max_{k}\mu_{k}$ and $\mu_{k}$. At the same time, by (\ref{eq:diffusion limit}),
the change to $q_{k}$ and $x_{k}$ over this time period is approximately
(henceforth we use $\pi_{k}$ as a shorthand for $\pi_{k}(s)$)
\begin{align*}
\Delta q_{k} & \approx\pi_{k}\Delta t;\quad\Delta x_{k}\approx\pi_{k}\mu_{k}\Delta t+\sigma_{k}\sqrt{\pi_{k}}\Delta W(t).
\end{align*}
Hence, up to a first order approximation, $V_{\pi}(s)$ satisfies
the recursion
\begin{equation}
V_{\pi}(s)\approx\mathbb{E}\left[\left.\left(\mu^{\max}(s)-\sum_{k}\mu_{k}(s)\pi_{k}\right)\cdot\Delta t+V_{\pi}\left(\{x_{k}+\Delta x_{k},q_{k}+\Delta q_{k}\}_{k},t+\Delta t\right)\right|s\right],\label{eq:recursion for V_pi}
\end{equation}
with the terminal condition $V_{\pi}(s)=0$ if $t=1$. 

Now, Ito's lemma implies that
\begin{align*}
 & \mathbb{E}\left[\left.V_{\pi}\left(\{x_{k}+\Delta x_{k},q_{k}+\Delta q_{k}\}_{k},t+\Delta t\right)-V_{\pi}(s)\right|s\right]\\
 & \approx\left(\partial_{t}V_{\pi}+\sum_{k}\left\{ \pi_{k}\partial_{q_{k}}V_{\pi}+\pi_{k}\mu_{k}(s)\partial_{x_{k}}V_{\pi}+\frac{1}{2}\pi_{k}\sigma_{k}^{2}\partial_{x_{k}}^{2}V_{\pi}\right\} \right)\Delta t.
\end{align*}
Thus, subtracting $V_{\pi}(s)$ from both sides of the recursion (\ref{eq:recursion for V_pi})
and dividing by $\Delta t$, we find that $V_{\pi}(\cdot)$ solves
the PDE 
\begin{align}
\partial_{t}V_{\pi}+\mu^{\max}(s)+\sum_{k}\pi_{k}(s)\left\{ -\mu_{k}(s)+L_{k}[V_{\pi}](s)\right\}  & =0\ \textrm{if }t<1,\label{eq:PDE- bayes risk}
\end{align}
with the terminal condition $V_{\pi}(s)=0$ if $t=1$. Here,
\[
L_{k}[f]:=\partial_{q_{k}}f+\mu_{k}(s)\partial_{x_{k}}f+\frac{1}{2}\sigma_{k}^{2}\partial_{x_{k}}^{2}f.
\]
denotes the infinitesimal generator of $\{x_{k}(\cdot),q_{k}(\cdot)\}$
for each $k$. It is the continuous time counterpart of the transition
density matrix for these state variables.

We can derive a similar characterization of the minimal Bayes risk,
$V_{\pi}^{*}(s):=\inf_{\pi(\cdot)\in\Pi}V_{\pi}(s)$, where $\Pi$
denotes the class of all possible policy rules. By the dynamic programming
principle, and in analogy with (\ref{eq:recursion for V_pi}), we
should have
\[
V^{*}(s)\approx\inf_{\pi\in[0,1]}\mathbb{E}\left[\left.\left(\mu^{\max}(s)-\sum_{k}\mu_{k}(s)\pi_{k}\right)\cdot\Delta t+V^{*}\left(\{x_{k}+\Delta x_{k},q_{k}+\Delta q_{k}\}_{k},t+\Delta t\right)\right|s\right],
\]
for any small time increment $\Delta t$, with the terminal condition
$V^{*}(s)=0$ if $t=1$. Then, by similar heuristic arguments as those
leading to (\ref{eq:PDE- bayes risk}), we obtain 
\begin{align}
\partial_{t}V^{*}+\mu^{\max}(s)+\min_{k}\left\{ -\mu_{k}(s)+L_{k}[V^{*}](s)\right\}  & =0\ \textrm{if }t<1,\label{eq:PDE_optimal_bayes_risk_General}\\
V^{*}(s) & =0\ \textrm{if }t=1.\nonumber 
\end{align}
As with PDE (\ref{eq:PDE- bayes risk}), PDE (\ref{eq:PDE_optimal_bayes_risk_General})
can be solved using knowledge only of $\{\sigma_{k}^{2}\}_{k}$. We
can thus characterize the minimal ex-ante Bayes risk as $V^{*}(0):=V^{*}(s_{0})$,
where $s_{0}:=\{\{x_{k}=0,q_{k}=0\}_{k},t=0\}$ is the initial state.

\subsubsection*{Discussion}

In the context of PDE (\ref{eq:PDE_optimal_bayes_risk_General}),
we can interpret $\mu_{k}(s)$ as the `exploitation-value' of arm
$k$, and $-L_{k}[V^{*}](s)$ as its `exploration-value' (the marginal
reduction to future regret from pulling arm $k$), so the `overall-value'
from pulling arm $k$ is $\mu_{k}(s)-L_{k}[V^{*}](s)$. Hence, (\ref{eq:PDE_optimal_bayes_risk_General})
describes an exploration-exploitation tradeoff: the regret payoffs
are one of $\{\mu^{\max}(s)-\mu_{k}(s)\}_{k}$ and always greater
than $0$, but as $\{q_{k}\}_{k}$ increase, the posterior collapses
to a point, in which case one chooses the optimal arm with certainty
and the instantaneous regret $\varpi(s):=\min_{k}\{\mu^{\max}(s)-\mu_{k}(s)\}$
becomes $0$. The DM thus faces a tradeoff between exploration, i.e,
pulling the arm enough times to increase $q_{k}$ and thereby reduce
$\varpi(s)$ in the future, and exploitation, i.e., choosing the best
action, $\argmax_{k}\mu_{k}(s)$, at the present. 

If a classical, i.e., twice continuously differentiable, solution,
$V^{*}(\cdot)$, to PDE (\ref{eq:PDE_optimal_bayes_risk_General})
exists, the optimal Bayes policy is $\pi^{*}(s)=\argmin_{k}\{L_{k}[V^{*}](s)-\mu_{k}(s)\}$,
i.e., it pulls the arm with the highest `overall-value'. While a classical
solution to (\ref{eq:PDE_optimal_bayes_risk_General}) is generally
impossible, one can always construct measurable policies whose Bayes
risk is arbitrarily close to $V^{*}(\cdot)$. One such construction
is provided in Section \ref{subsec:Optimal-and-approximately}. 

\subsubsection{A special case: one-armed bandits\label{subsec:A-special-case:}}

The one-armed bandit is a special case of the MAB problem with two
arms and with arm $0$ corresponding to a known outside option. We
normalize the reward from the outside option to $0$, i.e., $\mu_{0}=0$
and $\sigma_{0}=0$. The set of sufficient statistics can then be
reduced to $s\equiv\{x(t):=x_{1}(t),q(t):=q_{1}(t),t\}$. Let $\mu:=\mu_{1}$
and $\sigma^{2}:=\sigma_{1}$ denote the mean and variance of arm
1. For Bayesian analysis, we place a prior $m_{0}$ on the unknown
$\mu$. The PDE characterization of minimal Bayes risk, $V^{*}$,
then simplifies to
\begin{align}
\partial_{t}V^{*}+\mu^{+}(s)+\min\left\{ -\mu(s)+L[V^{*}](s),0\right\}  & =0\ \textrm{if }t<1,\label{eq:PDE optimal bayes risk}
\end{align}
with the terminal condition $V^{*}(s)=0$ if $t=1$, where $\mu^{+}(s):=\mathbb{E}_{\mu\vert s}[\max\{\mu,0\}]$,
$\mu(s):=\mathbb{E}_{\mu\vert s}[\mu]$ and $L[f]:=\partial_{q}f+\mu(s)\partial_{x}f+\frac{1}{2}\sigma^{2}\partial_{x}^{2}f$.
We make frequent reference to one-armed bandits in what follows as
the reduced state space enables us to describe our theoretical results
with minimal notational overhead while still preserving the essential
conceptual features of the MAB problem. 

For one-armed bandits, it is easily verified that the optimal policy
is a retirement policy, i.e., if the DM did not pull the arm at some
time $t$, she will not do so at any other time in the future.\footnote{This because the posterior remains unchanged while the arm is not
being pulled.} Also, it needs to be non-decreasing in $x$. These properties imply
$\pi^{*}(s)$ is of the form $\mathbb{I}\left\{ x>f(q,t)\right\} $. 

\subsection{Comparison with existing methods\label{subsec:Comparison-with-existing}}

\begin{figure}
\includegraphics[height=5cm]{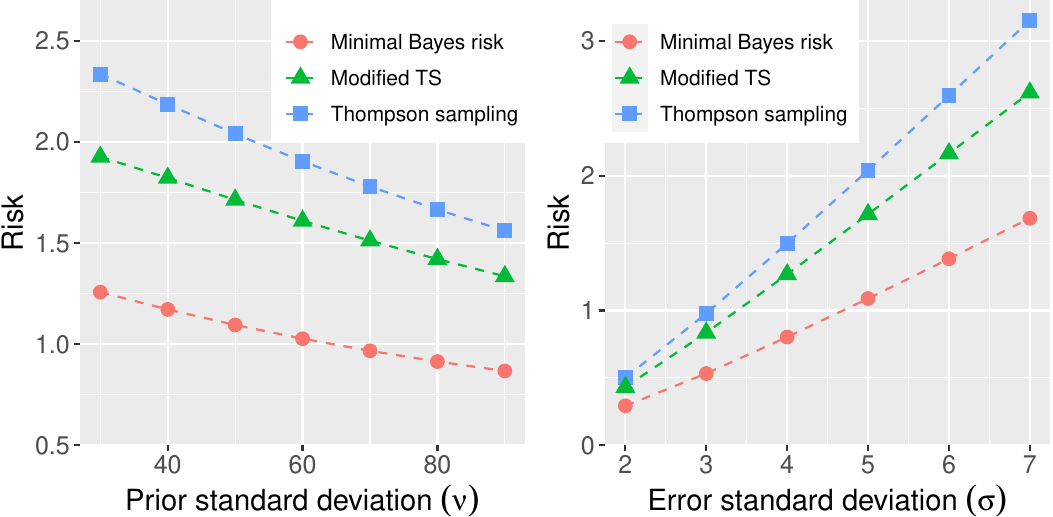}\\

\begin{raggedright}
{\scriptsize Note: The default parameter values are $\mu_{0}=0$, $\nu=50$
and $\sigma=5$. Modified TS refers to the Thompson sampling rule
modified so that $\pi=1$ whenever $x\ge0$. }{\scriptsize\par}
\par\end{raggedright}
\caption{Asymptotic risk of various policies under one-armed bandits\label{fig:Policy plots-1-1}}
\end{figure}

Perhaps the two most commonly used algorithms for MAB problems are
Thompson Sampling (TS) and UCB. The TS rule is $\pi_{k}^{\textrm{ts}}(s)=\mathbb{P}\left(\mu_{k}\ge\max_{k^{\prime}}\mu_{k^{\prime}}\vert s\right)$
and its asymptotic Bayes risk can be obtained by solving (\ref{eq:PDE- bayes risk}).
Figure \ref{fig:Policy plots-1-1} compares this with the corresponding
minimal Bayes risk for one-armed bandits under Gaussian priors, obtained
by solving PDE (\ref{eq:PDE optimal bayes risk}). For the numerical
comparison, we set the prior mean to $0$ and vary the prior and error
variances, $\nu^{2}$ and $\sigma^{2}$. To interpret the ranges of
$\nu^{2},\sigma^{2}$, note that the unscaled prior variance is $\nu^{2}/n$
(which is why $\nu>\sigma$) and all the policies considered here
are invariant to $\nu/\sigma$; for reference, our empirical application
in Section \ref{subsec:Two-armed-bandits} uses $\nu/\sigma\approx15$.
TS is inferior to the optimal Bayes policy across all parameter values
and substantially so - its Bayes risk is generally twice as high. 

\begin{figure}
\includegraphics[height=4cm]{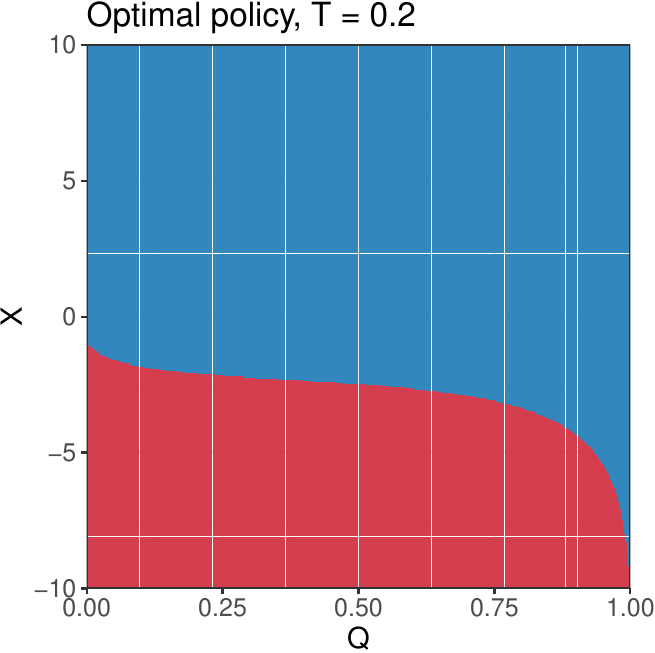}~\includegraphics[height=4cm]{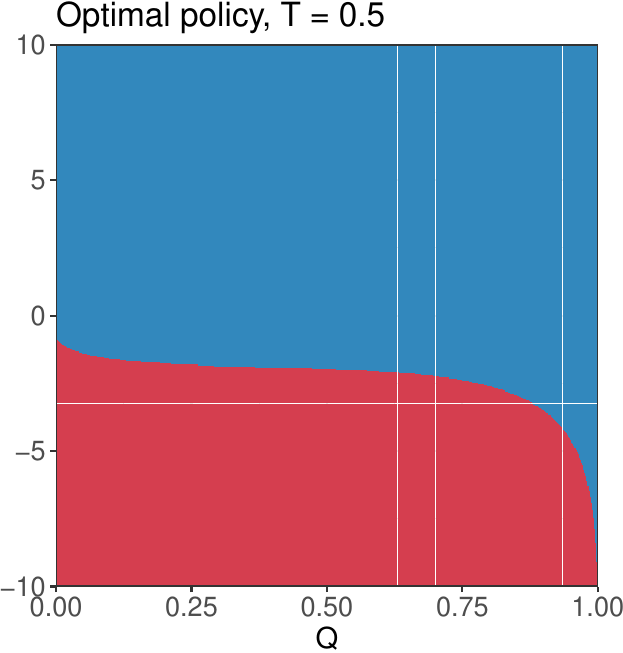}~\includegraphics[height=4cm]{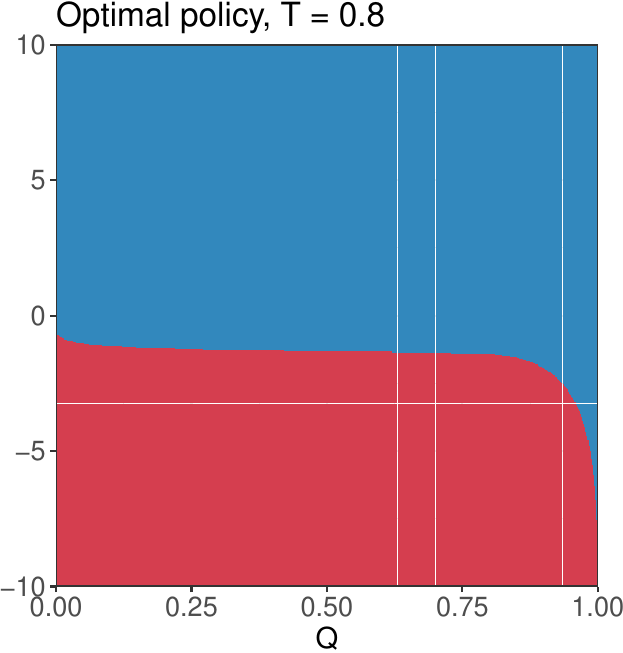}

\vspace*{0.2cm}$\quad$\includegraphics[height=4cm]{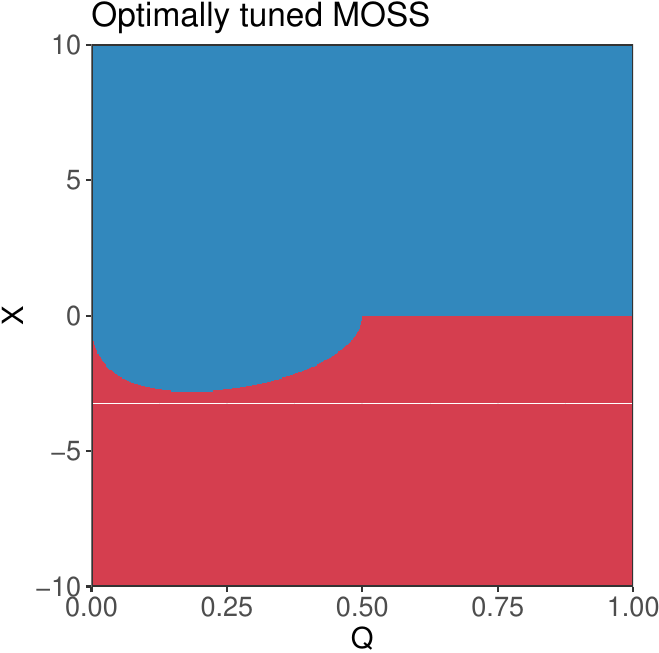}$\ $\includegraphics[height=4cm]{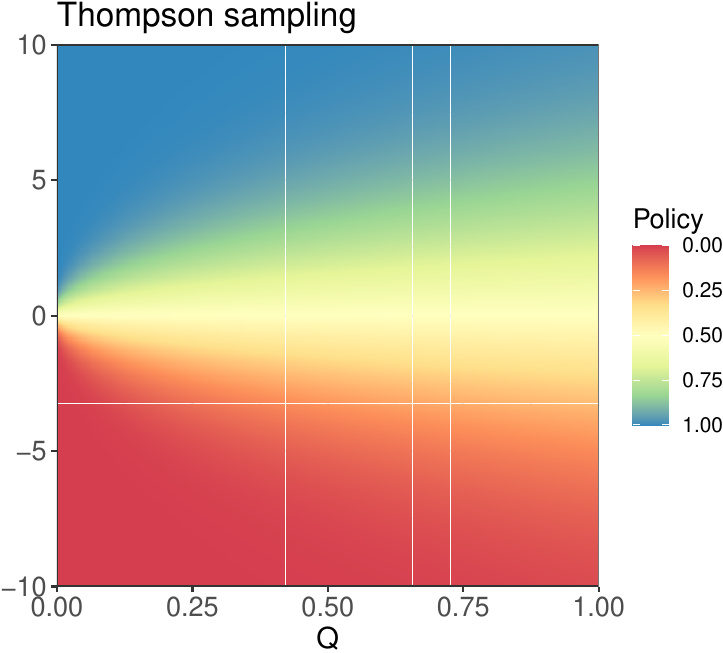}\\

\begin{raggedright}
{\scriptsize Note: The parameter values are $\mu_{0}=0$, $\nu=50$
and $\sigma=5$. Blue corresponds to pulling the arm while red corresponds
to not pulling it. The TS and MOSS policies do not change over time.
In comparing the optimal policy with the others, note that the regions
where $Q\ge T$ are not actually attainable (though the optimal policy,
as plotted, remains well-defined). }{\scriptsize\par}
\par\end{raggedright}
\caption{Policy maps for a one-armed bandit\label{fig:Policy plots-2}}
\end{figure}

Figure \ref{fig:Policy plots-2} plots the associated optimal policy
rule, under the parameter values $(\nu=50,\sigma=5)$, as a function
of $x,q$ at a few different snapshots in time. As conjectured earlier,
it is of the form $\pi^{*}(s)=\mathbb{I}\{x\ge f(q,t)\}$ with\textcolor{blue}{{}
$f(\cdot)$ }increasing in $t$ and decreasing in $q$. The policy
recommends pulling the arm for some $x<0$, even though this indicates
negative expected rewards, $\mu(s)$. This is an example of exploration.
The extent of exploration, i.e., the values of $x$ for which $\pi^{*}(s)=1$,
declines over time. The figure also plots a heat-map of, $\pi^{\textrm{ts}}(\cdot)$,
the TS rule. The reason why $\pi^{\textrm{ts}}(\cdot)$ is inferior
is simple: it over-explores. TS continuously attempts to trade-off
exploration and exploitation against each other, but these motives
are not always at odds. Indeed, when $x\ge0$, pulling the arm is
optimal for both exploitation (since the posterior mean is positive)
\textit{and} exploration. A simple modification to the TS rule, that
sets $\pi=1$ whenever $x\ge0$ but is otherwise equivalent to TS,
thus delivers 15-20\% lower Bayes risk under our one-armed bandit
setups, as Figure \ref{fig:Policy plots-1-1} illustrates. More generally,
for MABs, we can improve the Bayes risk of TS by modifying it as follows:
whenever there exist arms $k,k^{\prime}$ such that $\mu_{k}(s)<\mu_{k^{\prime}}(s)$
and $\textrm{Var}[\mu_{k}\vert s]\le\textrm{Var}[\mu_{k^{\prime}}\vert s]$,
we should transfer all the probability that TS assigns to $k$ to
$k^{\prime}$ (this can be repeated for all $k,k^{\prime}$). 

On the other hand, the optimal policy shares a number of similarities
with UCB algorithms. Note that $\hat{\mu}:=x/q\sqrt{n}$ is the MLE
estimator of the sample mean $\mu/\sqrt{n}$. Then, defining $F(q,t)=-f(q,t)/q$,
we find that the optimal policy for one-armed bandits has the form
$\pi^{*}=\mathbb{I}\left\{ \hat{\mu}+F(q,t)/\sqrt{n}\ge0\right\} $.
We can thus interpret $F(q,t)/\sqrt{n}$ as the optimal confidence
width in that setting. More generally, for the MAB problem, if the
prior is normal and independent across arms, $\mu_{k}(s)$ is a function
only of $x_{k},q_{k}$ and monotonically increasing in $\hat{\mu}_{k}=x_{k}/q_{k}\sqrt{n}$.
We can then rewrite the optimal policy in the UCB form $\pi^{*}=\argmax_{k}\{\hat{\mu}_{k}+F_{k}(s)/\sqrt{n}\}$,
even if, unlike a typical UCB, $F_{k}(\cdot)$ depends on all of $s$
instead of just $x_{k},q_{k}$. 

For correlated priors, however, the optimism principle fails and the
optimal policy may be very different from UCB. Indeed, it can then
even be optimal to pull an arm with a lower UCB than the others if
it is highly informative about the common parameter. The reason for
this difference is that while uncertainty over $\mu_{k}$ is just
one of many factors that determine the exploration-value, $-L_{k}[V^{*}](\cdot)$,
of an arm $k$, the confidence width used in UCBs is determined solely
by it. 

The vanilla UCB policy uses the confidence width $\sqrt{2\sigma^{2}\ln(1/\delta)/nq_{k}}$
for each $k$, where $\delta$ is a tuning parameter. But this is
far from optimal. For two-armed bandits, \citet{kalvit2021closer}
show that it converges (under diffusion asymptotics) to a fixed (i.e.,
non-adaptive) allocation rule that is independent of $\delta,\mu_{1},\mu_{0}$.
Thus, this class of UCBs over-explore, and their minimax rate of regret
is $O(\sqrt{\log n/n})$. 

The minimax optimal rate, $O(n^{-1/2})$, can be regained with a more
refined confidence width, as evidenced by the MOSS algorithm which
uses $\sqrt{2\sigma^{2}g(q_{k})/nq_{k}}$, where $g(q)\propto\ln(Kq)^{-1}$.
In fact, by \citet{lai1987adaptive}, this is an approximation, as
$q\to0$, of the optimal width, $F(q,t)/\sqrt{n}$, in the one-armed
setting under a flat prior (i.e., when $\nu^{2}\to\infty$). More
generally, with multiple arms and independent Gaussian priors, Section
\ref{sec:Computation-and-simulations} shows that while the standard
implementation of MOSS performs a lot worse than the optimal policy,
an optimally tuned MOSS, that uses the confidence width $\sqrt{\gamma\sigma^{2}g(q_{k})/nq_{k}}$,
comes close to attaining the risk lower bound. Our proposal for the
optimal $\gamma$ here is to choose the value that minimizes the local
asymptotic Bayes risk of MOSS under the given prior. Such an optimal
$\gamma$ is, however, highly sensitive to the prior parameters. Figure
\ref{fig:Policy plots-2} shows that the optimally tuned MOSS $(\gamma^{*}\approx1.72$)
under the parameter values $(\nu=50,\sigma=5)$ shares broad similarities
with the optimal policy, albeit being independent of time.

\subsection{Minimax risk\label{subsec:Minimax-risk}}

Following \citet{wald1945statistical}, we define minimax risk as
the value of a two player zero-sum game played between nature and
the DM. Nature's action consists of choosing a prior, $m_{0}\in\mathcal{P}$,
over $\bm{\mu}$, while the DM chooses the policy rule $\pi$. The
minimax risk $\bar{V}^{*}$ is defined as 
\begin{equation}
\bar{V}^{*}=\sup_{m_{0}\in\mathcal{P}}V^{*}(0;m_{0})=\sup_{m_{0}\in\mathcal{P}}\inf_{\pi\in\Pi}V_{\pi}(0;m_{0}),\label{eq:def of minimax risk}
\end{equation}
where $V_{\pi}(0;m_{0})$ and $V^{*}(0;m_{0})$ denote the ex-ante
Bayes risk under a policy $\pi$, and the minimal Bayes risk, when
the prior is $m_{0}$. The equilibrium action of nature is termed
the least-favorable prior, and that of the DM, the minimax policy.
Under a minimax theorem, which holds if there is a Nash equilibrium
to the game (with proper priors), the $\sup$ and $\inf$ operations
in (\ref{eq:def of minimax risk}) can be interchanged, so that 
\begin{equation}
\sup_{m_{0}\in\mathcal{P}}\inf_{\pi\in\Pi}V_{\pi}(0;m_{0})=\inf_{\pi\in\Pi}\sup_{m_{0}\in\mathcal{P}}V_{\pi}(0;m_{0})=\inf_{\pi\in\Pi}\sup_{\bm{\mu}}V_{\pi}(0;\bm{\mu}).\label{eq:game equivalence}
\end{equation}
Here, $V_{\pi}(0;\bm{\mu})$ denotes the frequentist risk of a policy
$\pi$ when the local parameter is $\bm{\mu}$. The last term, $\inf_{\pi\in\Pi}\sup_{\bm{\mu}}V_{\pi}(0;\bm{\mu})$,
is perhaps the more common definition of minimax risk. Thus, by (\ref{eq:game equivalence}),
the problem of computing minimax risk reduces to that of computing
Bayes risk under the least favorable prior. 

For one-armed bandits, we conjecture, and verify numerically by solving
the two player game, that the least favorable prior, $m_{0}^{*}$,
involves only two support points at $\{\text{\underbar{\ensuremath{\mu}}},\bar{\mu}\}$,
with $\text{\underbar{\ensuremath{\mu}}}<0$ and $\bar{\mu}>0$. This
is because both low and high values of $\vert\mu\vert$ are associated
with low risk, the former by definition, and the latter because the
DM quickly learns to always pull or never pull the arm. Indeed, for
$\sigma=1$, it turns out $m_{0}^{*}$ has a two point support at
$\text{\underbar{\ensuremath{\mu}} \ensuremath{\approx} -2.5 }$ and
$\bar{\mu}\approx1.7$ with $m_{0}^{*}(\bar{\mu})\approx0.415$. In
fact, it suffices to solve the game under $\sigma=1$ as we can always
rescale the rewards to have unit variance (the risk comparisons are
invariant to scale transformations). 

Based on the above analysis, we find that the sharp lower bound on
the (unscaled) minimax risk of any one-armed bandit algorithm is given
by $0.373\sigma\sqrt{n}$. By contrast, existing theoretical results
only demonstrate a $\sigma\sqrt{n}$ rate. 

Computing the least favorable prior when there are more than two arms
is a lot more demanding. We conjecture, however, that it has a discrete
support. 

\section{Formal properties under gaussian rewards\label{sec:Formal-properties} }

For simplicity, the results in this section are stated for the one-armed
bandit problem. However, all our results extend to the general MAB
problem with straightforward adjustments to the proofs, see Appendix
\ref{subsec:Multi-armed-bandits}.

\subsection{Existence and uniqueness of PDE solutions}

Equation (\ref{eq:PDE optimal bayes risk}) describes a nonlinear
$2^{\textrm{nd}}$-order PDE. It is well known that such PDEs do not
admit classical, i.e., twice continuously differentiable, solutions.
Instead, the relevant weak solution concept is that of a viscosity
solution (\citealt{crandall1992user}). 

\begin{thm}\label{theorem 1} \textbf{\textit{\citep[Theorem A.1]{barles2007error}}}
Suppose $\mu^{+}(\cdot),\mu(\cdot)$ are $\gamma$-H{\"o}lder continuous
for some $\gamma>0$. Then there exists a unique, $\gamma$-H{\"o}lder
continuous viscosity solution to PDE (\ref{eq:PDE optimal bayes risk}).
\end{thm}

\subsection{Convergence to the PDE solution\label{subsec:Convergence-to-PDE}}

In Section 2, we provided a heuristic derivation of PDE (\ref{eq:PDE optimal bayes risk}).
For a formal result, one would need to prove that a discrete analogue,
$V_{n}^{*}(\cdot)$, of $V^{*}(\cdot)$, defined for a fixed $n$,
converges to $V^{*}(\cdot)$ as $n\to\infty$. Define $\mathbb{I}_{n}=\mathbb{I}\{t\le1-1/n\}$
and $Y_{i}$ as the $i$-th realization of the rewards (corresponding
to the $i$-th pull of the arm). Let $V_{n}^{*}(\cdot)$ denote the
solution to the recursive equation 
\begin{align}
V_{n}^{*}\left(x,q,t\right) & =\min_{\pi\in[0,1]}\mathbb{E}\left[\left.\frac{\mu^{+}(s)-\pi\mu(s)}{n}+\mathbb{I}_{n}\cdot V_{n}^{*}\left(x+\frac{A_{\pi}Y_{nq+1}}{\sqrt{n}},q+\frac{A_{\pi}}{n},t+\frac{1}{n}\right)\right|s\right];\nonumber \\
 & \quad\textrm{if }t<1,\nonumber \\
V_{n}^{*}\left(x,q,t\right) & =0\quad\textrm{if }t=1.\label{eq:discrete approximation}
\end{align}
In (\ref{eq:discrete approximation}), $A_{\pi}\sim\textrm{Bernoulli}(\pi)$,
and the expectation is a joint one over $(Y_{nq+1},A_{\pi})$ given
$s$. Existence of a unique $V_{n}^{*}(\cdot)$ follows by backward
induction. Clearly, $V_{n}^{*}(\cdot)$ is the minimal Bayes risk
in the fixed $n$ setting under Gaussian rewards. We can thus interpret
(\ref{eq:discrete approximation}) as a discrete approximation to
PDE (\ref{eq:PDE optimal bayes risk}). As such, it falls under the
abstract framework of \citet{barles1991convergence} for showing convergence
to viscosity solutions. An application of their techniques proves
the following result (the proof is in Appendix \ref{subsec:Proof-of-Theorem-2}):
Denote $\varpi(s):=\min\left\{ \mu^{+}(s)-\mu(s),\mu^{+}(s)\right\} $. 

\begin{thm}\label{Thm: Convergence to minimal PDE} Suppose $\mu^{+}(\cdot),\mu(\cdot)$
are $\gamma$-H{\"o}lder continuous, $\sup_{s}\varpi(s)<\infty$
and the prior $m_{0}$ is such that $\mathbb{E}[\vert\mu\vert^{3}\vert s]<\infty$
at each $s$. Then, as $n\to\infty$, $V_{n}^{*}(\cdot)$ converges
locally uniformly to $V^{*}(\cdot)$, the unique viscosity solution
of PDE (\ref{eq:PDE optimal bayes risk}). \end{thm}

The assumptions are satisfied for Gaussian priors. Note also that
the theorem is proved without appealing to (\ref{eq:diffusion limit}).
In Appendix \ref{subsec:Rates-of-convergence} we derive a coarse
upper bound on the rate of convergence of $V_{n}^{*}(\cdot)$ to $V^{*}(\cdot)$
and provide simulation evidence suggesting that the quality of the
approximation is quite good in practice.

\subsection{Piece-wise constant policies and batched bandits\label{subsec:Optimal-and-approximately}}

While we are not able to characterize the optimal Bayes policy in
closed form, it is possible to construct (Lebesgue) measurable policies
whose Bayes risk is arbitrarily close to $V^{*}(\cdot)$. One way
to do so is using piece-wise constant policies. In fact, a bandit
experiment with such a policy is equivalent to a batched bandit experiment,
where the data is forced to be considered in batches. The results
in this section thus give an upper bound on the welfare loss due to
batching.

Let $\Delta t$ denote a small time increment, and $\mathcal{T}_{\Delta t}:=\{t_{1},\dots,t_{L}\}$
a set of grid points for time, where $t_{1}=0$, $t_{L}=1$ and $t_{l}-t_{l-1}=\Delta t$
for all $l$. The optimal piece-wise constant policy, $\pi_{\Delta t}^{*}:\mathcal{X}\times\mathcal{Q}\times\mathcal{T}_{\Delta t}\mapsto\{0,1\}$,
is allowed to change only at the time points on the grid $\mathcal{T}_{\Delta t}$.
In particular, suppose that $x=x_{l}$ and $q=q_{l}$ at the grid
point $t=t_{l}$. Then one computes $\pi_{\Delta t}^{*}(x_{l},q_{l},t_{l})\in\{0,1\}$
and holds this policy value fixed until the next time point $t_{l+1}$.
Define $V_{\Delta t,l}^{*}(x,q)$ as the Bayes risk, in the diffusion
regime, at state $(x,q,t_{L-l})$ under $\pi_{\Delta t}^{*}(\cdot).$
We then have the following recursion for $V_{\Delta t,l}^{*}(x,q)$:
\begin{align}
V_{\Delta t,l+1}^{*}(x,q) & =\min\left\{ S_{\Delta t}\left[V_{\Delta t,l}^{*}\right](x,q),V_{\Delta t,l}^{*}(x,q)+\Delta t\cdot\mu^{+}(x,q)\right\} ,\ l=0,\dots,L-1,\nonumber \\
V_{\Delta t,0}^{*}(x,q) & =0,\label{eq:piecewise-constant policy}
\end{align}
where the operator $S_{\Delta t}[\phi](x,q)$ denotes the solution
at $(x,q,\Delta t)$ of the linear second order PDE 
\begin{align}
-\partial_{t}f(s)+\mu^{+}(s)-\mu(s)+L[f](s) & =0,\ \textrm{if }t>0;\quad f=\phi,\ \textrm{if }t=0.\label{eq:linear PDE}
\end{align}

The following theorem assures that $V_{\Delta t,l}^{*}(\cdot)$ can
be made arbitrarily close to $V^{*}(\cdot,\cdot,t_{L-l})$ by letting
$\Delta t\to0$. 

\begin{thm}\label{Thm: Approximate control}\textbf{\textit{\citep[Theorem 2.1]{jakobsen2019improved}}}
Suppose $\mu^{+}(\cdot),\mu(\cdot)$ are Lipschitz continuous. Then,
there exists $C<\infty$ that depends only on the Lipschitz constants
of $\mu^{+}(\cdot),\mu(\cdot)$ such that $0\le\max_{l}\left\{ V_{\Delta t,l}^{*}(\cdot)-V^{*}(\cdot,t_{L-l})\right\} \le C(\Delta t)^{1/4}$
uniformly over $\mathcal{X}\times\mathcal{Q}$.\end{thm}

Note that $\pi_{\Delta t}^{*}(\cdot)$ is not required to converge
to some measurable $\pi^{*}(\cdot)$ as $\Delta t\to0$. Still, we
can employ $\pi_{\Delta t}^{*}(\cdot)$ in the fixed $n$ setting:
to apply, one simply sets $t=\left\lfloor i/n\right\rfloor $, where
$i$ is the current period. The following theorem asserts that employing
$\pi_{\Delta t}^{*}(\cdot)$ in this manner results in a Bayes risk
that is arbitrarily close to $V^{*}(0)$. 

\begin{thm}\label{Thm: Approximate control-2}Suppose $\mu^{+}(\cdot),\mu(\cdot)$
are Lipschitz continuous and $\sup_{s}\mu^{+}(s)<\infty$. Then, for
any fixed $\Delta t$, $\lim_{n\to\infty}\left|V_{\pi_{\Delta t}^{*},n}(0)-V^{*}(0)\right|\le C(\Delta t)^{1/4}$
. \end{thm}

\section{Algorithms and empirical illustrations\label{sec:Computation-and-simulations}}

\subsection{Algorithms}

We provide two empirical illustrations of bandit experiments to show
how our methods translate to real world practice. The first application
solves PDE (\ref{eq:PDE_optimal_bayes_risk_General}) using a finite-difference
(FD) scheme, which is very accurate but scales poorly with the number
of arms, while the second uses a Monte-Carlo method, which is less
accurate but scales linearly with the number of arms. The FD algorithm
is discussed in Appendix \ref{sec:Details-on-computation}. Here we
focus on the Monte-Carlo algorithm as it is arguably more useful in
practice with multiple arms. 

Algorithm 1 provides the pseudo-code for the Monte-Carlo method. The
basic elements of this approach are well-known and widely used for
solving PDEs of the HJB kind; our specific implementation is similar
to Approximate Value Iteration (\citealp{munos2008finite}). The general
steps are the following: (1) we discretize time into periods of length
$\Delta t$, (2) at each period $j$, we randomly draw a vector of
state variables, (3) starting from $j=T-1$ and going backwards, and
using the random draw of state variables at period $j$ as input,
we use forward simulation and prediction methods to obtain an estimate
of the action-value function, $V_{k,j}(\cdot)$, at period $j$ given
the (previously obtained) estimate of the value function, $\min_{k}V_{k,j+1}(\cdot)$,
in period $j+1$.\textcolor{blue}{{} }Care must be taken to ensure that
the distribution of state variables drawn is close to what would have
been observed under the optimal policy; as prediction methods minimize
expected MSE, we would like this expectation to be close to that induced
by the optimal policy. Hence, we draw the state variables using a
pilot policy, typically Thompson Sampling, and then run the algorithm
once again with the updated policy.\footnote{In principle, one could iterate this, but we found it to be unnecessary
in practice.} 

\noindent\begin{minipage}{\textwidth}
\renewcommand\footnoterule{}  
\begin{algorithm}[H]
\caption{Monte-Carlo algorithm for solving PDE (\ref{eq:PDE_optimal_bayes_risk_General})}\label{alg:Monte-Carlo alg}
\footnotesize
\begin{algorithmic}[1]
\Require{$K$ ($\#$ arms), $\Delta t$ (step size), $B$, $M$ (simulation draws), $T := 1/\Delta t$, $\pi^{\textrm{init}}$ (pilot policy)}
\vspace{0.3em}
\State Simulate $b=1,\dots,B$ sample paths $s^{(b)}(\cdot):= \{ x_{k}^{(b)}(\cdot),q_{k}^{(b)}(\cdot): k = 1,\dots ,K \}$ from $\pi^{\textrm{init}}$  
\State Save values at discrete time points:
\[
(\forall j=1,\dots,T):\ s_{j}^{(b)} =s^{(b)}( j\Delta t),\ x_{k,j}^{(b)}=x_{k}^{(b)}( j\Delta t),\ q_{k,j}^{(b)}=q_{k}^{(b)}(j\Delta t)
\]
\State Initialize period $T-1$ action-value and value functions:
\begin{align*} 
(\forall k):\ V_{k,T-1}(\cdot) & = \mu^{\textrm{max}}(\cdot)-\mu_{k}(\cdot)\\
V_{T-1}^{*}(\cdot) & = \min_{k}V_{k,T-1}(\cdot)
\end{align*}
\For{$j=T-2,T-3,\dots,1$:}
\State{$(\forall b,k)$: Compute $z_{k}^{(b)}$ as sample mean of $M$ simulation draws of
 \begin{align*} 
	& V_{k,j+1}^{*}\left(\left\{ x_{l,j}^{(b)}+ \mathbb{I}\{l=k\} \cdot e_{k,j}^{(b)}, \ q_{l,j}^{(b)}+ \mathbb{I}\{l=k\} \cdot \Delta t \right\} _{l=0}^{K-1}\right),\ \textrm{where }\\
    & e_{k,j}^{(b)} \sim \mathcal{N} \left(\mu_k \left(s_{j}^{(b)}\right) \cdot \Delta t, \sigma_{k} \cdot \Delta t \right)
\end{align*} 
}
\State{$(\forall k)$: Run prediction model of $\{z_{k}^{(b)}\}_{b=1}^{B}$ on $\{s_{j}^{(b)}\}_{b=1}^{B}$, output prediction function $\hat{f}_{k,j}(\cdot)$} 
\State{$(\forall k)$: Return as function
\begin{align*}
V_{k,j}(\cdot) & =\mu^{\textrm{max}}(\cdot)-\mu_k(\cdot)+\hat{f}_{k,j}(\cdot)\\
V_{j}^{*}(\cdot) & =\min_{k}V_{k,j}(\cdot)
\end{align*}
}
\EndFor
\State Return policy function $\pi(\cdot, t)=\argmin_{k}V_{k,\left\lfloor t/\Delta t\right\rfloor }(\cdot)$
\State \textbf{Repeat:} steps 1-9 with new pilot policy $\pi^{\textrm{init}}=\pi$
\end{algorithmic}
\footnotetext{Notes: Step 6 requires a prediction method, e.g., Random Forest. The algorithm assumes oracle knowledge of $\mu_k(\cdot), \mu^\textrm{max}(\cdot)$, which are policy independent and computed from the posterior (\ref{eq:posterior}). For Gaussian priors, closed-form expressions exist; otherwise, they can be computed numerically via MCMC/Laplace approximations, akin to the procedure for TS (which employs similar terms).}
\end{algorithm} 
\end{minipage}

\vspace{1.7em}Computation is generally fast; for the second empirical
illustration with 2 arms and Gaussian priors, it takes about 40 minutes.
As for the minimax policy under one-armed bandits, used in our first
application, it only needs to be computed once, as we already did
here. In future applications it can be employed straightaway after
simply rescaling the rewards to have unit variance.\footnote{Even with multiple arms, game-theoretic reasoning and the scale invariance
of Brownian motion suggests the minimax policy only needs to be computed
once under the case $\sigma_{k}=1\ \forall\ k$.}

\subsection{A one-armed bandit\label{subsec:A-one-armed-bandit}}

This illustration is based on a Google Analytics blog example on website
optimization.\footnote{The webpage describing the simulation study can be accessed \href{https://analytics.googleblog.com/2013/01/multi-armed-bandit-experiments.html\#:~:text=Google\%20Analytics\%20uses\%20a\%20multi,updated\%20as\%20the\%20experiment\%20progresses}{here}.}
Suppose that we currently have a website with a known conversion rate
of $p_{0}=0.05$.\footnote{The conversion rate is defined as the percentage of users who have
completed a desired action, e.g., clicking an ad.} We would like to experiment with a new version of the website whose
conversion rate, $p$, is unknown. Let $\tilde{Y}_{i}\sim\textrm{Bernoulli}(p)$
denote the outcome variable under the new website. As our setup normalizes
the reward from the known option to $0$, we redefine the outcomes
as $Y_{i}=(\tilde{Y}_{i}-p_{0})/\sqrt{p_{0}(1-p_{0})}$.\textcolor{blue}{{}
}Though $Y_{i}$ is not normally distributed, Section \ref{sec:General-parametric-models}
shows that the asymptotically sufficient statistics, $x(t),q(t)$,
are the same as in the normal setting with $\sigma^{2}=1$, and the
optimal policies also remain unchanged.\textcolor{blue}{{} }We report
results for different sample sizes $n$. For comparison, the blog
example used $n=6600$.

\begin{figure}
\includegraphics[height=5cm]{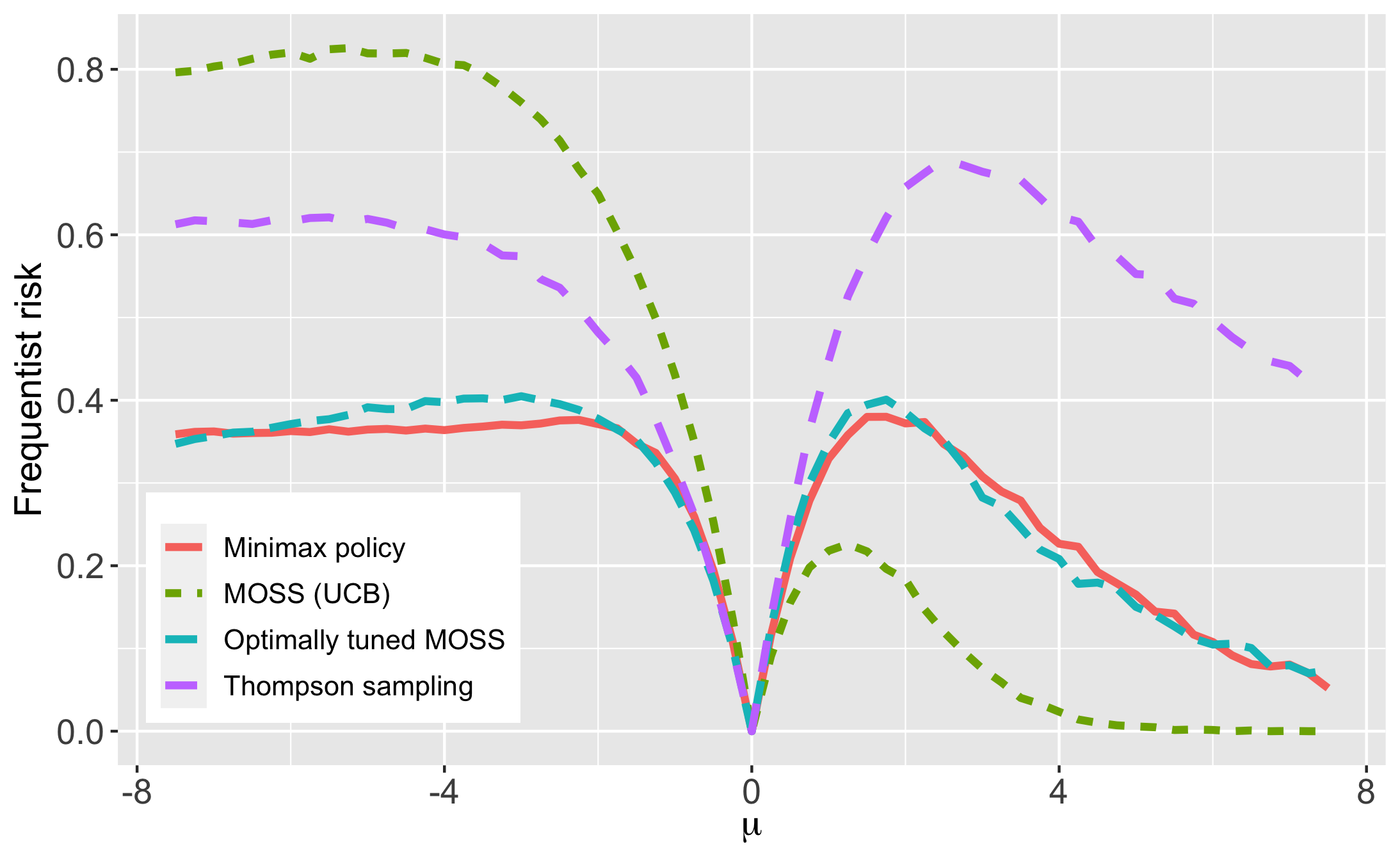}~\includegraphics[height=5cm]{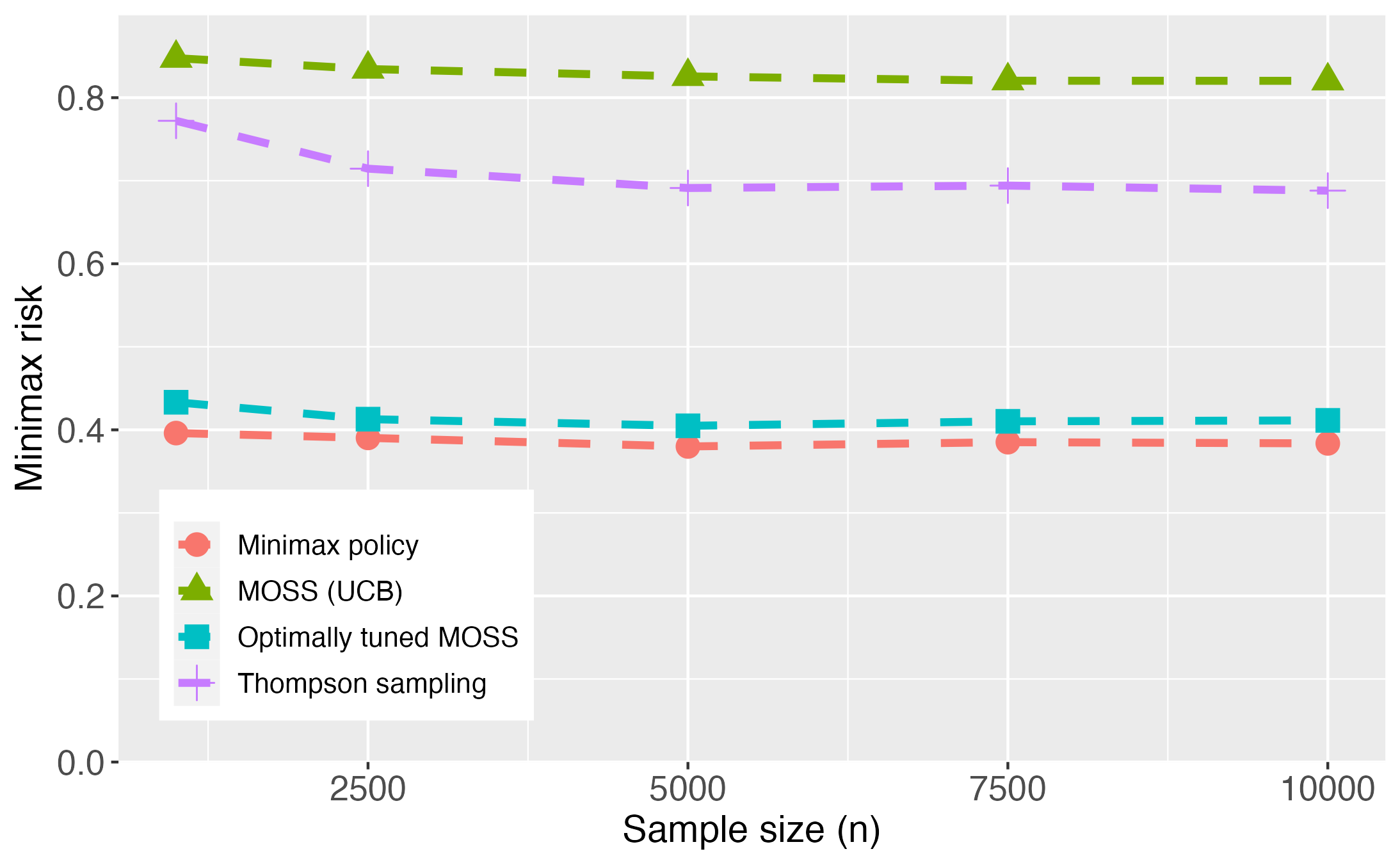}~

\begin{tabular}{>{\centering}p{7.5cm}>{\centering}p{7.5cm}}
{\scriptsize$\ $A: Risk profiles of various policies} & {\scriptsize$\ \ $B: Minimax risk vs n}\tabularnewline
\end{tabular}
\begin{raggedright}
{\scriptsize Note: Panel A shows the frequentist risk profiles of various
policies under $n=5000$. The x-axis represents the scaled mean $\mu=(p-p_{0})\cdot\sqrt{p_{0}(1-p_{0})/n}$
with $p_{0}=0.05$. Panel B shows how the minimax risk of the various
policies changes with $n$. For reference, the minimax lower bound
is $0.373$.}{\scriptsize\par}
\par\end{raggedright}
\caption{Empirical illustration - one-armed bandit\label{fig:Illustration 1}}
\end{figure}

For this illustration, we apply the minimax risk criterion, and compare
the minimax optimal estimator with Thompson sampling (TS) and MOSS
(see Section \ref{subsec:Comparison-with-existing}). For TS we employ
a beta-prior centered at $p_{0}$, with the prior variance optimally
tuned to minimize max risk.\footnote{The Google Analytics example employed TS updated every 100 observations.}
For MOSS, we employ two versions: the first, a textbook implementation
as in \citet{lattimore2020bandit}, and the second, an optimally tuned
version as described in Section \ref{subsec:Comparison-with-existing},
with $\gamma$ chosen to minimize max-risk. Figure \ref{fig:Illustration 1},
Panel A displays the frequentist risk profiles of the different policies,
for various values of rescaled mean rewards $\mu=(p-p_{0})\cdot\sqrt{p_{0}(1-p_{0})/n}$,
when $n=5000$ (this equivalent to a range of $[0.027,0.073]$ for
$p$). By way of comparison, the Google Analytics example set $p=0.4$,
which corresponds to $\mu=-3$ in the plot. Compared to the optimal
policy, the minimax risks of TS and the standard MOSS algorithm are
substantially higher, by about 80\% and 110\% respectively. On the
other hand, the optimally tuned MOSS comes within 7-10\% of the minimax
lower bound. These relationships are stable over $n$ as Panel B of
same figure illustrates. 

\subsection{Two-armed bandits\label{subsec:Two-armed-bandits}}

The second illustration is based on experiments conducted by The Washington
Post for selecting between two different images for the headline of
a news article. The goal was to choose the one with the highest click-through
rate (CTR).\footnote{More information on the experiments can be found \href{https://web.archive.org/web/20161013134841/https://developer.washingtonpost.com/pb/blog/post/2016/02/08/bandito-a-multi-armed-bandit-tool-for-content-testing/}{here}.}
Let $p_{0},p_{1}$ denote the CTRs for the two proposals. For this
illustration we employ a Bayesian approach with an independent Gaussian
prior $p_{k}\sim\mathcal{N}(p_{\textrm{ref}},\sigma_{\textrm{ref}}^{2}\nu^{2}/n)$
for $k\in\{0,1\}$, where $\sigma_{\textrm{ref}}:=p_{\textrm{ref}}(1-p_{\textrm{ref}})$
and $n$ is the number of periods of experimentation. In practice,
one would like to set $p_{\textrm{ref}}$ and $\nu^{2}$ based on
prior knowledge of the distribution of CTRs across all the news articles.
In the absence of this information, we set $p_{\textrm{ref}}=0.175$,
which is a typical CTR for media websites, along with $\nu=5$, and
vary $n$ between $1000$ and $5000$. When $n=2500$, our choice
implies that the 95\% range for the prior is $[0.1,0.25]$. For comparison,
in the Washington Post study, the actual CTRs turned out to be $0.117$
and $0.246$. Let $\tilde{Y}^{(k)}\sim\textrm{Bernoulli}(p_{k})$
denote the outcomes (i.e., clicks) under the options $k\in\{0,1\}$;
we rescale them to $Y^{(k)}=(\tilde{Y}^{(k)}-p_{\textrm{ref}})/\sigma_{\textrm{ref}}$.
Section \ref{sec:General-parametric-models} shows that the asymptotically
sufficient statistics $\{x_{k}(t),q_{k}(t)\}_{k=0,1}$ are then the
same as in the normal setting with $\sigma_{k}^{2}=1$, and the optimal
policies also remain unchanged.

\begin{figure}
\includegraphics[height=5cm]{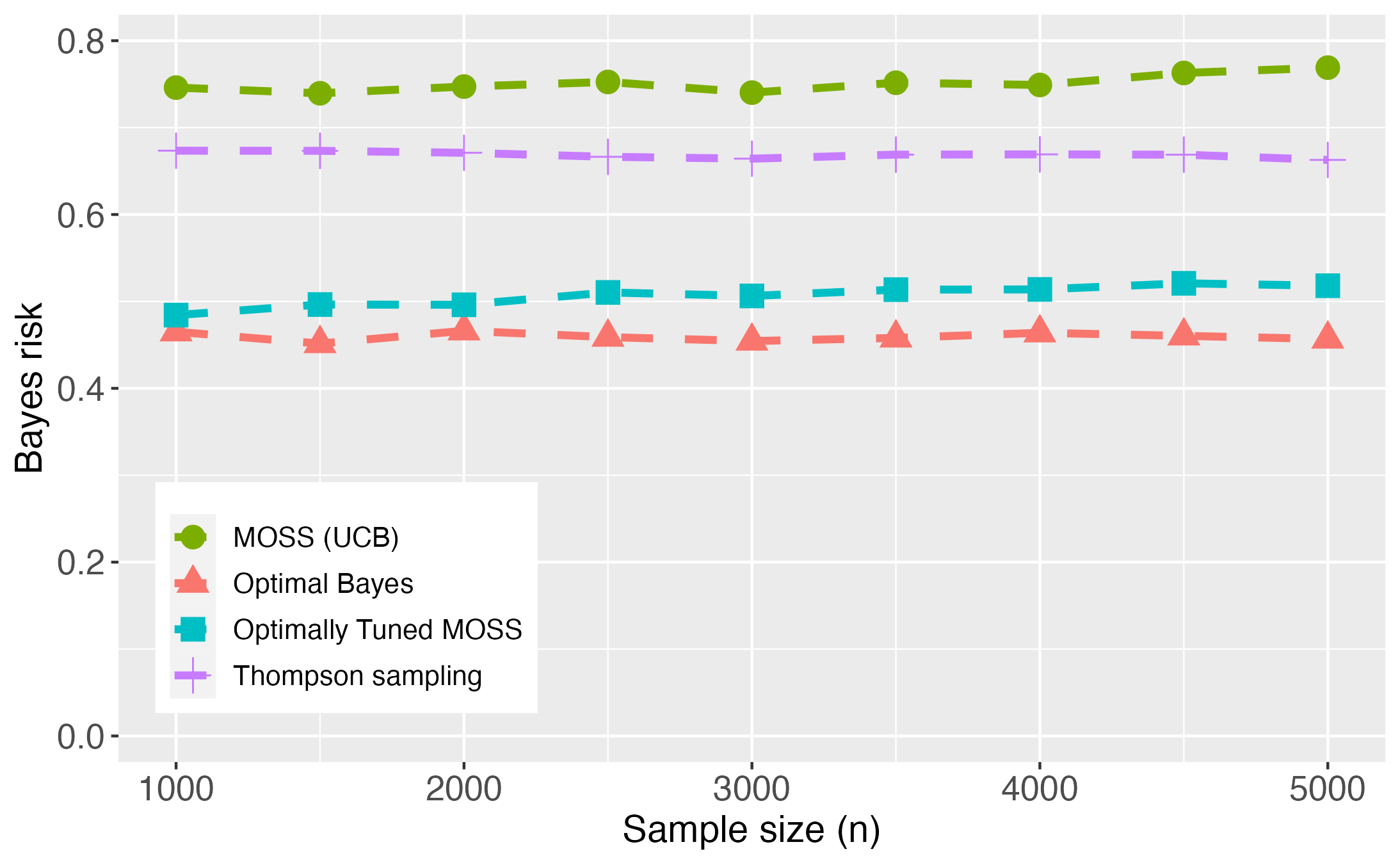}~\includegraphics[height=5cm]{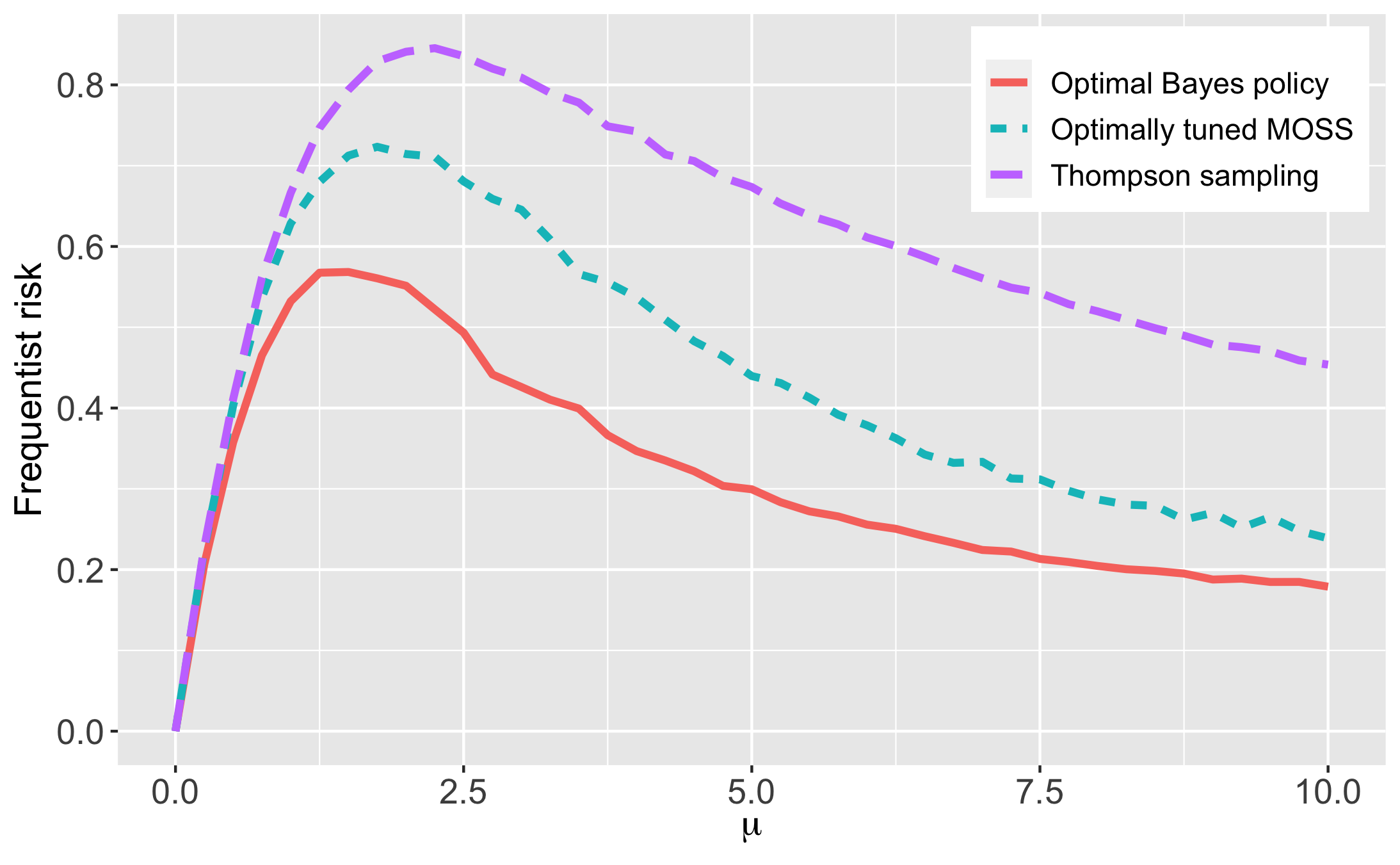}~

\begin{tabular}{>{\centering}p{7.5cm}>{\centering}p{7.5cm}}
{\scriptsize$\ $A: Bayes risk vs n} & {\scriptsize$\ \ $B: Frequentist risk profiles}\tabularnewline
\end{tabular}
\begin{raggedright}
{\scriptsize Note: Panel A shows the Bayes risk of the different algorithms
under various values of $n$. Panel B shows the risk profiles of the
various policies under $(p_{0},p_{1})\equiv(p_{\textrm{ref}}-\mu\sigma_{\textrm{ref}}/\sqrt{n},p_{\textrm{ref}}+\mu\sigma_{\textrm{ref}}/\sqrt{n})$
when $n=3000$, and we we vary $\mu$ between $0$ and $10$.}{\scriptsize\par}
\par\end{raggedright}
\caption{Empirical illustration - two-armed bandits\label{fig:Illustration 1-1}}
\end{figure}

The set of algorithms considered are the optimal Bayes algorithm,
TS (with the Gaussian prior) and MOSS with both the textbook and tuned
implementations. For the tuned version, we set the tuning parameter
to the value that minimizes Bayes risk. Figure \ref{fig:Illustration 1-1},
Panel A plots the Bayes risk of these policies under different $n$.
As in the first illustration, while the risk of TS and the standard
MOSS algorithm is substantially worse that that of the optimal Bayes
policy, the optimally tuned MOSS comes within $15\%$ of the lower
bound on risk. The actual Washington Post study employed a standard
UCB algorithm without any tuning; this performs even worse than MOSS.
Panel B of the same figure plots the frequentist risk profiles of
these policies under $(p_{0},p_{1})=(p_{\textrm{ref}}-\mu\sigma_{\textrm{ref}}/\sqrt{n},p_{\textrm{ref}}+\mu\sigma_{\textrm{ref}}/\sqrt{n})$,
with $n=3000$, and as we vary $\mu$ between 0 and $10$. Setting
$\mu=10$ gives a value of $(p_{0},p_{1})$ that is roughly the same
as that actually observed in the Washington Post study. Atleast for
this class of mean reward values, it is seen that the optimal Bayes
policy uniformly dominates all the existing algorithms.

\subsubsection{Implementation details}

We employ Algorithm 1 with $\Delta t=0.01$, $B=2000$ and $M=50$.\footnote{Setting $M=1$ is also fine and does not make much of a difference
in practice.} For the prediction model, we employ Random Forest (RF) as it is relatively
insensitive to tuning parameter selection.\footnote{We use 250 trees and left $\textrm{mtry}$ at the default value, but
changing these did not change the results.} As an alternative, MARS (multivariate adaptive regression splines)
delivers essentially the same results, but requires more fine-tuning.
In running the RF algorithm, we find that better predictive performance
(in terms of achieving lower prediction error with fewer $B$) could
be achieved by using $\{\mu_{k}(\cdot),q_{k}(\cdot)\}_{k},\mu^{\max}(\cdot)$
as inputs instead of $\{x_{k}(\cdot),q_{k}(\cdot)\}_{k}$; the former
is of course just a nonlinear transformation of the latter. 

\section{General parametric models\label{sec:General-parametric-models}}

We now relax the Gaussian assumption, and suppose that rewards are
distributed according to some parametric model $P_{\theta}$, with
$\theta$ unknown. In this setting, a dynamic-programming solution
to the optimal Bayes policy generally involves a state space of dimension
$O(n)$. However, we show that it is possible to reduce this asymptotically
to just two state variables per arm (apart from time): the number
of times the arm has been pulled, and the score process, i.e., the
cumulative sum of scores scaled by $n^{-1/2}$, corresponding to the
distribution of rewards for that arm. All our results previously derived
for Gaussian models then continue to apply after simply reinterpreting
$x_{k}(t)$ from before as the score process. Underlying these claims
is a posterior approximation result that states that the posterior
density of the parametric model can be uniformly approximated, at
every point in time, by that from a Gaussian model. 

For the rest of this section, we focus on the one-armed bandit for
simplicity. We start by assuming $\theta$ to be scalar to simplify
notation, but the vector case (discussed in Section \ref{subsec:Vector-valued})
does not otherwise present any new conceptual difficulties. The mean
rewards are denoted by $\mu(\theta)\equiv\mathbb{E}_{P_{\theta}}[X]$.
As in \citet{hirano2009asymptotics}, we focus on local perturbations
of the form $\{\theta_{n,h}\equiv\theta_{0}+h/\sqrt{n}:h\in\mathbb{R}\}$,
where $\theta_{0}$ is a reference parameter, chosen such that $\mu(\theta_{0})=0$.
This induces diffusion asymptotics. Indeed, under these perturbations,
$\mu_{n}(h):=\mu(\theta_{n,h})\approx\dot{\mu}_{0}h/\sqrt{n}$, where
$\dot{\mu}_{0}:=\mu^{\prime}(0)$. If instead, $\mu(\theta_{0})\neq0$,
the asymptotic risk is $0$ under all the policies considered here,
including TS, UCB and our PDE based proposals. Focusing on $\mu(\theta_{0})=0$
thus ensures that we are comparing policies under the hardest instances
of the bandit problem. For Bayesian analysis, we place a `non-negligible'
prior, $M_{0}$, on the local parameter $h$. In practice, this simply
involves translating a given prior on $\theta$ to one around $h$. 

Let $\nu:=\nu_{1}\times\nu_{2}$, where $\nu_{1}$ is a dominating
measure for $\{P_{\theta}:\theta\in\mathbb{R}\}$ and $\nu_{2}$ is
a dominating measure for the prior $M_{0}$ on $h$. Define $p_{\theta}=dP_{\theta}/d\nu$,
$m_{0}=dM_{0}/d\nu$ (in the sequel, we shorten the Radon-Nikodym
derivative $dP/d\nu$ to just $dP$). Also, let $P_{n,h}$ denote
the joint probability measure over the stacked rewards $\bm{y}_{n}:=\{Y_{i}\}_{i=1}^{n}$.
We assume $\{P_{\theta}:\theta\in\mathbb{R}\}$ is quadratic mean
differentiable (qmd), i.e., there exists a score $\psi(\cdot)\in L^{2}(P_{\theta_{0}})$
such that 
\begin{equation}
\int\left[\sqrt{p_{\theta_{0}+h}}-\sqrt{p_{\theta_{0}}}-\frac{1}{2}h\psi\sqrt{p_{\theta_{0}}}\right]^{2}d\nu=o(\vert h\vert^{2}).\label{eq:QMD}
\end{equation}
Among the many examples of qmd families are the Gaussian, Poisson,
and Bernoulli distributions, along with their shifted versions.\footnote{As we set the mean rewards from the known arm to $0$, many of these
distributions, including the Bernoulli, have to be shifted by a constant.
See Section \ref{sec:Computation-and-simulations} for an illustration.} The information matrix is $I:=\mathbb{E}_{P_{\theta_{0}}}\left[\psi^{2}\right]$
and we set $\sigma^{2}:=I^{-1}$. In addition, for $q\in[0,1]$, define
\[
x_{nq}:=\frac{\sigma^{2}}{\sqrt{n}}\sum_{i=1}^{\left\lfloor nq\right\rfloor }\psi(Y_{i})
\]
 as the (normalized) score process over $q$. 

\subsection{Heuristics\label{subsec:Heuristics}}

Our key assertion is that the posterior density of $h$ at time $t$
can be approximately characterized using just 2 state variables: the
number of times the arm has been pulled, $q(t):=n^{-1}\sum_{j=1}^{\left\lfloor nt\right\rfloor }\mathbb{I}(A_{j}=1)$,
and the score process over $t$, $x(t):=x_{nq(t)}$. We now provide
some intuition behind this. The ideas introduced here are applicable
more broadly to any sequential experiment. 

In the one-armed bandit setting, if the arm is pulled $q$ times we
will have observed the first $q$ elements of the stack $\bm{y}_{n}$,
denoted ${\bf y}_{nq}:=\{Y_{i}\}_{i=1}^{\left\lfloor nq\right\rfloor }$.
After $q$ pulls, the log-likelihood ratio process under the local
alternative $h$ is
\begin{align*}
\hat{\varphi}(h;q) & =\ln\frac{dP_{\theta_{0}+h^ {}/\sqrt{n}}}{dP_{\theta_{0}^ {}}}\left({\bf y}_{nq}\right):=\sum_{i=1}^{\left\lfloor nq\right\rfloor }\ln\frac{dP_{\theta_{0}+h/\sqrt{n}}}{dP_{\theta_{0}^ {}}}\left(Y_{i}\right).
\end{align*}
It may appear odd that the likelihood-ratio does not feature the past
actions, which are random, nor does it depend on the policy rule.
Note, however, that given any (possibly randomized) policy, the probability
of choosing an action depends only on the past outcomes, and is therefore
independent of $h$. Hence, these probabilities drop out of the likelihood-ratio.\footnote{We can also interpret this as a consequence of the strong likelihood
principle (see, e.g., \citealp[Chapter 7]{berger2013statistical}):
the likelihood-ratio of the data following $q$ pulls of the arm depends
solely on ${\bf y}_{nq}$, and the exact procedure taken to reach
it is immaterial.} In Appendix \ref{subsec:Supporting-lemmas-for}, we show that (\ref{eq:QMD})
implies the important Sequential Local Asymptotic Normality (SLAN)
property: for any given $h\in\mathbb{R}$, 
\begin{equation}
\hat{\varphi}(\bm{h};q)=\frac{h}{\sigma^{2}}x_{nq}-\frac{q}{2\sigma^{2}}h^{2}+o_{P_{n,\theta_{0}}}(1),\ \textrm{uniformly over }q.\label{eq:LAN}
\end{equation}
The SLAN property, which appears to be new in its current form, extends
the usual Local Asymptotic Normality (LAN) to sequential data.\footnote{Previously, an abstract version of it was stated as an assumption
for analyzing sequential experiments of the optimal stopping kind
in \citet[Chapter 13]{le1986asymptotic}.} 

The DM employs a sampling rule $\{\pi_{j}\}_{j}\equiv\{\pi_{\left\lfloor nt\right\rfloor }\}_{t}$
that prescribes the probability of pulling the arm at period $t$,
given the information set, $\mathcal{F}_{t}$, consisting of all the
actions and rewards until that time; formally, $\mathcal{F}_{t}$
is the $\sigma$-algebra generated by $\xi_{t}\equiv\{\{A_{j}\}_{j=1}^{\left\lfloor nt\right\rfloor },\{Y_{i}\}_{i=1}^{\left\lfloor nq(t)\right\rfloor }\}$.
Clearly, $\textrm{dim}(\mathcal{F}_{t})=n(t+q(t))$, so it is very
large and increasing in $n,t$. However, (\ref{eq:LAN}) suggests
a way to reduce this. Observe that if the rewards were Gaussian, the
log-likelihood ratio would have been exactly 
\[
\tilde{\varphi}(h;q):=\frac{h}{\sigma^{2}}x_{nq}-\frac{q}{2\sigma^{2}}h^{2},
\]
and the sufficient statistics would just be $x_{nq},q,t$. But by
(\ref{eq:LAN}), the true likelihood-ratio is close to that obtained
under Gaussian rewards anyway as $n\to\infty$.

The precise argument relies on the posterior. By Lemma \ref{Lemma 1}
in Appendix \ref{subsec:Supporting-lemmas-for}, the posterior density,
$p_{}(\cdot\vert\mathcal{F}_{t})$, of $h$ depends only on ${\bf y}_{nq(t)}$,
and is given by
\begin{align}
p_{n}(h\vert\mathcal{F}_{t})=p_{n}(h\vert{\bf y}_{nq(t)}) & \propto\left[\prod_{i=1}^{\left\lfloor nq(t)\right\rfloor }p_{\theta_{0}+h/\sqrt{n}}(Y_{i})\right]\cdot m_{0}(h)\nonumber \\
 & \equiv\left[\exp\left\{ \hat{\varphi}(h;q(t))\right\} dP_{nq(t),\theta_{0}}({\bf y}_{nq(t)})\right]\cdot m_{0}(h),\label{eq:True posterior}
\end{align}
where $dP_{nq,\theta_{0}}({\bf y}_{nq}):=\prod_{i=1}^{\left\lfloor nq\right\rfloor }p_{\theta_{0}}(Y_{i})\ \forall\ q\in[0,1]$.
Replacing $\hat{\varphi}(\cdot;\cdot)$ with $\tilde{\varphi}(\cdot;\cdot)$,
the SLAN property (\ref{eq:LAN}) suggests that the likelihood at
time $t$ - i.e.\! the term within $[\cdot]$ brackets in (\ref{eq:True posterior})
- can be uniformly approximated over all possible realizations of
$q(t)$ by a new likelihood, the density of the `tilted' measure $\Lambda_{nq(t),h}({\bf y}_{nq(t)})$,
defined as
\begin{equation}
d\Lambda_{nq,h}({\bf y}_{nq})=\exp\left\{ \tilde{\varphi}(h;q)\right\} dP_{nq,\theta_{0}}({\bf y}_{nq})\ \forall\ q\in[0,1].\label{eq:tilted measure}
\end{equation}
Replacing the actual likelihood in (\ref{eq:True posterior}) with
this approximation, we obtain an approximate posterior density $\tilde{p}_{n}(h\vert{\bf y}_{nq(t)})$,
where for any $q\in[0,1]$,\footnote{Formally, $\tilde{p}_{n}(h\vert{\bf y}_{nq})$ is defined via disintegration
of the product measure $d\Lambda_{nq,h}({\bf y}_{nq})\cdot m_{0}(h)$;
see the proof of Lemma \ref{Lem: approximation of posteriors} in
Appendix \ref{subsec:Supporting-lemmas-for}.}
\begin{align}
\tilde{p}_{n}(h\vert{\bf y}_{nq})\equiv\tilde{p}_{n}(h\vert x_{nq},q) & \propto d\Lambda_{nq,h}({\bf y}_{nq})\cdot m_{0}(h)\nonumber \\
 & \propto\tilde{p}_{q}(x_{nq}\vert h)\cdot m_{0}(h);\quad\tilde{p}_{q}(\cdot\vert h)\equiv\mathcal{N}(\cdot\vert qh,q\sigma^{2}).\label{eq:posterior - general parametric models}
\end{align}
In Appendix \ref{subsec:Supporting-lemmas-for}, we show that the
total variation distance between $p_{n}(\cdot\vert{\bf y}_{nq})$
and $\tilde{p}_{n}(\cdot\vert{\bf y}_{nq})$ converges to $0$ uniformly
over $q\in[0,1]$. Hence, the true posterior can be approximated arbitrarily
well by one that is obtained under Gaussian rewards.

\subsection{Formal results\label{subsec:Formal-results}}

Define $s:=(x,q,t)$, $\mu^{+}(s):=\dot{\mu}_{0}\tilde{\mathbb{E}}\left[\left.h\mathbb{I}(\dot{\mu}_{0}h\ge0)\right|s\right]$,
$\mu(s):=\dot{\mu}_{0}h(s)$ and $h(s):=\tilde{\mathbb{E}}\left[\left.h\right|s\right]$,
where $\tilde{\mathbb{E}}[\cdot\vert{\bf y}_{nq}]\equiv\tilde{\mathbb{E}}[\cdot\vert s]$
is the expectation corresponding to the approximate posterior density
$\tilde{p}_{n}(\cdot\vert{\bf y}_{nq})\equiv\tilde{p}_{n}(\cdot\vert x_{nq},q)$.
It will be shown that the minimal asymptotic Bayes risk in the parametric
regime is again characterized by (\ref{eq:PDE optimal bayes risk}),
but the infinitesimal generator is now modified slightly to\footnote{The difference is that $\partial_{x}f$ is multiplied by $h(s)$ as
opposed to $\mu(s)=\dot{\mu}_{0}h(s)$.} 
\begin{equation}
L[f]:=\partial_{q}f+h(s)\partial_{x}f+\frac{1}{2}\sigma^{2}\partial_{x}^{2}f.\label{eq:infinitesimal generator - parametric}
\end{equation}
We impose the following assumptions:

\begin{asm1} (i) The class $\{P_{\theta}\}$ is differentiable in
quadratic mean as in (\ref{eq:QMD}). (ii) $\mathbb{E}_{P_{\theta_{0}}}[\exp\vert\psi(Y)\vert]<\infty$.
(iii) There exists $\dot{\mu}_{0}<\infty$ and $\delta_{n}\to0$ such
that $\sqrt{n}\mu_{n}(h)=\dot{\mu}_{0}h+\delta_{n}\vert h\vert^{2}\ \forall\ h$.
(iv) The support of $m_{0}(\cdot)$ is a compact set $\{h:\vert h\vert\le\Gamma\}$
for some $\Gamma<\infty$. (v) $\mu(\cdot)$ and $\mu^{+}(\cdot)$
are H{\"o}lder continuous. Additionally, $\sup_{s}\varpi(s)\le C<\infty$.
\end{asm1}

Assumptions 1(i), (iii) and (v) are standard. Assumption 1(ii) is
restrictive, but is related to the fact $\Lambda_{nq,h}$ approximates
the true likelihood rather coarsely when $h$ is large. One could
consider replacing $\exp\left\{ \tilde{\varphi}(h;q)\right\} $ in
its definition with $g(\tilde{\varphi}(h;q))$, where $g(z)=\exp(z)+o(z^{3})$
for small $z$ and bounded for large $z$, e.g., $g(z)=\min\{2,\max\{1+z+z^{2}/2,0\}\}$.
We conjecture that Assumption 1(ii) could then be weakened to $\mathbb{E}_{P_{\theta_{0}}}\left[\vert\psi(Y)\vert^{3}\right]<\infty$.
Assumption 1(iv), which is also employed in \citet[Proposition 6.4.4]{le2000asymptotics},
requires the prior to have a compact support. It is possible to drop
this assumption under some additional conditions, e.g., if the prior
has finite $1+\alpha$ moments, $\alpha>0$, and Assumption 1(iii)
is strengthened to $\vert\mu(P_{\theta_{0}+h})\vert\le C\vert h\vert\ \forall\ h$.
Assumptions 1(ii) \& (iv) are therefore not the most general possible,
but they lead to relatively transparent proofs.

For the theorem below, let $\Pi$ denote the class of all policies
sequentially measurable wrt $\{\mathcal{F}_{j}\}_{j}$, and $\Pi^{\mathcal{S}}\subset\Pi$
the subset of it consisting of policies that depend only on $s=(x,q,t)$.
For a fixed $n$ and $\pi\in\Pi$, the ex-ante Bayes risk is $V_{\pi,n}(0)=\mathbb{E}_{({\bf y}_{n},h)}\left[\sum_{j=1}^{n}R(Y_{j},\pi_{j},h)\right]$,
where $\mathbb{E}_{({\bf y}_{n},h})[\cdot]$ is the expectation under
the joint density $\left\{ \prod_{i=1}^{n}p_{\theta_{0}+h/\sqrt{n}}(Y_{i})\right\} \cdot m_{0}(h).$
The minimal ex-ante Bayes risk is $V_{n}^{*}(0)=\inf_{\pi\in\Pi_{\mathcal{}}}V_{\pi,n}(0)$,
and we also define $V_{n}^{\mathcal{S}*}(0):=\inf_{\pi\in\Pi_{\mathcal{}}^{\mathcal{S}}}V_{\pi,n}(0)$.
Lastly, $\pi_{\Delta t}^{*}$ is the optimal piece-wise constant policy
with $\Delta t$ increments as in Section \ref{subsec:Optimal-and-approximately}.

\begin{thm} \label{Thm: General parametric families}Suppose Assumption
1 holds. Then: (i) $\lim_{n\to\infty}\left|V_{n}^{*}(0)-V_{n}^{\mathcal{S}*}(0)\right|=0$.
(ii) $\textrm{\ensuremath{\lim}}_{n\to\infty}V_{n}^{*}(0)=V^{*}(0)$,
where $V^{*}(\cdot)$ solves PDE (\ref{eq:PDE optimal bayes risk})
with the infinitesimal generator (\ref{eq:infinitesimal generator - parametric}).
(iii) If, further, $\mu(\cdot)$, $\mu^{+}(\cdot)$ are Lipschitz
continuous, $\lim_{n\to\infty}\vert V_{\pi_{\Delta t}^{*},n}(0)-V^{*}(0)\vert\lesssim\Delta t{}^{1/4}$
for any fixed $\Delta t$.\end{thm}

Part (i) states that it is sufficient to restrict attention to just
3 state variables $s=(x,q,t)$. Part (ii) asserts that the minimal
Bayes risk is characterized by PDE (\ref{eq:PDE optimal bayes risk}),
while part (iii) implies piece-wise constant policies can attain this
bound. 

\subsection{Vector valued $\theta$\label{subsec:Vector-valued}}

The vector case can be analyzed in the same manner as the scalar setting,
so we only describe the results. Let $\psi(\cdot)$ denote the score
function, $\Sigma^{-1}=\mathbb{E}_{P_{\theta_{0}}}\left[\psi\psi^{\intercal}\right]$
the information matrix, and $x(t)=n^{-1/2}\sum_{i=1}^{\left\lfloor nq(t)\right\rfloor }\Sigma\psi(Y_{i})$,
the normalized score process. The asymptotically sufficient state
variables are still $s(t)=(x(t),q(t),t)$. Given a prior $m_{0}(\cdot)$
on $h$, the approximate posterior density is $\tilde{p}_{n}(h\vert x,q)\propto\mathcal{N}(x\vert qh,q\Sigma)\cdot m_{0}(h)$.
Define $h(s)=\tilde{\mathbb{E}}\left[\left.h\right|s\right]$, $\mu^{+}(s)=\mathbb{\tilde{E}}\left[\left.\dot{\mu}_{0}^{\intercal}h\mathbb{I}(\dot{\mu}_{0}^{\intercal}h\ge0)\right|s\right]$
and $\mu(s)=\dot{\mu}_{0}^{\intercal}h(s)$, where $\tilde{\mathbb{E}}[\cdot\vert s]$
is the expectation corresponding to $\tilde{p}_{n}(\cdot\vert x,q)$
and $\dot{\mu}_{0}:=\nabla\mu(\theta_{0})$. With these definitions,
the minimal Bayes risk is still characterized by PDE (\ref{eq:PDE optimal bayes risk}),
but with the infinitesimal generator now being
\begin{equation}
L[f]:=\partial_{q}f+h(s)^{\intercal}D_{x}f+\frac{1}{2}\textrm{Tr}\left[\Sigma\cdot D_{x}^{2}f\right].\label{eq:vector infinitesimal generator}
\end{equation}

\subsection{Lower bound on minimax risk\label{subsec:Lower-bound-on}}

Let $V_{n,\pi}(0;h)$ denote the fixed-$n$ frequentist risk of policy
$\pi$ when the local parameter is $h$, and write $V^{*}(0)$ as
$V^{*}(0;m_{0})$ to make explicit its dependence on the prior $m_{0}$.
In Appendix \ref{sec:lower bounds on minimax risk}, we use Theorem
\ref{Thm: General parametric families} to derive a lower bound on
asymptotic minimax risk as
\begin{equation}
\lim_{n\to\infty}\inf_{\pi\in\Pi}\sup_{\vert h\vert\le\Gamma}V_{n,\pi}(0;h)\ge\sup_{m_{0}\in\mathcal{P}}V^{*}(0;m_{0})=\bar{V}^{*},\label{eq:minimax result - parametrics}
\end{equation}
where $\mathcal{P}$ is the set of all compactly supported distributions,
and $\bar{V}^{*}$ is just the asymptotic minimax risk in the Gaussian
setting as in (\ref{eq:def of minimax risk}). Proving the sharpness
of the lower bound (\ref{eq:minimax result - parametrics}) is more
involved, however, and left for future research. 

\section{The non-parametric setting\label{sec:The-nonparametric-setting}}

Very often we do not have any a-priori information about the distribution
of the rewards. In this section, we show that our characterization
of Bayes and minimax risk also applies in such a non-parametric regime
after we replace the score process with the cumulative sum process
of the rewards. In short, there is no loss in simply pretending that
the outcomes are Gaussian.

Our formal analysis of the non-parametric regime follows \citet{van2000asymptotic}.
Let $\mathcal{P}$ denote the class of probability distributions with
bounded variance and dominated by some measure $\nu$. We then fix
a reference $P_{0}\in\mathcal{P}$, and surround it with various smooth
one-dimensional parametric sub-models, $\{P_{t,\bm{h}}:t\le\eta\}$,
whose score function is $\hm{h}$ and that pass through $P_{0}$ at
$t=0$ (i.e., $P_{0,\bm{h}}=P_{0}$). To obtain non-trivial risk bounds,
we suppose $\mu(P_{0})=0$, where $\mu(P):=\int xdP(x)$ denotes the
mean rewards under $P$. The rationale is akin to setting $\mu(\theta_{0})=0$
in the parametric setting: it focuses attention on the hardest instances
of the bandit problem. The formal definition of $\{P_{t,\bm{h}}:t\le\eta\}$
is given in Appendix \ref{subsec:Additional details for non-parametric},
we just note here that the only requirements on $\bm{h}$ are $\int\bm{h}dP_{0}=0$
and $\int\bm{h}^{2}dP_{0}<\infty$. The set of all such functions
$\bm{h}$ is termed the tangent space $T(P_{0})$. 

Denote $\left\langle f_{1},f_{2}\right\rangle =\int f_{1}f_{2}dP_{0}$.
For any regular functional $\mu(\cdot)$ on $\mathcal{P}$ (and not
just the mean), we say that $\psi(\cdot)$ is the efficient influence
function corresponding to it if
\begin{equation}
\frac{\mu(P_{t,\bm{h}})-\mu(P_{0})}{t}-\left\langle \psi,\bm{h}\right\rangle =\frac{\mu(P_{t,\bm{h}})}{t}-\left\langle \psi,\bm{h}\right\rangle =o(t)\ \forall\ \bm{h}\in T(P_{0}).\label{eq:influence function}
\end{equation}
For mean-estimation, $\psi(x)=x$. Now, (\ref{eq:influence function})
implies $\mu(P_{1/\sqrt{n},\bm{h}})\approx\left\langle \psi,\bm{h}\right\rangle /\sqrt{n}$.
This suggests that for non-trivial notions of Bayes and minimax risk
under a $n^{-1/2}$ scaling of mean rewards, we should place `non-negligible'
priors on the set of probability distributions $\mathcal{P}_{n}:=\{P_{1/\sqrt{n},\bm{h}}:\bm{h}\in T(P_{0})\}$.\footnote{Note that priors in the non-parametric regime are probability distributions
over the space of candidate distributions for the rewards.} This is in turn equivalent to a prior, $\rho_{0}$ (say), on $T(P_{0})$.
We impose two restrictions on $\rho_{0}$. First, while $T(P_{0})$
is infinite dimensional, $\rho_{0}$ should be supported on a finite
dimensional sub-space of it (i.e., on a sub-space spanned by a finite
number of basis functions from $T(P_{0})$). Second, it should be
possible to decompose $\rho_{0}=m_{0}\times\lambda$, where $m_{0}$
is a prior on $h_{0}:=\left\langle \psi,\bm{h}\right\rangle $ and
$\lambda$ is a prior over the part of $T(P_{0})$ that is orthogonal
to $\psi$. 

The first restriction on $\rho_{0}$ is for mathematical convenience,
but also follows the standard approach of defining minimax risk through
finite dimensional sub-models \citep[Chapter 25]{van2000asymptotic}.
As for the second restriction, the rationale behind product priors
is two-fold: First, they suffice for obtaining a lower bound on minimax
risk. Second, and more importantly, our welfare criterion depends
on $\hm{h}$ only through $h_{0}$, which determines the mean reward.
Invariance considerations would then suggest restricting attention
to policies that deliver the same frequentist risk for any $\bm{h}_{1},\bm{h}_{2}\in T(P_{0})$
such that $\mu(\bm{h}_{1})=\mu(\bm{h}_{2})$. Product priors achieve
this as they ensure the posterior of $h_{0}$ is independent of $\lambda$,
the component of the prior placing beliefs over the part of $\bm{h}$
that is orthogonal to mean-estimation. Incidentally, the above considerations
also apply to parametric models with vector $\theta$. Using product
priors there then leads to a further dimension reduction\textcolor{blue}{:
}we can replace the score process, $x(t)$, with its univariate projection
$\dot{\mu}_{0}^{\intercal}\Sigma^{-1}x(t)$. See Appendix \ref{subsec:Heuristics-non-parametrics}
for the intuition.

While the focus in this paper is on mean rewards, the theory itself
is more general and applies to any regular functional $\mu(\cdot)$
of $\mathcal{P}$. For instance, $\mu(\cdot)$ could be the median,
in which case the risk criterion would be the cumulative sum of median
outcomes. All our results go through unchanged after simply reinterpreting
$\psi(\cdot)$ as the efficient influence function corresponding to
$\mu(\cdot)$. 

Let $V_{n}^{*}(0;\rho_{0})$ denote the minimal Bayes risk in the
one-armed bandit setting, when the prior is $\rho_{0}=m_{0}\times\lambda$.
We show that $V_{n}^{*}(0;\rho_{0})$ converges to $V^{*}(0;m_{0})$,
where $V^{*}(\cdot;m_{0})$ solves PDE (\ref{eq:PDE optimal bayes risk})
under the prior $m_{0}$. The asymptotically sufficient state variables
are still $(x_{nq},q,t)$ as before, but $x_{nq}=n^{-1/2}\sigma^{2}\sum_{i=1}^{\left\lfloor nq\right\rfloor }\psi(Y_{i})$
is now the efficient influence function process, with $\sigma^{2}:=\textrm{Var}[P_{0}]$.
The intuition behind this result, and the assumptions required for
it, are described in Appendix \ref{subsec:Additional details for non-parametric}. 

\begin{thm} \label{Thm: Non-parametric models}Suppose Assumption
2 in Appendix \ref{subsec:Additional details for non-parametric}
holds. Then:

(i) $\textrm{\ensuremath{\lim}}_{n\to\infty}V_{n}^{*}(0;\rho_{0})=V^{*}(0;m_{0})$. 

(ii) If, further, $\mu(\cdot)$, $\mu^{+}(\cdot)$ are Lipschitz continuous,
$\lim_{n\to\infty}\vert V_{\pi_{\Delta t}^{*},n}(0;\rho_{0})-V^{*}(0;m_{0})\vert\lesssim\Delta t{}^{1/4}$
for any fixed $\Delta t$, where $V_{\pi_{\Delta t}^{*},n}(0;\rho_{0})$
is defined in Section \ref{subsec:Optimal-and-approximately}.\end{thm}

As with parametric models, Theorem \ref{Thm: Non-parametric models}
can be used to derive a lower bound on minimax risk. Let $V_{n,\pi}(0;\bm{h})$
denote the fixed $n$ (ex-ante) frequentist risk of a policy $\pi$
under $P_{1/\sqrt{n},\bm{h}}$. Suppose that $\mathbb{E}[\exp\vert Y\vert]<\infty$
and $\mathcal{P}$ is the set of all compactly supported $m_{0}$.
Then, Theorem \ref{Thm: Non-parametric models} implies
\begin{equation}
\sup_{I\in\mathbb{N}}\lim_{n\to\infty}\inf_{\pi\in\Pi}\sup_{\bm{h}\in\mathcal{H}_{I}}V_{n,\pi}(0;\bm{h})\ge\sup_{m_{0}\in\mathcal{P}}V^{*}(0;m_{0})=\bar{V}^{*},\label{eq:minimax result - non-parametrics}
\end{equation}
where $\sup_{\bm{h}\in\mathcal{H}_{I}}$ denotes the supremum over
all finite, $I$-dimensional subspaces, $H_{I}$, of the tangent space
$T(P_{0})$, with $I\in\mathbb{N}$. By \citet[Theorem 25.21]{van2000asymptotic},
the left hand side of (\ref{eq:minimax result - non-parametrics})
is the value of minimax risk. The right hand side of (\ref{eq:minimax result - non-parametrics})
is simply the lower bound on minimax risk under Gaussian rewards,
as in (\ref{eq:def of minimax risk}). 

\section{Conclusion\label{sec:Conclusion}}

In this article, we derive sharp lower bounds for Bayes and minimax
risk of bandit algorithms under diffusion asymptotics and suggest
ways to numerically compute the corresponding optimal policies. Our
local asymptotic analysis of Bayes risk is substantially different
from existing approaches and is arguably more powerful, as it enables
us to rank various policies which were previously were indistinguishable
on the basis of their large-deviation regret properties. We show that
all bandit problems, be they parametric or non-parametric, are asymptotically
equivalent to Gaussian bandits. Furthermore, it is asymptotically
sufficient to restrict attention to just two state variables per arm.
For minimax risk, the paper only proves a lower bound. While we believe
the bound is tight, further work is needed to show this. The work
also raises a number of additional avenues for future research, a
few of which are discussed below: 

\textit{Unknown $\sigma$.} A drawback of diffusion asymptotics, and
of first-order efficiency criteria more generally, is that replacing
unknown variances with consistent estimates has no effect on asymptotic
risk. One could in principle achieve optimal risk by (say) sampling
all arms equally for $\bar{n}:=n^{\rho}$ periods, $\rho\in(0,1)$,
obtaining estimates of $\sigma$, and applying the optimal policies
based on those estimates from $\bar{n}$ onwards. But in finite samples,
the choice of $\rho$ will matter and further work is needed to choose
this efficiently. 

\textit{Other sequential experiments.} \citet{adusumilli2022minimax}
applies insights from this paper to derive the minimax optimal policy
for best-arm identification with two arms, while \citet{adusumilli2022Wald}
does the same for the problem of costly sampling. 

\bibliographystyle{IEEEtranSN}
\bibliography{Optimal_sequential_policies}

\appendix

\section{Proofs\label{sec:Appendix:A}}

\subsection{Proof of Theorem \ref{Thm: Convergence to minimal PDE}\label{subsec:Proof-of-Theorem-2}}

For this proof, we make the time change $\tau:=1-t$. Let $s:=(x,q,\tau)$,
$\mathbb{I}_{n}\equiv\{\tau<1/n\}$ and denote the domain of $s$
by $\mathcal{S}$. Also, let $C^{\infty}(\mathcal{S})$ denote the
set of test functions, i.e., the set of all infinitely differentiable
functions $\phi:\mathcal{S}\to\mathbb{R}$ such that $\sup_{q\ge0}\vert D^{q}\phi\vert\le M$
for some $M<\infty$. 

Following the time change, we can alternatively represent the solution,
$V_{n}^{*}(\cdot)$, to (\ref{eq:discrete approximation}) as the
solution (over the set of all possible functions $\phi:\mathcal{S}\to\mathbb{R}$)
to the approximation scheme 
\begin{align}
S_{n}(s,\phi(s),[\phi]) & =0\ \textrm{for \ensuremath{\tau>0}};\quad\phi(x,q,0)=0,\label{eq:scheme definition}
\end{align}
where for any $u\in\mathbb{R}$ and $\phi_{2}:\mathcal{S}\to\mathbb{R}$,
\begin{align*}
 & S_{n}(s,u,[\phi_{2}])\\
 & :=-\min_{\pi\in[0,1]}\left\{ \frac{\mu^{+}(s)-\pi\mu(s)}{n}+\mathbb{E}\left[\left.\mathbb{I}_{n}\cdot\phi_{2}\left(x+\frac{A_{\pi}Y_{nq+1}}{\sqrt{n}},q+\frac{A_{\pi}}{n},\tau-\frac{1}{n}\right)-u\right|s\right]\right\} .
\end{align*}
The notation $[\phi_{2}]$ in $S_{n}(s,u,[\phi_{2}])$ refers to the
fact that it is a functional argument. Define 
\[
F(D^{2}\phi,D\phi,s)=\partial_{\tau}\phi-\mu^{+}(s)-\min\left\{ -\mu(s)+L[\phi](s),0\right\} ,
\]
as the left-hand side of PDE (\ref{eq:PDE optimal bayes risk}) after
the time change. \citet{barles1991convergence} show that the solution,
$V_{n}^{*}(\cdot)$, of (\ref{eq:scheme definition}) converges to
the solution, $V^{*}(\cdot)$, of $F(D^{2}\phi,D\phi,s)=0$ with the
boundary condition $\phi(x,q,0)=0$ if the scheme $S_{n}(\cdot)$
satisfies the properties of monotonicity, stability and consistency.

Monotonicity requires $S_{n}(s,u,[\phi_{1}])\le S_{n}(s,u,[\phi_{2}])$
for all $s\in\mathcal{S}$, $u\in\mathbb{R}$ and $\phi_{1}\ge\phi_{2}$.
This is clearly satisfied. 

Stability requires (\ref{eq:scheme definition}) to have a unique
solution, $V_{n}^{*}(\cdot)$, that is uniformly bounded. That a unique
solution exists follows from backward induction. To obtain an upper
bound, note that following a state $s$, the DM may choose to pull
the arm in all subsequent periods. This results in a risk of $\tau\left(\mu^{+}(s)-\mu(s)\right)$.
Alternatively, if DM chooses not to pull the arm in all subsequent
periods, the resulting risk is $\tau\mu^{+}(s)$. Hence, by definition
of $V_{n}^{*}(\cdot)$ as the risk under an optimal policy, 
\begin{equation}
0\le V_{n}^{*}(s)\le\tau\min\left\{ \mu^{+}(s)-\mu(s),\mu^{+}(s)\right\} \le C\tau.\label{eq:pf:Thm1:upper bound}
\end{equation}

Finally, consistency requires that for all $\phi\in C^{\infty}(\mathcal{S})$,
and $s\equiv(x,q,\tau)\in\mathcal{S}$ such that $\tau>0$, 
\begin{align}
\limsup_{\substack{n\to\infty\\
\rho\to0\\
z\to s
}
}nS_{n}(z,\phi(z)+\rho,[\phi+\rho]) & \le F(D^{2}\phi(s),D\phi(s),s),\textrm{ and }\label{eq:upper consistency}\\
\liminf_{\substack{n\to\infty\\
\rho\to0\\
z\to s
}
}nS_{n}(z,\phi(z)+\rho,[\phi+\rho]) & \ge F(D^{2}\phi(s),D\phi(s),s).\label{eq:lower consistency}
\end{align}
It suffices to restrict attention to $\tau>0$ because (\ref{eq:pf:Thm1:upper bound})
implies that for any $s$ on the boundary, i.e., of the form $(x,q,0)$,
\[
\limsup_{\substack{n\to\infty\\
z\to s
}
}V_{n}^{*}(z)=0=\liminf_{\substack{n\to\infty\\
z\to s
}
}V_{n}^{*}(z).
\]
When the above holds, an analysis of the proof of \citet[Theorem 2.1]{barles1991convergence}
shows that we only need prove (\ref{eq:upper consistency}) and (\ref{eq:lower consistency})
for interior values of $s$, i.e., when $\tau>0$.

We now show (\ref{eq:upper consistency}). The argument for (\ref{eq:lower consistency})
is similar. Since any $z\equiv(\tilde{x},\tilde{q},\tilde{\tau})$
converging to $s\equiv(x,q,\tau)$ with $\tau>0$ will eventually
satisfy $\tilde{\tau}>1/n$, we can drop $\mathbb{I}_{n}$ in the
definition of $S_{n}(\cdot)$ while taking the $\limsup$ operation
in (\ref{eq:upper consistency}). Now, for any $s\in\mathcal{S}$,
a third order Taylor expansion gives
\begin{align*}
 & n\mathbb{E}\left[\left.\phi\left(x+\frac{\mathbb{I}(A_{\pi}=1)Y_{nq+1}}{\sqrt{n}},q+\frac{\mathbb{I}(A_{\pi}=1)}{n},\tau-\frac{1}{n}\right)-\phi(s)\right|s\right]\\
 & =\mathbb{E}\left[\left.\sqrt{n}\mathbb{I}(A_{\pi}=1)Y_{nq+1}\right|s\right]\partial_{x}\phi+\frac{1}{2}\mathbb{E}\left[\left.\mathbb{I}(A_{\pi}=1)Y_{nq+1}^{2}\right|s\right]\partial_{x}^{2}\phi\\
 & \qquad+\mathbb{E}\left[\left.\mathbb{I}(A_{\pi}=1)\right|s\right]\partial_{q}\phi-\partial_{\tau}\phi+\frac{R(s)}{\sqrt{n}}
\end{align*}
where $R(s)$ is a continuous function of $\mu(s)$, $\mathbb{E}\left[\left.\mu{}^{2}\right|s\right]$
and $\mathbb{E}\left[\left.\vert Y_{nq+1}\vert^{3}\right|s\right]$
that is bounded at each $s$ as long as these three functions are
also bounded. Because $A_{\pi}\sim\textrm{Bernoulli}(\pi)$ for any
given $\pi\in[0,1]$, we have $\mathbb{E}\left[\left.\sqrt{n}\mathbb{I}(A_{\pi}=1)Y_{nq+1}\right|s\right]=\pi\mu(s)$,
$\mathbb{E}\left[\left.\mathbb{I}(A_{\pi}=1)Y_{nq+1}^{2}\right|s\right]=\pi(\sigma^{2}+n^{-1}\mathbb{E}[\mu^{2}\vert s])$
and $\mathbb{E}\left[\left.\mathbb{I}(A_{\pi}=1)\right|s\right]=\pi$.
Furthermore, recalling that $Y\vert\mu\sim\mathcal{N}(\mu/\sqrt{n},\sigma^{2})$,
the properties of the Gaussian distribution imply 
\[
\mathbb{E}\left[\left.\vert Y_{nq+1}\vert^{3}\right|s\right]=\mathbb{E}\left[\left.\mathbb{E}\left[\vert Y_{nq+1}\vert^{3}\vert\mu\right]\right|s\right]\apprle n^{-3/2}\mathbb{E}\left[\left.\vert\mu\vert^{3}\right|s\right]<\infty
\]
under the stated assumptions. Based on the above, we obtain 
\begin{align*}
 & nS_{n}(z,\phi(z)+\rho,[\phi+\rho])\\
 & =-\min_{\pi\in[0,1]}\left\{ \left(\mu^{+}(z)-\pi\mu(z)\right)+\pi L[\phi](z)-\partial_{\tau}\phi(z)+\frac{R(z)}{\sqrt{n}}+\partial_{x}^{2}\phi(z)\frac{\mathbb{E}[\mu^{2}\vert z]}{n},0\right\} \\
 & \le-\min_{\pi\in[0,1]}\left\{ \left(\mu^{+}(z)-\pi\mu(z)\right)+\pi L[\phi](z)-\partial_{\tau}\phi(z),0\right\} +\frac{\vert R(z)\vert}{\sqrt{n}}+\frac{M\mathbb{E}[\mu^{2}\vert z]}{n}\\
 & =\partial_{\tau}\phi(z)-\mu^{+}(z)-\min\left\{ -\mu(z)+L[\phi](z),0\right\} +\frac{\vert R(z)\vert}{\sqrt{n}}+\frac{M\mathbb{E}[\mu^{2}\vert z]}{n}.
\end{align*}
Because $\limsup_{z\to s}\{\vert R(z)\vert+\mathbb{E}[\mu^{2}\vert z]\}<\infty$,
$\phi\in C^{\infty}(\mathcal{S})$ and $\mu^{+}(\cdot),\mu(\cdot)$
are continuous functions, 
\begin{align*}
 & \limsup_{\substack{n\to\infty\\
\rho\to0\\
z\to s
}
}nS_{n}(z,\phi(z)+\rho,[\phi+\rho])\\
 & \le\limsup_{\substack{z\to s}
}\partial_{\tau}\phi(z)-\mu^{+}(z)-\min\left\{ -\mu(z)+L[\phi](z),0\right\} \\
 & =F(D^{2}\phi(s),D\phi(s),s).
\end{align*}
This completes the proof of consistency.

\subsection{Proof of Theorem \ref{Thm: Approximate control-2}}

For this proof, we use $\vert f\vert$ to represent the sup norm of
$f$. Let $V_{\Delta t,n,l}^{*}(x,q)$ denote the Bayes risk in the
fixed $n$ setting at state $(x,q,t_{L-l})$ under $\pi_{\Delta t}^{*}(\cdot).$
Then $V_{\Delta t,n,0}^{*}(x,q)=0,$ and $V_{\Delta t,n,l+1}^{*}(\cdot)$
satisfies 
\begin{align}
V_{\Delta t,n,l+1}^{*}(x,q) & =\tilde{\Gamma}_{\Delta t}\left[V_{\Delta t,n,l}^{*}\right](x,q);\ l=0,\dots,L-1,\ \textrm{where}\label{eq:piecewise-constant policy-1}\\
\tilde{\Gamma}_{\Delta t}[\phi](x,q) & :=\min\left\{ \tilde{S}_{\Delta t}\left[\phi\right](x,q),\phi(x,q)+\Delta t\cdot\mu^{+}(x,q)\right\} ,\nonumber 
\end{align}
and $\tilde{S}_{\Delta t}\left[\phi\right](x,q)$ denotes the solution
at $(x,q,\Delta t)$ of the recursive equation
\begin{align}
f\left(x,q,\tau\right) & =\mathbb{E}\left[\left.\frac{\mu^{+}(x,q)-\mu(x,q)}{n}+f\left(x+\frac{Y}{\sqrt{n}},q+\frac{1}{n},\tau-\frac{1}{n}\right)\right|s\right];\ \tau>0\nonumber \\
f\left(x,q,0\right) & =\phi(x,q).\label{eq:discrete approximation-piece-wise-constant}
\end{align}
In other words, $\tilde{S}_{\Delta t}\left[\phi\right](x,q)$ is the
discrete time counterpart of the operator $S_{\Delta t}[\cdot]$ defined
in Section \ref{subsec:Optimal-and-approximately}. 

For any $k>0$, it can be seen from the recursive definitions of $V_{\Delta t,n,l}^{*}$
and $V_{\Delta t,l}^{*}$, 
\begin{align*}
\vert V_{\Delta t,n,l+1}^{*}-V_{\Delta t,l+1}^{*}\vert & \le\left|\tilde{\Gamma}_{\Delta t}\left[V_{\Delta t,n,l}^{*}\right]-\tilde{\Gamma}_{\Delta t}\left[V_{\Delta t,l}^{*}\right]\right|+\left|\tilde{S}_{\Delta t}\left[V_{\Delta t,l+1}^{*}\right]-S_{\Delta t}\left[V_{\Delta t,l+1}^{*}\right]\right|.
\end{align*}
Recall that $\tilde{S}_{\Delta t}\left[\phi\right]$ denotes the solution
to (\ref{eq:discrete approximation-piece-wise-constant}), while $S_{\Delta t}[\phi]$
denotes the solution to (\ref{eq:linear PDE}), when the initial condition
in both cases is $\phi$. Hence, by \citet[Theorem 3.1]{barles2007error},
the regularity conditions of which can be verified as in Appendix
\ref{subsec:Rates-of-convergence}, we have $\left|\tilde{S}_{\Delta t}\left[V_{\Delta t,l+1}^{*}\right]-S_{\Delta t}\left[V_{\Delta t,l+1}^{*}\right]\right|\lesssim n^{-1/14}$.
Additionally, it is straightforward to verify $\left|\tilde{\Gamma}_{\Delta t}\left[\phi_{1}\right]-\tilde{\Gamma}_{\Delta t}\left[\phi_{2}\right]\right|\le\vert\phi_{1}-\phi_{2}\vert$
for all $\phi_{1},\phi_{2}$. Together, these results imply
\[
\vert V_{\Delta t,n,l+1}^{*}-V_{\Delta t,l+1}^{*}\vert\lesssim\vert V_{\Delta t,n,l}^{*}-V_{\Delta t,l}^{*}\vert+n^{-1/14}\lesssim l\cdot n^{-1/14},
\]
where the last inequality follows by iterating on $l$. Since $L$
is finite under a fixed $\Delta t$, we have thereby shown $\lim_{n\to\infty}\vert V_{\Delta t,n,l+1}^{*}-V_{\Delta t,l+1}^{*}\vert=0$
for all $l=0,\dots,L-1$. The claim follows by combining this result
with Theorem \ref{Thm: Approximate control}. 

\subsection{Proof outline of Theorem \ref{Thm: General parametric families}\label{subsec:Proof-sketch-of-Theorem 5}}

\!\!\!\footnote{See Appendix \ref{subsec:Proof-of-Theorem-4} for the full details.}
We may suppose without loss of generality that $\Pi$ consists only
of deterministic policies as this restriction is immaterial for Bayes
risk. We start by writing $V_{\pi,n}(0)$ in a convenient form. Define
$q_{j}:=q(j/n)$. The regret payoff (\ref{eq:regret payoff defn})
can be expanded as
\[
R(Y,\pi,h)=\frac{\mu_{n}(h)}{\sqrt{n}}\left\{ \mathbb{I}(\mu_{n}(h)\ge0)-\pi\right\} +\frac{\epsilon}{\sqrt{n}}\left\{ \mathbb{I}(\mu_{n}(h)\ge0)-\pi\right\} ,
\]
where $\epsilon:=Y-\mu_{n}(h)$ is mean $0$ conditional on $\pi,h$
(we have used $\pi$ in place of $A$ as they are equivalent for deterministic
policies). For any $\bar{\pi}\in\{0,1\}$, set 
\[
R_{n}(h,\bar{\pi}):=n\mathbb{E}\left[R(Y,\bar{\pi},h)\vert\bar{\pi},h\right]=\sqrt{n}\mu_{n}(h)\left\{ \mathbb{I}(\mu_{n}(h)\ge0)-\bar{\pi}\right\} .
\]
Now, $\pi_{j+1}$ is a deterministic function of ${\bf y}_{nq_{j}}$
for deterministic policies. Then, by the definition of $V_{\pi,n}(0)$
given in Section \ref{subsec:Formal-results}, and the law of iterated
expectations,
\begin{align}
V_{\pi,n}(0) & =\mathbb{E}_{({\bf y}_{n},h)}\left[\frac{1}{n}\sum_{j=1}^{n}R_{n}(h,\pi_{j})\right]=\mathbb{E}_{{\bf y}_{n}}\left[\frac{1}{n}\sum_{j=0}^{n-1}\mathbb{E}\left[\left.R_{n}(h,\pi_{j+1})\right|{\bf y}_{nq_{j}(\pi)}\right]\right],\label{eq:proof-sketch-eq0}
\end{align}
where we write $q_{j}(\pi)$ to make explicit the dependence of $q_{j}$
on the policy $\pi$. 

In Section \ref{subsec:Heuristics}, we used the approximate likelihood
$\Lambda_{nq,h}({\bf y}_{nq})$ to obtain an approximation, $\tilde{p}_{n}(\cdot\vert{\bf y}_{nq})\equiv\tilde{p}_{n}(\cdot\vert x_{nq},q)$,
to the true posterior density. In a similar vein, we can approximate
the true marginal density, $d\bar{P}_{n}({\bf y}_{n}):=\int p_{n,\theta_{0}+h/\sqrt{n}}({\bf y}_{n})\cdot m_{0}(h)d\nu(h)$,
with $d\tilde{\bar{P}}_{n}({\bf y}_{n}):=\int d\Lambda_{n,h}({\bf y}_{n})\cdot m_{0}(h)d\nu(h)$.
Let $\tilde{\mathbb{E}}[\cdot\vert{\bf y}_{nq}]$, $\tilde{\mathbb{E}}_{n}[\cdot]$
denote the expectations corresponding to $\tilde{p}_{n}(\cdot\vert{\bf y}_{nq})$
and $d\tilde{\bar{P}}_{n}$. Define $\tilde{V}_{\pi,n}(0)$ as the
quantity obtained by replacing the inner and outer expectations in
(\ref{eq:proof-sketch-eq0}) with their approximations $\tilde{\mathbb{E}}[\cdot\vert{\bf y}_{nq}]$
and $\tilde{\mathbb{E}}_{n}[\cdot]$, i.e., 
\begin{equation}
\tilde{V}_{\pi,n}(0):=\tilde{\mathbb{E}}_{n}\left[\frac{1}{n}\sum_{j=0}^{n-1}\tilde{\mathbb{E}}\left[\left.R_{n}(h,\pi_{j+1})\right|{\bf y}_{nq_{j}(\pi)}\right]\right].\label{eq:proof-sketch-eq1}
\end{equation}
From the SLAN property (\ref{eq:LAN}), we can show that $\tilde{p}_{n}(\cdot\vert{\bf y}_{nq}),\tilde{\bar{P}}_{n}({\bf y}_{n})$
converge uniformly over $q$ in the total-variation metric to $p_{n}(\cdot\vert{\bf y}_{nq}),\bar{P}_{n}({\bf y}_{n})$;
see \ref{eq:bound on A_n original}-\ref{bound on A_n approximate}
in Appendix \ref{subsec:Proof-of-Theorem-4} for the precise claim.
This in turn implies that
\begin{equation}
\lim_{n\to\infty}\sup_{\pi\in\Pi}\left|V_{\pi,n}(0)-\tilde{V}_{\pi,n}(0)\right|=0.\label{eq:pf:proof-sketch-eq2}
\end{equation}

Now, $\tilde{\bar{P}}_{n}$ is not a probability measure, even as
it integrates to 1 asymptotically. We therefore modify $\tilde{\mathbb{E}}_{n}[\cdot]$
slightly to make it a `true' expectation, leading to another approximation,
$\breve{V}_{\pi,n}(0)$, of $\tilde{V}_{\pi,n}(0)$, such that $\lim_{n\to\infty}\sup_{\pi\in\Pi}\left|\breve{V}_{\pi,n}(0)-\tilde{V}_{\pi,n}(0)\right|=0$
(see step 2 in Appendix \ref{subsec:Proof-of-Theorem-4}). Following
this adjustment and using dynamic-programming, the optimization problem
$\inf_{\pi\in\Pi}\breve{V}_{\pi,n}(0)$ can written in a recursive
form akin to (\ref{eq:discrete approximation}), see (\ref{eq:pf:Thm:5:recursive problem})
in Appendix \ref{subsec:Proof-of-Theorem-4}. Inspection of this recursive
form shows $\inf_{\pi\in\Pi}\breve{V}_{\pi,n}(0)=\inf_{\pi\in\Pi^{\mathcal{S}}}\breve{V}_{\pi,n}(0)$.
Intuitively, this is because $\tilde{\mathbb{E}}[\cdot\vert{\bf y}_{nq}]$
is a function only of $x_{nq},q$, while $\Lambda_{n,h}({\bf y}_{n})$,
which was used to define the approximate marginal $\tilde{\bar{P}}_{n}({\bf y}_{n})$,
has a similar form to a Gaussian likelihood that depends only on $x_{nq},q$
as well. This proves the first claim. For the second claim, similar
arguments as in the proof of Theorem \ref{Thm: Convergence to minimal PDE}
show that the solution to the recursive problem converges to the solution
of PDE (\ref{eq:PDE optimal bayes risk}). 

For the last claim, observe that (\ref{eq:pf:proof-sketch-eq2}) also
implies $\lim_{n\to\infty}V_{\pi_{\Delta t}^{*},n}(0)-\tilde{V}_{\pi_{\Delta t}^{*}n}(0)=0$.
We then approximate $\tilde{V}_{\pi_{\Delta t}^{*}n}(0)$ with $\breve{V}_{\pi_{\Delta t}^{*}n}(0)$,
write the latter again in recursive form, and argue as in the proof
of Theorem \ref{Thm: Approximate control-2} that $\lim_{n\to\infty}\vert\breve{V}_{\pi_{\Delta t}^{*}n}(0-V^{*}(0)\vert\lesssim\Delta t{}^{1/4}$. 

\newpage{}

\section*{Supplementary appendix}

\section{Rates of convergence to the PDE solution\label{subsec:Rates-of-convergence}}

The results of \citet[Theorem 3.1]{barles2007error} provide a bound
on the rate of convergence of $V_{n}^{*}(\cdot)$ to $V^{*}(\cdot)$.
The technical requirements to obtain this are described in their Assumptions
A2 and S1-S3. Assumptions A2 and S1-S2 are straightforward to verify
using the regularity conditions given for Theorem \ref{Thm: Convergence to minimal PDE}
with the additional requirement $\sup_{s}\vert\mu^{+}(s)\vert<\infty$. 

Assumption S3 of \citet{barles2007error} is a strengthening of the
consistency requirement in (\ref{eq:upper consistency}) and (\ref{eq:lower consistency}).
Suppose that the test function $\phi\in\mathcal{C}^{\infty}(\mathcal{S})$
is such that $\left|\partial_{t}^{\beta_{0}}D_{(x,q)}^{\beta}\phi(x,q,t)\right|\le K\varepsilon^{1-2\beta_{0}-\left\Vert \beta\right\Vert }$
for all $\beta_{0}\in\mathbb{N},\beta\in\mathbb{N}\times\mathbb{N}$.
Then by a third order Taylor expansion as in the proof of Theorem
\ref{Thm: Convergence to minimal PDE} and some tedious but straightforward
algebra,
\[
\left|nS_{n}(z,\phi(z)+\rho,[\phi+\rho])-F(D^{2}\phi(s),D\phi(s),s)\right|\le E(n,\varepsilon)\equiv\frac{\bar{K}}{n^{1/2}\varepsilon^{2}},
\]
where $\bar{K}$ depends only on $K$, defined above, and the upper
bounds on $\mu^{+}(\cdot),\mu(\cdot)$. The above suffices to verify
the Assumption S3 of \citet{barles2007error}; note that the definition
of $S(\cdot)$ in that paper is equivalent to $nS_{n}(\cdot)$ here. 

\begin{figure}
\includegraphics[height=5cm]{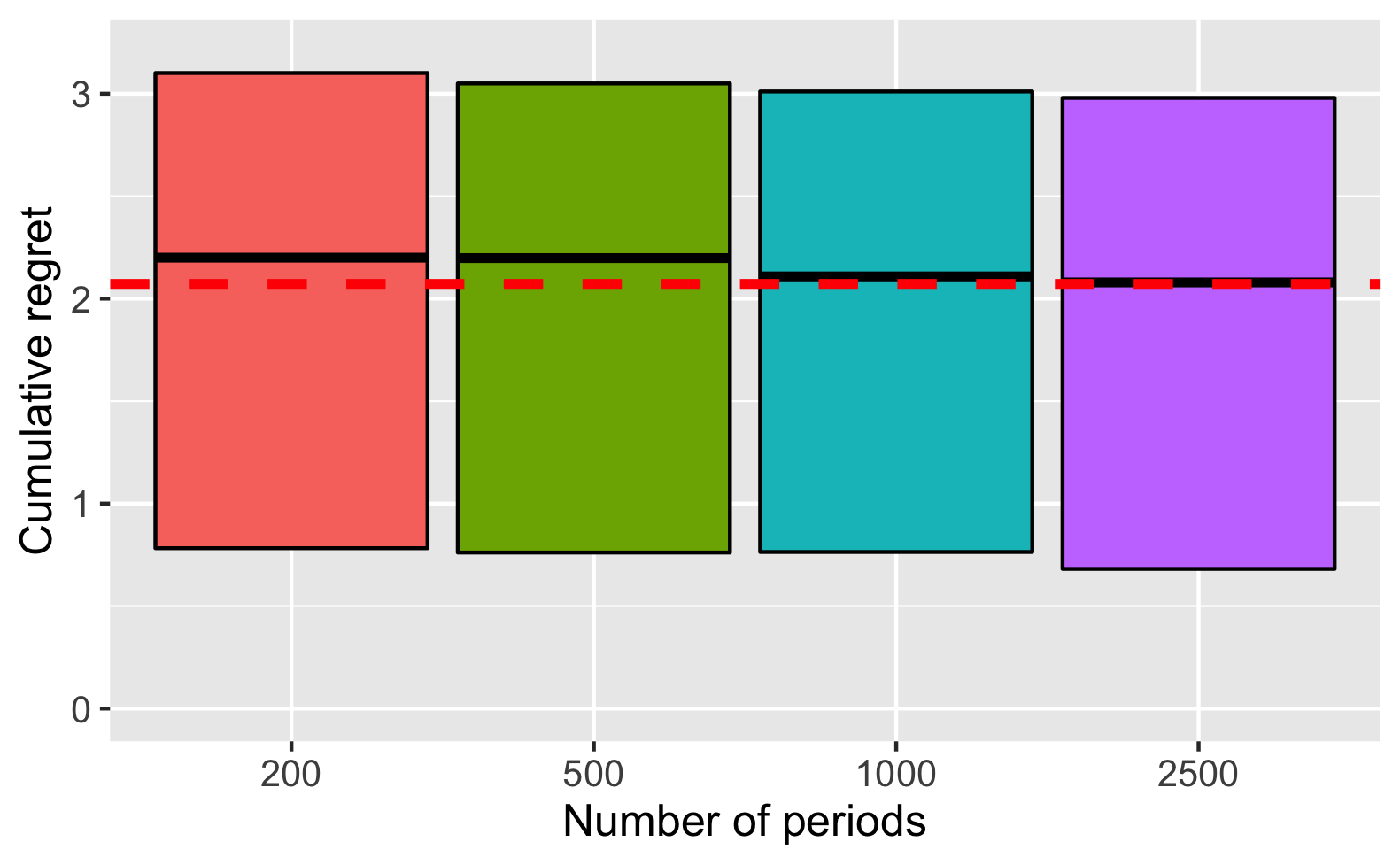}~\includegraphics[height=5cm]{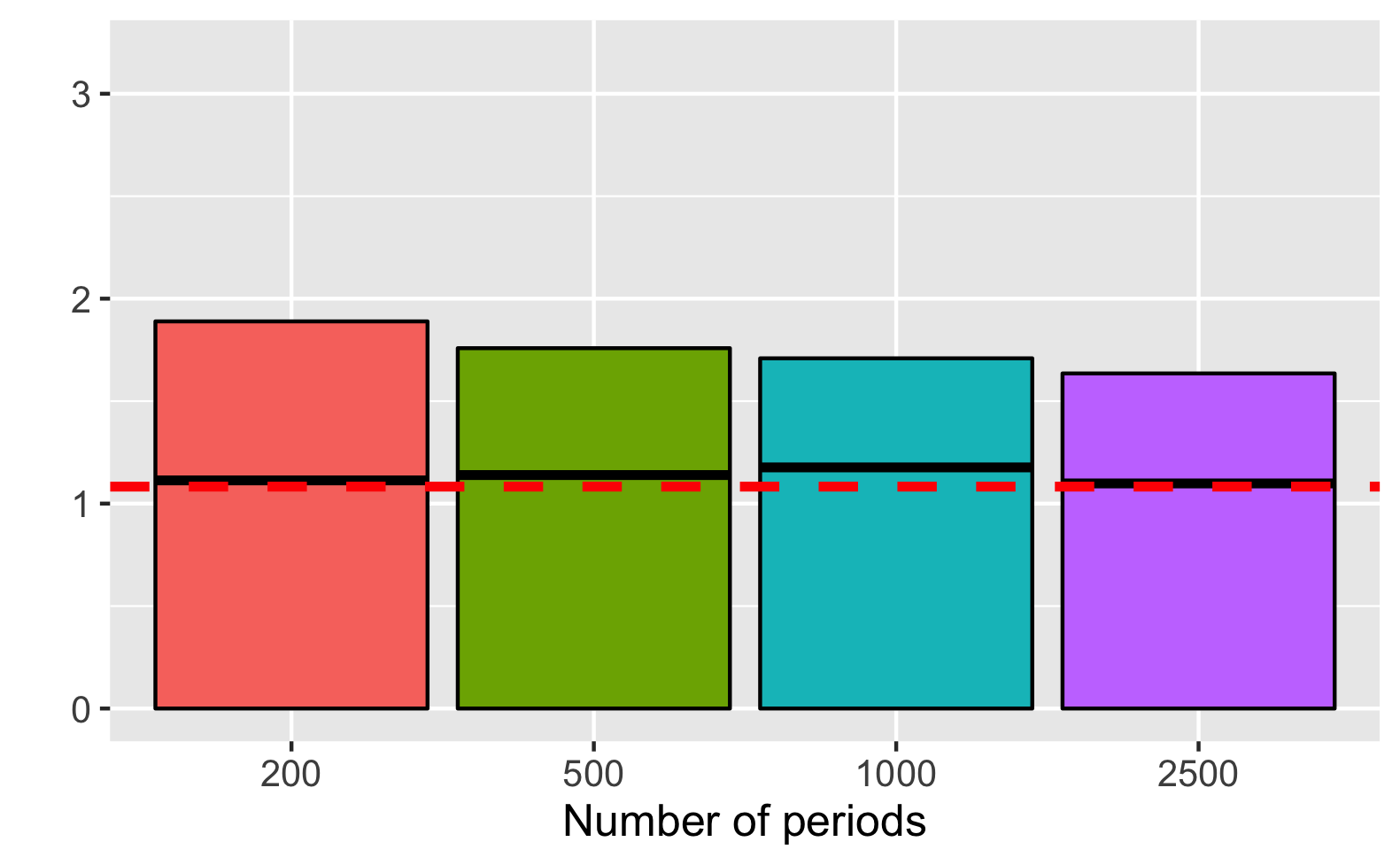}~

\begin{tabular}{>{\centering}p{7.5cm}>{\centering}p{7.5cm}}
{\scriptsize A: Thompson sampling} & {\scriptsize B: Optimal Bayes policy}\tabularnewline
\end{tabular}
\begin{raggedright}
{\scriptsize Note: The parameter values are $\mu_{0}=0$, $\nu=50$
and $\sigma=5$. The dashed red lines denote the values of asymptotic
Bayes risk. Black lines within the bars denote the Bayes risk in finite
samples. The bars describe the interquartile range of regret.}{\scriptsize\par}
\par\end{raggedright}
\caption{Monte-Carlo simulations\label{fig:Monte Carlo}}
\end{figure}

Under the above conditions, \citet[Theorem 3.1]{barles2007error}
implies
\begin{align}
V^{*}-V_{n}^{*} & \apprle\sup_{\varepsilon}\left(\varepsilon+E(n,\varepsilon)\right)\apprle n^{-1/6}\ \textrm{and }\nonumber \\
V_{n}^{*}-V^{*} & \apprle\sup_{\varepsilon}\left(\varepsilon^{1/3}+E(n,\varepsilon)\right)\apprle n^{-1/14}.\label{eq:approximation rate}
\end{align}
The asymmetry of the rates is an artifact of the techniques of \citet{barles2007error}.
The rates are also far from optimal. The results of \citet{barles2007error},
while being relatively easy to apply, do not exploit any regularity
properties of the approximation scheme. There do exist approximation
schemes for PDE (\ref{eq:PDE optimal bayes risk}) that converge at
the faster $n^{-1/2}$ rates. While it is unknown whether (\ref{eq:discrete approximation})
is one of them, we do find that in practice the quality of approximation
of $V^{*}$ with $V_{n}^{*}$ is far better than what (\ref{eq:approximation rate})
appears to suggest; the Monte-Carlo simulation in Figure \ref{fig:Monte Carlo}
attests to this (the simulation employs a normal prior $\mu\sim\mathcal{N}(0,50^{2})$
with $\sigma=5$).

\section{Lower bounds on minimax risk\label{sec:lower bounds on minimax risk}}

Recall the definition of $V_{n,\pi}(0;h)$ from Section \ref{subsec:Lower-bound-on}
as the frequentist risk under some $\pi\in\Pi$. We also make the
dependence of $V_{n}^{*}(0),V^{*}(0)$ on the priors $m_{0}$ explicit
by writing them as $V_{n}^{*}(0;m_{0}),V^{*}(0;m_{0})$. Clearly,
$\inf_{\pi\in\Pi}\sup_{\vert h\vert\le\Gamma}V_{n,\pi}(0;h)\ge V_{n}^{*}(0;m_{0})$
for any prior $m_{0}$ supported on $\vert h\vert\le\Gamma$. So,
Theorem \ref{Thm: General parametric families} implies 
\[
\lim_{n\to\infty}\inf_{\pi\in\Pi}\sup_{\vert h\vert\le\Gamma}V_{n,\pi}(0;h)\ge\sup_{m_{0}\in\mathcal{P}}V^{*}(0;m_{0})
\]
where $\mathcal{P}$ is the set of all compactly supported distributions.
We now claim that 
\begin{equation}
\sup_{m_{0}\in\mathcal{P}}V^{*}(0;m_{0})=\bar{V}^{*},\label{eq:minimax lower bound derivation}
\end{equation}
where $\bar{V}^{*}$ is the asymptotic minimax risk in the Gaussian
setting. The above is easily shown for scalar $\theta$ by transforming
the state variable $x$ to $\dot{\mu}_{0}x$ and replacing $\sigma^{2}$
with $\dot{\mu}_{0}^{2}\sigma^{2}$, following which the infinitesimal
generator (\ref{eq:infinitesimal generator - parametric}) becomes
equivalent to the one in (\ref{eq:PDE optimal bayes risk}) since
$\mu(s)=\dot{\mu}_{0}h(s)$. The argument for vector $\theta$ is
given below.

\subsubsection{Proof of (\ref{eq:minimax lower bound derivation}) for vector $\theta$}

We employ the same notation as in Section \ref{subsec:Vector-valued}.
It is without loss of generality to suppose $\Sigma=I$, otherwise,
we can perform the subsequent analysis after applying the transformations
$h\leftarrow\Sigma^{-1/2}h,x\leftarrow\Sigma^{-1/2}x$ and $\dot{\mu}_{0}\leftarrow\Sigma^{1/2}\dot{\mu_{0}}$.
Consider the class, $\bar{\mathcal{P}}$, of priors, $m_{0}$, over
$h$ supported on $\mu\cdot\dot{\mu}_{0}/(\dot{\mu}_{0}^{\intercal}\mu_{0})$,
where $\mu\in\mathbb{R}$ can take on various values (so $m_{0}$
is, in essence, a prior on $\mu$). For these priors, $\dot{\mu}_{0}^{\intercal}h=\mu$.
Recall that under the approximate posterior, $\tilde{p}_{n}(h\vert x,q)\propto\mathcal{N}(x\vert qh,q\Sigma)\cdot m_{0}(h)$.
It is then easily verified that, for the class $\bar{\mathcal{P}}$,
$\tilde{p}_{n}(h\vert x,q)$ depends on $x$ only through $\dot{\mu}_{0}^{\intercal}x$.
Furthermore, we also have $h(s)=\mu(s)\cdot\dot{\mu}_{0}/(\dot{\mu}_{0}^{\intercal}\mu_{0})$,
where $\mu(s),h(s)$ are the posterior means of $\mu,h$ under $\tilde{p}_{n}(\cdot\vert x,q)$. 

Choose $\{\phi_{i}\}_{i=1}^{d-1}$ such that $\{\dot{\mu}_{0}/\dot{\mu}_{0}^{\intercal}\mu_{0},\phi_{1},\dots,\phi_{d-1}\}$
are orthonormal and span $\mathbb{R}^{d}$. Suppose we transform the
state variables $x$ to $z$ as $z=Px$, where $P^{\intercal}=[\dot{\mu}_{0},\phi_{1},\dots,\phi_{d-1}]$.
Clearly, $P$ is invertible, and the first component of $z$ is $\bar{x}:=\dot{\mu}_{0}^{\intercal}x$.
Consider the generator $L[\cdot]$ in (\ref{eq:vector infinitesimal generator}).
Following the transformation of variables,
\begin{align*}
h(s)^{\intercal}D_{x}f & =\frac{\mu(s)}{\dot{\mu}_{0}^{\intercal}\dot{\mu}_{0}}\dot{\mu}_{0}^{\intercal}\cdot P^{\intercal}D_{z}f=\mu(s)\cdot\left[1,\bm{0}_{1\times(d-1)}\right]\cdot D_{z}f=\mu(s)\partial_{\bar{x}}f,
\end{align*}
and $\textrm{Tr}\left[D_{x}^{2}f\right]=\textrm{Tr}\left[PP^{\intercal}\cdot D_{z}^{2}f\right]$.
Clearly, $PP^{\intercal}$ is block diagonal, with diagonal entries
$\dot{\mu}_{0}^{\intercal}\dot{\mu}_{0}$ and $I_{(d-1)}$. Hence,
we can write $\textrm{Tr}\left[D_{x}^{2}f\right]=(\dot{\mu}_{0}^{\intercal}\dot{\mu}_{0})\cdot\partial_{\bar{x}}^{2}f+\textrm{Tr}\left[D_{\tilde{x}}^{2}f\right]$
where $\tilde{x}$ is the part of $z$ excluding the first component.
Combining the above, and defining $\sigma^{2}:=\dot{\mu}_{0}^{\intercal}\dot{\mu}_{0}$
(more generally, for $\Sigma\neq I$, this would be $\dot{\mu}_{0}^{\intercal}\Sigma\dot{\mu}_{0}$),
we have thus shown $L[f](s)=\partial_{q}f+\mu(s)\partial_{\bar{x}}f+\frac{1}{2}\sigma^{2}\partial_{\bar{x}}^{2}f+\frac{1}{2}\textrm{Tr}\left[D_{\tilde{x}}^{2}f\right].$

The minimal Bayes risk, $V^{*}(s;m_{0})$, solves the PDE: 
\begin{align*}
\partial_{t}f(s)+\mu^{+}(s)+\min\left\{ -\mu(s)+L[f](s),0\right\}  & =0\ \textrm{if }t<1;\quad f(s)=0\ \textrm{if }t=1.
\end{align*}
Now, $\tilde{p}_{n}(h\vert x,q)$ depends on $x$ only though $\bar{x}$,
so $\mu(s)\equiv\mathbb{\tilde{E}}[\mu\vert s],\mu^{+}(s)\equiv\mathbb{\tilde{E}}[\mu\mathbb{I}\{\mu\ge0\}\vert s]$
are functions only of $\bar{x},q$. Hence, by similar viscosity solution
arguments as in the proof of Theorem \ref{Thm: Non-parametric models}
(Appendix \ref{subsec:Additional details for non-parametric}), it
follows that $V^{*}(s;m_{0})$ solves 
\begin{align*}
\partial_{t}f(\bar{s})+\mu^{+}(\bar{s})+\min\left\{ -\mu(\bar{s})+\bar{L}[f](\bar{s}),0\right\}  & =0\ \textrm{if }t<1;\quad f(\bar{s})=0\ \textrm{if }t=1,
\end{align*}
where $\bar{s}:=(\bar{x},q,t)$ and $\bar{L}[f](\bar{s})=\partial_{q}f+\mu(\bar{s})\partial_{\bar{x}}f+\frac{1}{2}\sigma^{2}\partial_{\bar{x}}^{2}f$.
But the above has the same form as PDE (\ref{eq:PDE optimal bayes risk})
in the Gaussian setting if we interpret $m_{0}$ as a prior on $\mu$.
Hence, $\sup_{m_{0}\in\bar{\mathcal{P}}}V^{*}(0;m_{0})=\bar{V}^{*}$,
the minimax risk in the Gaussian regime. 

Since $\bar{\mathcal{P}}\subset\mathcal{P}$, the set of all compactly
supported priors on $\bm{h}$, we have thereby derived a lower bound
on minimax risk. As an aside, we note that our proof also goes through
after replacing $\bar{\mathcal{P}}$ with the class of product priors
defined in Section \ref{sec:The-nonparametric-setting}; the argument
would then be similar to the proof of Theorem \ref{Thm: Non-parametric models},
see Appendix \ref{subsec:Additional details for non-parametric}.

\section{Proof of Theorem \ref{Thm: General parametric families}\label{subsec:Proof-of-Theorem-4}}

Recall that ${\bf y}_{i}=\{Y_{k}\}_{k=1}^{i}$ denotes the rewards
after $i$ pulls of the arms. Denote by $\mathbb{E}_{({\bf y}_{n},h})[\cdot]$
the expectation under the `true' joint density $dS_{n}({\bf y}_{n},h):=\left\{ \prod_{i=1}^{n}p_{\theta_{0}+h/\sqrt{n}}(Y_{i})\right\} \cdot m_{0}(h).$
Let $\nu({\bf y}_{n}):=\prod_{i=1}^{n}\nu(Y_{i})$, $p_{n,\theta}({\bf y}_{n}):=\prod_{k=1}^{n}p_{\theta}(Y_{k})$
and $\bar{P}_{n}$ be the probability measure corresponding to the
`true' marginal density $d\bar{P}_{n}({\bf y}_{n}):=\int p_{n,\theta_{0}+h/\sqrt{n}}({\bf y}_{n})\cdot m_{0}(h)d\nu(h)$.
We use $\bar{\mathbb{E}}_{n}[\cdot]$ to denote its corresponding
expectation. As first defined in Appendix \ref{subsec:Proof-sketch-of-Theorem 5},
let $\tilde{\bar{P}}_{n}$ denote the measure (but not necessarily
a probability) corresponding to the density $d\tilde{\bar{P}}_{n}({\bf y}_{n}):=\int d\Lambda_{n,h}({\bf y}_{n})\cdot m_{0}(h)d\nu(h)$.
In what follows, we denote $d\Lambda_{n,h}({\bf y}_{n})$ by $\lambda_{n,h}({\bf y}_{n})$
for ease of notation, and note that
\[
\lambda_{n,h}({\bf y}_{n}):=d\Lambda_{n,h}({\bf y}_{n})\equiv\frac{d\Lambda_{n,h}({\bf y}_{n})}{d\nu({\bf y}_{n})}=\exp\left\{ \frac{1}{\sigma^{2}}hx_{n}-\frac{1}{2\sigma^{2}}h^{2}\right\} p_{n,\theta_{0}}({\bf y}_{n}).
\]
Finally, $\left\Vert \cdot\right\Vert _{\textrm{TV}}$ denotes the
total variation metric between two measures. 

The proof follows the basic outline established in Appendix \ref{subsec:Proof-sketch-of-Theorem 5}.
Recall the notation used there, as well as the expressions for $V_{\pi,n}(0),\tilde{V}_{\pi,n}(0)$
given in (\ref{eq:proof-sketch-eq0}) and (\ref{eq:proof-sketch-eq1}). 

\subsubsection*{Step 1 (Approximation of $V_{\pi,n}(0)$ with $\tilde{V}_{\pi,n}(0)$):}

We start by proving some convergence properties of $\tilde{\bar{P}}_{n}$
and $\tilde{p}_{n}(\cdot\vert{\bf y}_{nq})$ to $\bar{P}_{n}$ and
$p_{n}(\cdot\vert{\bf y}_{nq})$. The proofs here make heavy use of
the SLAN property (\ref{eq:LAN}) established in Lemma \ref{Lem: Uniform LAN}.
Let $A_{n}$ denote the event $\left\{ {\bf y}_{n}:\sup_{q}\vert x_{nq}\vert\le M\right\} $.
For any measure $P$, define $P\cap A_{n}$ as the restriction of
$P$ to the set $A_{n}$. By Lemma \ref{Lem: Properties of approimate likelihood}
in Appendix \ref{subsec:Supporting-lemmas-for}, for any $\epsilon>0$
there exists $M<\infty$ such that
\begin{align}
\lim_{n\to\infty}\bar{P}_{n}\left(A_{n}^{c}\right) & \le\epsilon,\label{eq:bound on A_n original}\\
\lim_{n\to\infty}\left\Vert \bar{P}_{n}\cap A_{n}-\tilde{\bar{P}}_{n}\cap A_{n}\right\Vert _{\textrm{TV}} & =0,\ \textrm{and}\label{eq:TV bound for marginals}\\
\lim_{n\to\infty}\sup_{q}\bar{\mathbb{E}}_{n}\left[\mathbb{I}_{A_{n}}\left\Vert p_{n}(\cdot\vert{\bf y}_{nq})-\tilde{p}_{n}(\cdot\vert{\bf y}_{nq})\right\Vert _{\textrm{TV}}\right] & =0.\label{eq:TV bound for posteriors}
\end{align}

The measures $\Lambda_{n,h}(\cdot),\tilde{\bar{P}}_{n}(\cdot)$ are
not probabilities as they need not integrate to $1$. But Lemma \ref{Lem: Properties of approimate likelihood}
also shows the following: $\Lambda_{n,h}(\cdot),\tilde{\bar{P}}_{n}(\cdot)$
are $\sigma$-finite and contiguous with respect to $P_{n,\theta_{0}}$,
and letting $\mathcal{Y}_{n}$ denote the sample space of ${\bf y}_{n}$,
\begin{equation}
\text{\ensuremath{\lim_{n\to\infty}\tilde{\bar{P}}_{n}}(\ensuremath{\mathcal{Y}_{n}})}=1\quad\textrm{and }\lim_{n\to\infty}\tilde{\bar{P}}_{n}(A_{n}^{c})\le\epsilon.\label{bound on A_n approximate}
\end{equation}
The first result in (\ref{bound on A_n approximate}) implies that
$\tilde{\bar{P}}_{n}$ is almost a probability measure. 

Based on the above, we show that 
\begin{equation}
\lim_{n\to\infty}\sup_{\pi\in\Pi}\left|V_{\pi,n}(0)-\tilde{V}_{\pi,n}(0)\right|=0\label{eq:uniform convergence of risks}
\end{equation}
by bounding each term in the following expansion:
\begin{align}
 & V_{\pi,n}(0)-\tilde{V}_{\pi,n}(0)\nonumber \\
 & =\bar{\mathbb{E}}_{n}\left[\mathbb{I}_{A_{n}^{c}}\frac{1}{n}\sum_{j=0}^{n-1}\mathbb{E}\left[\left.R_{n}(h,\pi_{j+1})\right|{\bf y}_{nq_{j}(\pi)}\right]\right]+\tilde{\mathbb{E}}_{n}\left[\mathbb{I}_{A_{n}^{c}}\frac{1}{n}\sum_{j=0}^{n-1}\tilde{\mathbb{E}}\left[\left.R_{n}(h,\pi_{j+1})\right|{\bf y}_{nq_{j}(\pi)}\right]\right]\nonumber \\
 & \quad+\left(\bar{\mathbb{E}}_{n}-\tilde{\mathbb{E}}_{n}\right)\left[\mathbb{I}_{A_{n}}\frac{1}{n}\sum_{j=0}^{n-1}\tilde{\mathbb{E}}\left[\left.R_{n}(h,\pi_{j+1})\right|{\bf y}_{nq_{j}(\pi)}\right]\right]\nonumber \\
 & \quad+\bar{\mathbb{E}}_{n}\left[\mathbb{I}_{A_{n}}\frac{1}{n}\sum_{j=0}^{n-1}\left\{ \mathbb{E}\left[\left.R_{n}(h,\pi_{j+1})\right|{\bf y}_{nq_{j}(\pi)}\right]-\tilde{\mathbb{E}}\left[\left.R_{n}(h,\pi_{j+1})\right|{\bf y}_{nq_{j}(\pi)}\right]\right\} \right].\label{eq:step 2: decomposition}
\end{align}
Because of the compact support of the prior, the posteriors $p_{n}(\cdot\vert{\bf y}_{nq}),\tilde{p}_{n}(\cdot\vert{\bf y}_{nq})$
are also compactly supported on $\vert h\vert\le\Gamma$ for all $q$.
On this set $\vert R_{n}(h,\pi_{j})\vert\le b\Gamma$ for some $b<\infty$
by Assumption 1(iii). The first two quantities in (\ref{eq:step 2: decomposition})
are therefore bounded by $b\Gamma\bar{P}_{n}(A_{n}^{c})$ and $b\Gamma\tilde{\bar{P}}_{n}(A_{n}^{c})$.
By (\ref{eq:bound on A_n original}) and (\ref{bound on A_n approximate}),
these can be made arbitrarily small by choosing a suitably large $M$
in the definition of $A_{n}$. The third term in (\ref{eq:step 2: decomposition})
is bounded by $b\Gamma\left\Vert \bar{P}_{n}\cap A_{n}-\tilde{\bar{P}}_{n}\cap A_{n}\right\Vert _{\textrm{TV}}$.
By (\ref{eq:TV bound for marginals}) it converges to $0$ as $n\to\infty$.
The expression within $\{\}$ brackets in the fourth term of (\ref{eq:step 2: decomposition})
is smaller than $b\Gamma\left\Vert p_{n}(\cdot\vert{\bf y}_{nq_{j}(\pi)})-\tilde{p}_{n}(\cdot\vert{\bf y}_{nq_{j}(\pi)})\right\Vert _{\textrm{TV}}$.
Hence, by the linearity of expectations, the term overall is bounded
(uniformly over $\pi\in\Pi$) by 
\[
b\Gamma\sup_{q}\bar{\mathbb{E}}_{n}\left[\mathbb{I}_{A_{n}}\left\Vert p_{n}(\cdot\vert{\bf y}_{nq})-\tilde{p}_{n}(\cdot\vert{\bf y}_{nq})\right\Vert _{\textrm{TV}}\right],
\]
which is $o(1)$ because of (\ref{eq:TV bound for posteriors}). We
have thus shown (\ref{eq:uniform convergence of risks}).

\subsubsection*{Step 2 (Approximating $V_{n}^{*}(0)$ with a recursive formula):}

The measure, $\tilde{\bar{P}}_{n}$ , used in the outer expectation
in the definition of $\tilde{V}_{\pi,n}(0)$ is not a probability.
This can be rectified as follows: First, note that the density $\lambda_{n,h}(\cdot)$
can be written as 
\begin{equation}
\lambda_{n,h}({\bf y}_{n})=\prod_{i=1}^{n}\left\{ \exp\left\{ \frac{h}{\sqrt{n}}\psi(Y_{i})-\frac{h^{2}}{2\sigma^{2}n}\right\} p_{\theta_{0}}(Y_{i})\right\} =\prod_{i=1}^{n}\tilde{p}_{n}(Y_{i}\vert h),\label{eq:expansion of lambda_n}
\end{equation}
where\footnote{Despite the notation, $\tilde{p}_{n}(Y_{i}\vert h)$ is not a probability
density.}
\[
\tilde{p}_{n}(Y_{i}\vert h):=\exp\left\{ \frac{h}{\sqrt{n}}\psi(Y_{i})-\frac{h^{2}}{2\sigma^{2}n}\right\} p_{\theta_{0}}(Y_{i}).
\]
Using (\ref{eq:expansion of lambda_n}), Lemma \ref{Lem: Disintegration}
shows that $\tilde{\bar{P}}_{n}$ can be disintegrated as
\begin{align}
d\tilde{\bar{P}}_{n}({\bf y}_{n}) & =\prod_{i=1}^{n}\left\{ \int\tilde{p}_{n}(Y_{i}\vert h)\tilde{p}_{n}(h\vert{\bf y}_{i-1})d\nu(h)\right\} ,\label{eq:disintegration of tilde P_n}
\end{align}
with $\tilde{p}_{n}(h\vert{\bf y}_{0}):=m_{0}(h)$. Now define $c_{n,i}:=\int\left\{ \int\tilde{p}_{n}(Y_{i}\vert h)d\nu(Y_{i})\right\} \tilde{p}_{n}(h\vert{\bf y}_{i-1})d\nu(h)$,
and let $\tilde{\mathbb{P}}_{n}$ denote the probability measure
\begin{align}
\tilde{\mathbb{P}}_{n}({\bf y}_{n}) & =\prod_{i=1}^{n}\tilde{\mathbb{P}}_{n}(Y_{i}\vert{\bf y}_{i-1}),\ \textrm{where}\nonumber \\
d\tilde{\mathbb{P}}_{n}(Y_{i}\vert{\bf y}_{i-1}) & :=\frac{1}{c_{n,i}}\int\tilde{p}_{n}(Y_{i}\vert h)\tilde{p}_{n}(h\vert{\bf y}_{i-1})d\nu(h).\label{eq:disintegration result}
\end{align}
Note that $c_{n,i}$ is a random (because it depends on ${\bf y}_{i-1}$)
integration factor ensuring $\tilde{\mathbb{P}}_{n}(y_{i+1}\vert{\bf y}_{i})$,
and therefore $\tilde{\mathbb{P}}_{n}$, is a probability. In Lemma
\ref{Lem: Approximation of tilde P_n}, it is shown that there exists
some non-random $C<\infty$ such that
\begin{equation}
\sup_{i}\vert c_{n,i}-1\vert\le Cn^{-c}\ \textrm{for any }c<3/2,\label{eq:bound on integration factor}
\end{equation}
and furthermore, $\left\Vert \tilde{\mathbb{P}}_{n}-\tilde{\bar{P}}_{n}\right\Vert _{\textrm{TV}}\to0$
as $n\to\infty$. Hence, letting
\[
\breve{V}_{\pi,n}(0):=\mathbb{E}_{\tilde{\mathbb{P}}_{n}}\left[\frac{1}{n}\sum_{j=0}^{n-1}\tilde{\mathbb{E}}\left[\left.R_{n}(h,\pi_{j+1})\right|{\bf y}_{nq_{j}(\pi)}\right]\right],
\]
where $\mathbb{E}_{\tilde{\mathbb{P}}_{n}}[\cdot]$ is the expectation
with respect to $\tilde{\mathbb{P}}_{n}$, one obtains the approximation
\begin{equation}
\sup_{\pi\in\Pi}\left|\tilde{V}_{\pi,n}(0)-\breve{V}_{\pi,n}(0)\right|\le b\Gamma\left\Vert \tilde{\mathbb{P}}_{n}-\tilde{\bar{P}}_{n}\right\Vert _{\textrm{TV}}\to0.\label{eq:step 4 - approximation}
\end{equation}
See the arguments following (\ref{eq:step 2: decomposition}) for
the definition of $b$. 

Since $\tilde{p}_{n}(h\vert{\bf y}_{i-1})\equiv\tilde{p}_{n}(h\vert x=x_{i-1},q=(i-1)/n)$
by (\ref{eq:posterior - general parametric models}) with $\tilde{p}_{n}(h\vert x=0,q=0):=m_{0}(h)$,
it follows from (\ref{eq:disintegration result}) that $\tilde{\mathbb{P}}_{n}(Y_{i}\vert{\bf y}_{i-1})\equiv\tilde{\mathbb{P}}_{n}(Y_{i}\vert x=x_{i-1},q=(i-1)/n)$.

Define $\breve{V}_{n}^{*}(0)=\inf_{\pi\in\Pi}\breve{V}_{\pi,n}(0)$.
Recall that for a given $\pi\in\{0,1\}$, $\tilde{\mathbb{E}}\left[\left.R_{n}(h,\pi)\right|{\bf y}_{nq_{j}}\right]\equiv\tilde{\mathbb{E}}\left[\left.R_{n}(h,\pi)\right|x_{nq_{j}},q_{j}\right]$
by (\ref{eq:posterior - general parametric models}). Furthermore,
we have noted above that the conditional distribution of the future
values of the rewards, $\tilde{\mathbb{P}}_{n}(Y_{nq_{j}+1}\vert{\bf y}_{nq_{j}})$,
also depends only on $(x_{nq_{j}},q_{j})$. Based on this, standard
backward induction/dynamic programming arguments imply $\breve{V}_{n}^{*}(0)$
can be obtained as the solution at $(x,q,t)=(0,0,0)$ of the recursive
problem
\begin{align}
\breve{V}_{n}^{*}\left(x,q,t\right) & =\min_{\pi\in\{0,1\}}\left\{ \frac{\tilde{\mathbb{E}}\left[\left.R_{n}(h,\pi)\right|x,q\right]}{n}+\mathbb{E}_{\tilde{\mathbb{P}}_{n}}\left[\left.\mathbb{I}_{n}\cdot\breve{V}_{n}^{*}\left(x+\frac{\pi\sigma^{2}\psi(Y_{nq+1})}{\sqrt{n}},q+\frac{\pi}{n},t+\frac{1}{n}\right)\right|s\right]\right\} ;\nonumber \\
 & \quad\textrm{if }t<1,\nonumber \\
\breve{V}_{n}^{*}\left(x,q,1\right) & =0,\label{eq:pf:Thm:5:recursive problem}
\end{align}
where $\mathbb{E}_{\tilde{\mathbb{P}}_{n}}\left[\left.\cdot\right|s\right]$
denotes the expectation under $\tilde{\mathbb{P}}_{n}(Y_{nq+1}\vert{\bf y}_{nq})\equiv\tilde{\mathbb{P}}_{n}(Y_{nq+1}\vert x=x_{nq},q)$
and $\mathbb{I}_{n}=\mathbb{I}\{t\le1-1/n\}$. 

Now, Step 2 and (\ref{eq:step 4 - approximation}) imply $\lim_{n\to\infty}\vert V_{n}^{*}(0)-\breve{V}_{n}^{*}(0)\vert=0$.
But, the value $\pi^{*}\in\{0,1\}$ that attains the minimum in (\ref{eq:pf:Thm:5:recursive problem})
depends only on $s$. We would have thus obtained the approximation,
$\breve{V}_{n}^{*}(0)$, to $V_{n}^{*}(0)$ even if we restricted
the policy class to $\Pi^{\mathcal{S}}$. This proves the first claim
of the theorem.

\subsubsection*{Step 3 (Auxiliary results for showing PDE approximation of (\ref{eq:pf:Thm:5:recursive problem})):}

We now state a couple of results that will be used to show that the
solution, $\breve{V}_{n}^{*}(\cdot)$, to (\ref{eq:pf:Thm:5:recursive problem})
converges to the solution of a PDE. 

The first result is that, for any given $\pi\in\{0,1\}$, $\tilde{\mathbb{E}}\left[\left.R_{n}(h,\pi)\right|x,q\right]$
can be approximated by $\mu^{+}(s)-\pi\mu(s)$ uniformly over $(x,q)$.
To this end, denote $\bar{R}(h,\pi)=\dot{\mu}_{0}h\left(\mathbb{I}(\dot{\mu}_{0}h>0)-\pi\right)$.
Assumption 1(iii) implies $\sup_{\vert h\vert\le\Gamma}\vert\mu_{n}(h)-\dot{\mu}_{0}h/\sqrt{n}\vert\le\Gamma^{2}\delta_{n}/\sqrt{n}$.
Combining this with Lipschitz continuity of $x\mathbb{I}(x>0)-\pi x$
gives
\[
\sup_{\vert h\vert\le\Gamma;\pi\in\{0,1\}}\left|R_{n}(h,\pi)-\bar{R}(h,\pi)\right|\le2\Gamma^{2}\delta_{n}.
\]
Recalling the definitions of $\mu^{+}(s),\mu(s)$ from the main text,
the above implies
\begin{equation}
\sup_{(x,q);\pi\in\{0,1\}}\left|\tilde{\mathbb{E}}\left[\left.R_{n}(h,\pi)\right|x,q\right]-\left(\mu^{+}(s)-\pi\mu(s)\right)\right|\le2\Gamma^{2}\delta_{n}\to0.\label{eq:pf:Thm5:Step 4: approximation of rewards}
\end{equation}

The next result is given as Lemma \ref{Lem: Auxiliary results for Step 5}
in Appendix \ref{subsec:Supporting-lemmas-for}. It states that there
exists $\xi_{n}\to0$ independent of both $s$ and $\pi\in\{0,1\}$
such that 
\begin{align}
\sqrt{n}\sigma^{2}\mathbb{E}_{\tilde{\mathbb{P}}_{n}}\left[\left.\pi\psi(Y_{nq+1})\right|s\right] & =\pi h(s)+\xi_{n},\ \textrm{and}\label{eq:pf:Thm5:exp under posterior}\\
\sigma^{4}\mathbb{E}_{\tilde{\mathbb{P}}_{n}}\left[\left.\pi\psi^{2}(Y_{nq+1})\right|s\right] & =\pi\sigma^{2}+\xi_{n}.\label{eq:pf:Thm5:variance under post approximation}
\end{align}
Furthermore, 
\begin{equation}
\mathbb{E}_{\tilde{\mathbb{P}}_{n}}\left[\left.\left|\psi(Y_{nq+1})\right|^{3}\right|s\right]<\infty.\label{eq:pf:Thm5:Step 4: bound on 4th moment}
\end{equation}

\subsubsection*{Step 4 (PDE approximation of (\ref{eq:pf:Thm:5:recursive problem})):}

The unique solution, $\breve{V}_{n}^{*}(s)$, to (\ref{eq:pf:Thm:5:recursive problem})
converges locally uniformly to $V_{n}^{*}(s)$, the viscosity solution
to PDE (\ref{eq:PDE optimal bayes risk}). This follows by similar
arguments as in the proof of Theorem \ref{Thm: Convergence to minimal PDE}:

Clearly the scheme defined in (\ref{eq:pf:Thm:5:recursive problem})
is monotonic. Assumption 1(iii) implies there exists $b<\infty$ such
that $\sup_{\pi,\vert h\vert\le\Gamma}\vert R_{n}(h,\pi)\vert\le b\Gamma$.
Hence, the solution to (\ref{eq:pf:Thm:5:recursive problem}) is uniformly
bounded, with $\vert\breve{V}_{n}^{*}(s)\vert\le b\Gamma$ independent
of $s$ and $n$. This proves stability. Finally, consistency of the
scheme follows by similar arguments as in the proof of Theorem \ref{Thm: Convergence to minimal PDE},
after making use of (\ref{eq:pf:Thm5:Step 4: approximation of rewards})
and (\ref{eq:pf:Thm5:exp under posterior}) - (\ref{eq:pf:Thm5:Step 4: bound on 4th moment}). 

This completes the proof of the second claim of the theorem.

\subsubsection*{Step 5 (Proof of the third claim):}

Steps 1 and 2 imply $\lim_{n\to\infty}V_{\pi_{\Delta t}^{*},n}(0)-\breve{V}_{\pi_{\Delta t}^{*}n}(0)=0$.
In addition, we can follow the arguments in Step 2 to express $\breve{V}_{\pi_{\Delta t}^{*}n}(0)$
in recursive form, in a manner similar to the definition of $V_{\Delta t,n,l}^{*}(\cdot)$
in the proof of Theorem \ref{Thm: Approximate control-2}; the only
difference is that the operator $\tilde{S}_{\Delta t}\left[\phi\right](x,q)$
in that proof should now read as the solution at $(x,q,\Delta t)$
of the recursive equation
\begin{align*}
f\left(x,q,\tau\right) & =\frac{\tilde{\mathbb{E}}\left[\left.R_{n}(h,1)\right|x,q\right]}{n}+\mathbb{E}_{\tilde{\mathbb{P}}_{n}}\left[\left.f\left(x+\frac{\sigma^{2}\psi(Y_{nq+1})}{\sqrt{n}},q+\frac{1}{n},\tau-\frac{1}{n}\right)\right|s\right];\ \tau>0\\
f\left(x,q,0\right) & =\phi(x,q).
\end{align*}
Now, an application of \citet[Theorem 3.1]{barles2007error}, using
(\ref{eq:pf:Thm5:Step 4: approximation of rewards}) - (\ref{eq:pf:Thm5:Step 4: bound on 4th moment})
to verify the requirements (cf.\ Appendix \ref{subsec:Rates-of-convergence}),
gives $\left|\tilde{S}_{\Delta t}\left[V_{\Delta t,l+1}^{*}\right]-S_{\Delta t}\left[V_{\Delta t,l+1}^{*}\right]\right|\lesssim\min\left\{ n^{-1/14},\xi_{n},\delta_{n}\right\} $.
The rest of the proof is analogous to that of Theorem \ref{Thm: Approximate control-2}.

\section{Supporting lemmas for the proof of Theorem \ref{Thm: General parametric families}\label{subsec:Supporting-lemmas-for}}

We implicitly assume Assumption 1 for all the results in this section
apart from Lemma \ref{Lemma 1}. 

\begin{lem} \label{Lemma 1}Let $p(Y\vert h)$ denote the likelihood
of $Y$ given some parameter $h$ with prior distribution $m_{0}(h)$.
Under the one-armed bandit experiment, the posterior distribution,
$p_{n}(\cdot\vert\mathcal{F}_{t})$, of $h$ given all information
until time $t$ satisfies
\begin{equation}
p_{n}(h\vert\mathcal{F}_{t})\propto\left\{ \prod_{i=1}^{\left\lfloor nq(t)\right\rfloor }p(Y_{i}\vert h)\right\} \cdot m_{0}(h).\label{eq:True posterior-1}
\end{equation}
In particular, the posterior distribution is independent of the past
values of actions. \end{lem} 
\begin{proof}
Note that $\mathcal{F}_{t}$ is the sigma-algebra generated by $\xi_{t}\equiv\{\{A_{j}\}_{j=1}^{\left\lfloor nt\right\rfloor },\{Y_{i}\}_{i=1}^{\left\lfloor nq(t)\right\rfloor }\}$;
here, $j$ refers to the time period while $i$ refers to number of
pulls of the arm. The claim is shown using induction. Clearly, it
is true for $t=1$. For any $t>1$, we can think of $p_{n}(h\vert\xi_{t-1})$
as the revised prior for $\mu$. Suppose that $A_{t}=1$. Then $nq(t)=nq(t-1)+1$,
and 
\begin{align*}
p_{n}(h\vert\xi_{t}) & \propto p(Y_{t},A_{t}=1\vert\xi_{t},h)\cdot p_{n}(h\vert\xi_{t-1})\\
 & \propto\pi(A_{t}=1\vert\xi_{t-1})\cdot p(Y_{t}\vert h)\cdot p_{n}(h\vert\xi_{t-1})\\
 & \propto p(Y_{t}\vert h)\cdot p_{n}(h\vert\xi_{t-1})=\left\{ \prod_{i=1}^{\left\lfloor nq(t)\right\rfloor }p(Y_{i}\vert h)\right\} \cdot m_{0}(h).
\end{align*}
Alternatively, suppose $A_{t}=0$. Then, $nq(t)=nq(t-1)$, and $p(A_{t}=0\vert\xi_{t},h)=\pi(A_{t}=0\vert\xi_{t})$
is independent of $h$, so
\begin{align*}
p_{n}(h\vert\xi_{t}) & \propto p(A_{t}=0\vert\xi_{t},h)\cdot p_{n}(h\vert\xi_{t-1})\\
 & \propto p_{n}(h\vert\xi_{t-1})=\left\{ \prod_{i=1}^{\left\lfloor nq(t)\right\rfloor }p(Y_{i}\vert h)\right\} \cdot m_{0}(h).
\end{align*}
Thus the induction step holds under both possibilities, and the claim
follows.
\end{proof}
\begin{lem} \label{Lem: Uniform LAN} Suppose $P_{\theta}$ is quadratic
mean differentiable as in (\ref{eq:QMD}). Then $P_{\theta}$ satisfies
the SLAN property as defined in (\ref{eq:LAN}). \end{lem} 
\begin{proof}
The proof builds on \citet[Theorem 7.2]{van2000asymptotic}. Set $p_{n}:=dP_{\theta_{0}+h/\sqrt{n}}/d\nu$,
$p_{0}:=dP_{\theta_{0}}/d\nu$ and $W_{ni}:=2\left[\sqrt{p_{n}/p_{0}}(Y_{i})-1\right]$.
We use $E[\cdot]$ to denote expectations with respect to $P_{n,\theta_{0}}.$
Quadratic mean differentiability implies $E[\psi(Y_{i})]=0$ and $E[\psi^{2}(Y_{i})]=1/\sigma^{2}$,
see \citet[Theorem 7.2]{van2000asymptotic}.

It is without loss of generality for this proof to take the domain
of $q$ to be $\{0,1/n,2/n,\dots,1\}$. For any such $q$, 
\[
E\left[\sum_{i=1}^{nq}W_{ni}\right]=2nq\left(\int\sqrt{p_{n}\cdot p_{0}}d\nu-1\right)=-nq\int\left(\sqrt{p_{n}}-\sqrt{p_{0}}\right)^{2}d\nu.
\]
Now, (\ref{eq:QMD}) implies there exists $\epsilon_{n}\to0$ such
that
\[
\left|n\int\left(\sqrt{p_{n}}-\sqrt{p_{0}}\right)^{2}d\nu-\frac{h^{2}}{4\sigma^{2}}\right|\lesssim\epsilon_{n}h^{2}.
\]
Hence, for any given $h$, 
\begin{equation}
\sup_{q}\left|E\left[\sum_{i=1}^{nq}W_{ni}\right]-\frac{qh^{2}}{4\sigma^{2}}\right|\to0.\label{eq:pf:LAN-2}
\end{equation}

Next, denote $Z_{ni}=W_{ni}-h\psi(Y_{i})/\sqrt{n}-E[W_{ni}]$ and
$S_{nq}=\sum_{i=1}^{nq}Z_{ni}$. Observe that $E[Z_{ni}]=0$ since
$E[\psi(Y_{i})]=0$. Furthermore, by (\ref{eq:QMD}), 
\begin{equation}
\textrm{Var}[\sqrt{n}Z_{ni}]=E\left[\left(\sqrt{n}W_{ni}-h\psi(Y_{i})\right)^{2}\right]\lesssim\epsilon_{n}h^{2}\to0.\label{eq:pf:LAN-1}
\end{equation}
Now, an application of Kolmogorov's maximal inequality for partial
sum processes gives 
\[
P\left(\sup_{q}\left|S_{nq}\right|\ge\lambda\right)\le\frac{1}{\lambda^{2}}\textrm{Var}\left[\sum_{i=1}^{n}Z_{ni}\right]=\frac{1}{\lambda^{2}}\textrm{Var}[\sqrt{n}Z_{ni}].
\]
Combined with (\ref{eq:pf:LAN-2}) and (\ref{eq:pf:LAN-1}), the above
implies 
\begin{equation}
\sum_{i=1}^{nq}W_{ni}=\frac{h}{\sqrt{n}}\sum_{i=1}^{nq}\psi(Y_{i})-\frac{qh^{2}}{4\sigma^{2}}+o_{P_{n,\theta_{0}}}(1)\ \textrm{uniformly over q}.\label{eq:pf:LAN-3}
\end{equation}

We now employ a Taylor expansion of the logarithm $\ln(1+x)=x-\frac{1}{2}x^{2}+x^{2}R(2x)$
where $R(x)\to0$ as $x\to0$, to expand the log-likelihood as
\begin{align}
\ln\prod_{i=1}^{nq}\frac{p_{n}}{p_{0}}(Y_{i}) & =2\sum_{i=1}^{nq}\ln\left(1+\frac{1}{2}W_{ni}\right)\nonumber \\
 & =\sum_{i=1}^{nq}W_{ni}-\frac{1}{4}\sum_{i=1}^{nq}W_{ni}^{2}+\frac{1}{2}\sum_{i=1}^{nq}W_{ni}^{2}R(W_{ni}).\label{eq:pf:LAN-4}
\end{align}
Because of (\ref{eq:pf:LAN-1}), we can write $\sqrt{n}W_{ni}=h\psi(Y_{i})+C_{ni}$
where $E[\vert C_{ni}\vert^{2}]\to0$. Defining $A_{ni}:=2h\psi(Y_{i})C_{ni}+C_{ni}^{2}$,
some straightforward algebra then gives $nW_{ni}^{2}=h^{2}\psi^{2}(Y_{i})+A_{ni}$
with $E[\vert A_{ni}\vert]\to0$. Now, by the uniform law of large
numbers for partial sum processes, see e.g., \citet{bass1984strong},
$n^{-1}\sum_{i=1}^{nq}h^{2}\psi^{2}(Y_{i})$ converges uniformly in
$P_{n,\theta_{0}}$-probability to $qh^{2}/\sigma^{2}$. Furthermore,
$E\left[\sup_{q}n^{-1}\sum_{i=1}^{nq}\vert A_{ni}\vert\right]\le E\left[n^{-1}\sum_{i=1}^{n}\vert A_{ni}\vert\right]=E[\vert A_{ni}\vert]\to0$
and therefore $n^{-1}\sum_{i=1}^{nq}A_{ni}$ converges uniformly in
$P_{n,\theta_{0}}$-probability to $0$. These results yield
\[
\sum_{i=1}^{nq}W_{ni}^{2}=\frac{qh^{2}}{\sigma^{2}}+o_{P_{n,\theta_{0}}}(1)\ \textrm{uniformly over q}.
\]
Next, by the triangle inequality and Markov's inequality
\begin{align*}
nP_{n,\theta_{0}}\left(\vert W_{ni}\vert>\varepsilon\sqrt{2}\right) & \le nP_{n,\theta_{0}}\left(h^{2}\psi^{2}(Y_{i})>n\varepsilon^{2}\right)+nP_{n,\theta_{0}}\left(\vert A_{ni}\vert>n\varepsilon^{2}\right)\\
 & \le\varepsilon^{-2}h^{2}E\left[\psi^{2}(Y_{i})\mathbb{I}\left\{ \psi^{2}(Y_{i})>n\varepsilon^{2}\right\} \right]+\varepsilon^{-2}E\left[\vert A_{ni}\vert\right]\to0
\end{align*}
for any given $h$. The above implies $\max_{1\le i\le n}\vert W_{ni}\vert=o_{P_{n,\theta_{0}}}(1)$
and consequently, $\max_{1\le i\le n}\vert R(W_{ni})\vert=o_{P_{n,\theta_{0}}}(1)$.
The last term on the right hand side of (\ref{eq:pf:LAN-4}) is bounded
by $\max_{1\le i\le n}\vert R(W_{ni})\vert\cdot\sum_{i=1}^{n}W_{ni}^{2}$
and is therefore $o_{P_{n,\theta_{0}}}(1)$ by the above results.
We thus conclude 
\[
\ln\prod_{i=1}^{nq}\frac{p_{n}}{p_{0}}(Y_{i})=\sum_{i=1}^{nq}W_{ni}-\frac{qh^{2}}{4\sigma^{2}}+o_{P_{n,\theta_{0}}}(1)\ \textrm{uniformly over q}.
\]
The claim follows by combining the above with (\ref{eq:pf:LAN-3}).
\end{proof}
\begin{lem} \label{Lem: bound on A_n}For any $\epsilon>0$, there
exist $M(\epsilon),N(\epsilon)<\infty$ such that $M\ge M(\epsilon)$
and $n\ge N(\epsilon)$ implies $\bar{P}_{n}(A_{n}^{c})<\epsilon$.
Furthermore, letting $A_{n}^{q}=\left\{ {\bf y}_{nq}:\sup_{\tilde{q}\le q}\vert x_{n\tilde{q}}\vert<M\right\} $,
and $\mathbb{E}_{n,0}[\cdot]$, the expectation under $P_{n,\theta_{0}}$,
\[
\sup_{q}\mathbb{E}_{n,0}\left[\mathbb{I}_{A_{n}^{q}}\left\Vert \frac{dP_{nq,\theta_{0}+h/\sqrt{n}}}{dP_{nq,\theta_{0}}}({\bf y}_{nq})-\frac{d\Lambda_{nq,h}}{dP_{nq,\theta_{0}}}({\bf y}_{nq})\right\Vert \right]=o(1)\ \forall\ \{h:\vert h\vert\le\Gamma\}.
\]
 \end{lem} 
\begin{proof}
Set $A_{n,M}=\left\{ {\bf y}_{n}:\sup_{q}\vert x_{nq}\vert<M\right\} $
and $P_{nq,h}=P_{nq,\theta_{0}+h/\sqrt{n}}$. Note that $x_{nq}$
is a partial sum process with mean $0$ under $P_{n,0}:=P_{n,\theta_{0}}$.
By Kolmogorov's maximal inequality, $P_{n,0}\left(\sup_{q}\vert x_{nq}\vert\ge M\right)\le M^{-1}\textrm{Var}[x_{n}]=M^{-1}\sigma^{2}$.
Hence, $P_{n.0}(A_{n,M_{n}}^{c})\to0$ for any $M_{n}\to\infty$.
But by (\ref{eq:LAN}) and standard arguments involving Le Cam's first
lemma, $P_{n,h}$ is contiguous to $P_{n,0}$ for all $h$. This implies
$\bar{P}_{n}:=\int P_{n,h}dm_{0}(h)$ is also contiguous to $P_{n,0}$
(this can be shown using the dominated convergence theorem; see also,
\citeauthor[p.138]{le2000asymptotics}). Consequently, $\bar{P}_{n}(A_{n,M_{n}}^{c})\to0$
for any $M_{n}\to\infty$. The first claim is a straightforward consequence
of this. 

For the second claim, we follow \citet[Proposition 6.2]{le2000asymptotics}:

We first argue that $P_{nq_{n},h}$ is contiguous to $P_{nq_{n},0}$
for any deterministic sequence $\{q_{n}\}$ such that $q_{n}\to\bar{q}\in[0,1]$.
We have 
\begin{align}
\ln\frac{dP_{nq_{n},h}}{dP_{nq_{n},0}} & =\frac{1}{\sigma^{2}}hx_{nq_{n}}-\frac{q_{n}}{2\sigma^{2}}h^{2}+o_{P_{n,0}}(1)\nonumber \\
 & \xrightarrow[P_{n,0}]{d}N\left(-\frac{\bar{q}h^{2}}{2\sigma^{2}},\frac{\bar{q}h^{2}}{\sigma^{2}}\right),\label{eq:pf:Bound on A_n - 1}
\end{align}
where the equality follows from (\ref{eq:LAN}), and the weak convergence
limit follows from: (i) weak convergence of $x_{nq}$ under $P_{n,0}$
to a Brownian motion process $W(q)$, see e.g., \citet[Chapter 2.12]{van1996weak},
and (ii) the extended continuous mapping theorem, see \citet[Theorem 1.11.1]{van1996weak}.
Since $E_{P_{n,0}}[f({\bf y}_{nq_{n}})]=E_{P_{nq_{n},0}}[f({\bf y}_{nq_{n}})]$
for any $f(\cdot)$, we conclude from (\ref{eq:pf:Bound on A_n - 1})
and the definition of weak convergence that 
\[
\ln\frac{dP_{nq_{n},h}}{dP_{nq_{n},0}}\xrightarrow[P_{nq_{n},0}]{d}N\left(-\frac{\bar{q}h^{2}}{2\sigma^{2}},\frac{\bar{q}h^{2}}{\sigma^{2}}\right).
\]
An application of Le Cam's first lemma then implies $P_{nq_{n},h}$
is contiguous to $P_{nq_{n},0}$. 

Now, let $q_{n}\in[0,1]$ denote a quantity such that
\[
\sup_{q}\mathbb{E}_{n,0}\left[\mathbb{I}_{A_{n}^{q}}\left\Vert \frac{dP_{nq,h}}{dP_{nq,0}}-\frac{d\Lambda_{nq,h}}{dP_{nq,0}}\right\Vert \right]\le\mathbb{E}_{n,0}\left[\mathbb{I}_{A_{n}^{q_{n}}}\left\Vert \frac{dP_{nq_{n},h}}{dP_{nq_{n},0}}-\frac{d\Lambda_{nq_{n},h}}{dP_{nq_{n},0}}\right\Vert \right]+\epsilon
\]
for some arbitrarily small $\epsilon\ge0$ (such a $q_{n},\epsilon$
always exist by the definition of the supremum). Without loss of generality,
we may assume $q_{n}$ converges to some $\bar{q}\in[0,1]$; otherwise
we can employ a subsequence argument since $q_{n}$ lies in a bounded
set. Define 
\[
G_{n}(q):=\mathbb{I}_{A_{n}^{q_{n}}}\left\Vert \frac{dP_{nq,h}}{dP_{nq,0}}-\frac{d\Lambda_{nq,h}}{dP_{nq,0}}\right\Vert .
\]
The claim follows if we show $\mathbb{E}_{n,0}\left[G_{n}(q_{n})\right]\to0$.
By Lemma \ref{Lem: Uniform LAN} and the definition of $\Lambda_{nq,h}(\cdot)$,
\[
G_{n}(q)=\mathbb{I}_{A_{n}^{q_{n}}}\cdot\exp\left\{ \frac{1}{\sigma^{2}}hx_{nq}-\frac{q}{2\sigma^{2}}h^{2}\right\} \left(\exp\delta_{n,q}-1\right),
\]
where $\sup_{q}\vert\delta_{n,q}\vert=o(1)$ under $P_{n,0}$. Since
$\mathbb{I}_{A_{n}^{q_{n}}}\cdot\exp\left\{ \frac{1}{\sigma^{2}}hx_{nq_{n}}-\frac{q_{n}}{2\sigma^{2}}h^{2}\right\} $
is bounded for $\vert h\vert\le\Gamma$ by the definition of $\mathbb{I}_{A_{n}^{q}}$,
this implies $G_{n}(q_{n})=o(1)$ under $P_{n,0}$. Next, we argue
$G_{n}(q_{n})$ is uniformly integrable. The term $\mathbb{I}_{A_{n}^{q_{n}}}\cdot d\Lambda_{nq_{n},h}/dP_{nq_{n},0}$
in the definition of $G_{n}(q_{n})$ is bounded, and therefore uniformly
integrable, for $\vert h\vert\le\Gamma$. We now prove uniform integrability
of $dP_{nq_{n},h}/dP_{nq_{n},0}$, and thereby that of the remaining
term, $\mathbb{I}_{A_{n}^{q_{n}}}\cdot dP_{nq_{n},h}/dP_{nq_{n},0}$,
in the definition of $G_{n}(q_{n})$. For any $b<\infty$, 
\begin{align*}
\mathbb{E}_{n,0}\left[\frac{dP_{nq_{n},h}}{dP_{nq_{n},0}}\mathbb{I}\left\{ \frac{dP_{nq_{n},h}}{dP_{nq_{n},0}}>b\right\} \right] & =\int\frac{dP_{nq_{n},h}}{dP_{nq_{n},0}}\mathbb{I}\left\{ \frac{dP_{nq_{n},h}}{dP_{nq_{n},0}}>b\right\} dP_{nq_{n},0}\\
 & \le P_{nq_{n},h}\left(\frac{dP_{nq_{n},h}}{dP_{nq_{n},0}}>b\right).
\end{align*}
But, 
\[
P_{nq_{n},0}\left(\frac{dP_{nq_{n},h}}{dP_{nq_{n},0}}>b\right)\le b^{-1}\int\frac{dP_{nq_{n},h}}{dP_{nq_{n},0}}dP_{nq_{n},0}\le b^{-1},
\]
so the contiguity of $P_{nq_{n},h}$ with respect to $P_{nq_{n},0}$
implies we can choose $b$ and $\bar{n}$ large enough such that 
\[
\limsup_{n\ge\bar{n}}P_{nq_{n},h}\left(\frac{dP_{nq_{n},h}}{dP_{nq_{n},0}}>b\right)<\epsilon
\]
for any arbitrarily small $\epsilon$. These results demonstrate uniform
integrability of $G_{n}(q_{n})$ under $P_{n,0}$. Since convergence
in probability implies convergence in expectation for uniformly integrable
random variables, we have thus shown $\mathbb{E}_{n,0}\left[G_{n}(q_{n})\right]\to0$,
which concludes the proof.
\end{proof}
\begin{lem} \label{Lem: approximation of marginals} $\lim_{n\to\infty}\left\Vert \bar{P}_{n}\cap A_{n}-\tilde{\bar{P}}_{n}\cap A_{n}\right\Vert _{\textrm{TV}}=0$.
\end{lem} 
\begin{proof}
Set $P_{n,h}:=P_{n,\theta_{0}+h/\sqrt{n}}.$ By the properties of
the total variation metric, contiguity of $\bar{P}_{n}$ with respect
to $P_{n,0}$ and the absolute continuity of $\Lambda_{n,h}$ with
respect to $P_{n,0}$,
\begin{align*}
 & \lim_{n\to\infty}\left\Vert \bar{P}_{n}\cap A_{n}-\tilde{\bar{P}}_{n}\cap A_{n}\right\Vert _{\textrm{TV}}\\
 & =\frac{1}{2}\lim_{n\to\infty}\int\left\{ \int\mathbb{I}_{A_{n}}\left|\frac{dP_{n,h}}{dP_{n,0}}({\bf y}_{n})-\frac{d\Lambda_{n,h}}{dP_{n,0}}({\bf y}_{n})\right|dP_{n,0}({\bf y}_{n})\right\} m_{0}(h)d\nu(h).
\end{align*}
In the last expression, denote the term within the $\{\}$ brackets
by $f_{n}(h)$. By Lemma \ref{Lem: bound on A_n}, $f_{n}(h)\to0$
for each $h$. Additionally, $\mathbb{I}_{A_{n}}\cdot\left(d\Lambda_{n,h}/dP_{n,0}\right)$
is bounded because of the definition of $A_{n}$ and the fact $\vert h\vert\le\Gamma$,
while 
\[
\int\mathbb{I}_{A_{n}}\left|\frac{dP_{n,h}}{dP_{n,0}}\right|dP_{n,0}\le\int\frac{dP_{n,h}}{dP_{n,0}}dP_{n,0}\le1.
\]
Hence, $f_{n}(h)$ is dominated by a (suitably large) constant for
all $n$. The dominated convergence theorem then implies $\int f_{n}(h)m_{0}(h)d\nu(h)\to0$.
This proves the claim.
\end{proof}
\begin{lem} \label{Lem: approximation of posteriors} $\sup_{q}\bar{\mathbb{E}}_{n}\left[\mathbb{I}_{A_{n}}\left\Vert p_{n}(\cdot\vert{\bf y}_{nq})-\tilde{p}_{n}(\cdot\vert{\bf y}_{nq})\right\Vert _{\textrm{TV}}\right]=o(1)$.
\end{lem} 
\begin{proof}
Set $P_{n,h}=P_{n,\theta_{0}+h/\sqrt{n}}$, $p_{nq,h}({\bf y}_{nq})=dP_{nq,h}({\bf y}_{nq})/d\nu$,
$\lambda_{nq,h}({\bf y}_{nq})=d\Lambda_{nq,h}({\bf y}_{nq})/d\nu$,
$\bar{p}_{nq}({\bf y}_{nq})=d\bar{P}_{nq}({\bf y}_{nq})/d\nu$ and
$\tilde{\bar{p}}_{nq}({\bf y}_{nq})=d\tilde{\bar{P}}_{nq}({\bf y}_{nq})/d\nu$.
Let $S_{nq}$ and $\tilde{S}_{nq}$ denote joint measures over $({\bf y}_{nq},h)$,
corresponding to $dS_{nq}({\bf y}_{nq},h)=p_{nq,h}({\bf y}_{nq})\cdot m_{0}(h)$
and $d\tilde{S}_{nq}({\bf y}_{nq},h)=\lambda_{nq,h}({\bf y}_{nq})\cdot m_{0}(h)$. 

In the main text, we introduced the approximate posterior $\tilde{p}_{n}(h\vert{\bf y}_{nq})$.
Formally, this is defined via the disintegration $d\tilde{S}_{nq}({\bf y}_{nq},h)=\tilde{p}_{n}(h\vert{\bf y}_{nq})\cdot d\tilde{\bar{P}}_{n}({\bf y}_{nq})$,
where $d\tilde{\bar{P}}_{n}({\bf y}_{nq}):=\int\left\{ d\tilde{S}_{nq}({\bf y}_{nq},h)\right\} d\nu(h)$.
Such a conditional probability always exists, see, e.g., \citet[p. 136]{le2000asymptotics}.
In a similar vein, we can disintegrate $dS_{nq}=p_{n}(h\vert{\bf y}_{nq})\cdot\bar{p}_{nq}({\bf y}_{nq})$.
Since $p_{n}(h\vert{\bf y}_{nq}),\tilde{p}_{n}(h\vert{\bf y}_{nq})$
are both conditional probabilities, we obtain $\bar{p}_{nq}({\bf y}_{nq})=\int p_{nq,h}({\bf y}_{nq})m_{0}(h)d\nu(h)$
and $\tilde{\bar{p}}_{nq}({\bf y}_{nq})=\int\lambda_{nq,h}({\bf y}_{nq})m_{0}(h)d\nu(h)$. 

Define $\Omega_{n}\equiv\{{\bf y}_{n}:p_{n,0}({\bf y}_{n})\neq0\}$.
Since the total variation metric is bounded by $1$ and $\bar{P}_{n}$
is contiguous with respect to $P_{n,0}$, 
\[
\sup_{q}\bar{\mathbb{E}}_{n}\left[\mathbb{I}_{A_{n}}\left\Vert p_{n}(\cdot\vert{\bf y}_{nq})-\tilde{p}_{n}(\cdot\vert{\bf y}_{nq})\right\Vert _{\textrm{TV}}\right]=\sup_{q}\bar{\mathbb{E}}_{n}\left[\mathbb{I}_{A_{n}\cap\Omega_{n}}\left\Vert p_{n}(\cdot\vert{\bf y}_{nq})-\tilde{p}_{n}(\cdot\vert{\bf y}_{nq})\right\Vert _{\textrm{TV}}\right]+o(1).
\]
Now, by the properties of the total variation metric and the disintegration
formula, 
\begin{align*}
2\left\Vert p_{n}(\cdot\vert{\bf y}_{nq})-\tilde{p}_{n}(\cdot\vert{\bf y}_{nq})\right\Vert _{\textrm{TV}} & =\int\left|p_{n}(h\vert{\bf y}_{nq})-\tilde{p}_{n}(h\vert{\bf y}_{nq})\right|d\nu(h)\\
 & =\int\left|\frac{p_{nq,h}({\bf y}_{nq})\cdot m_{0}(h)}{\bar{p}_{nq}({\bf y}_{nq})}-\frac{\lambda_{nq,h}({\bf y}_{nq})\cdot m_{0}(h)}{\tilde{\bar{p}}_{nq}({\bf y}_{nq})}\right|d\nu(h).
\end{align*}
Hence, 
\begin{align*}
 & 2\bar{\mathbb{E}}_{n}\left[\mathbb{I}_{A_{n}\cap\Omega_{n}}\left\Vert p_{n}(\cdot\vert{\bf y}_{nq})-\tilde{p}_{n}(\cdot\vert{\bf y}_{nq})\right\Vert _{\textrm{TV}}\right]\\
 & \le\bar{\mathbb{E}}_{n}\left[\mathbb{I}_{A_{n}\cap\Omega_{n}}\int\frac{\left|p_{nq,h}({\bf y}_{nq})-\lambda_{nq,h}({\bf y}_{nq})\right|}{\bar{p}_{nq}({\bf y}_{nq})}m_{0}(h)d\nu(h)\right]\\
 & \qquad+\bar{\mathbb{E}}_{n}\left[\mathbb{I}_{A_{n}\cap\Omega_{n}}\int\lambda_{nq,h}({\bf y}_{nq})\left|\frac{1}{\bar{p}_{nq}({\bf y}_{nq})}-\frac{1}{\tilde{\bar{p}}_{nq}({\bf y}_{nq})}\right|m_{0}(h)d\nu(h)\right]\\
 & :=B_{1n}(q)+B_{2n}(q)
\end{align*}

We start by bounding $\sup_{q}B_{1n}(q)$. Recall the definition of
$A_{n}^{q}\supseteq A_{n}$ from the statement of Lemma \ref{Lem: bound on A_n}.
By Fubini's theorem and the definition of $\bar{p}_{nq}(\cdot)$ as
the density of $\bar{P}_{nq}$, 
\begin{align}
B_{1n}(q) & \le\int\left\{ \int\mathbb{I}_{A_{n}^{q}\cap\Omega_{n}}\left|p_{nq,h}({\bf y}_{nq})-\lambda_{nq,h}({\bf y}_{nq})\right|d\nu({\bf y}_{nq})\right\} m_{0}(h)d\nu(h)\nonumber \\
 & \le\int\left\{ \int\mathbb{I}_{A_{n}^{q}}\left|\frac{dP_{nq,h}}{dP_{nq,0}}({\bf y}_{nq})-\frac{d\Lambda_{nq,h}}{dP_{nq,0}}({\bf y}_{nq})\right|dP_{nq,0}({\bf y}_{nq})\right\} m_{0}(h)d\nu(h),\label{eq:pf:lem:posterior approx: 1}
\end{align}
the change of measure to $P_{nq,0}$ in the last inequality being
allowed under $\Omega_{n}$. Hence, 
\[
\sup_{q}B_{1n}(q)\le\int\left\{ \sup_{q}\int\mathbb{I}_{A_{n}^{q}}\left|\frac{dP_{nq,h}}{dP_{nq,0}}({\bf y}_{nq})-\frac{d\Lambda_{nq,h}}{dP_{nq,0}}({\bf y}_{nq})\right|dP_{nq,0}({\bf y}_{nq})\right\} m_{0}(h)d\nu(h).
\]
In the above expression, denote the term within the $\{\}$ brackets
by $g_{n}(h)$. By Lemma \ref{Lem: bound on A_n}, $g_{n}(h)\to0$
for each $h$. Furthermore, by similar arguments as in the proof of
Lemma \ref{Lem: approximation of marginals}, $g_{n}(h)$ is bounded
by a constant for all $n$ (it is easy to see that the bound derived
there applies uniformly over all $q$). The dominated convergence
theorem then gives $\int g_{n}(h)m_{0}(h)d\nu(h)\to0$, and therefore,
$\sup_{q}B_{1n}(q)=o(1)$. 

We now turn to $B_{2n}(q)$. The disintegration formula implies $\lambda_{nq,h}({\bf y}_{nq})\cdot m_{0}(h)=\tilde{\bar{p}}_{nq}({\bf y}_{nq})\cdot\tilde{p}_{n}(h\vert{\bf y}_{nq})$.
So, 
\begin{align}
B_{2n}(q) & =\bar{\mathbb{E}}_{n}\left[\mathbb{I}_{A_{n}\cap\Omega_{n}}\int\tilde{p}_{n}(h\vert{\bf y}_{nq})\left|\frac{\tilde{\bar{p}}_{nq}({\bf y}_{nq})-\bar{p}_{nq}({\bf y}_{nq})}{\bar{p}_{nq}({\bf y}_{nq})}\right|d\nu(h)\right]\nonumber \\
 & =\bar{\mathbb{E}}_{n}\left[\mathbb{I}_{A_{n}\cap\Omega_{n}}\left|\frac{\tilde{\bar{p}}_{nq}({\bf y}_{nq})-\bar{p}_{nq}({\bf y}_{nq})}{\bar{p}_{nq}({\bf y}_{nq})}\right|\right]\nonumber \\
 & \le\int\mathbb{I}_{A_{n}^{q}\cap\Omega_{n}}\left|\tilde{\bar{p}}_{nq}({\bf y}_{nq})-\bar{p}_{nq}({\bf y}_{nq})\right|d\nu({\bf y}_{nq}).\label{eq:pf:lem:posterior approx: 2}
\end{align}
By the integral representation for $\tilde{\bar{p}}_{nq}({\bf y}_{nq}),\bar{p}_{nq}({\bf y}_{nq})$
the right hand side of (\ref{eq:pf:lem:posterior approx: 2}) equals
\begin{align}
 & \int\mathbb{I}_{A_{n}^{q}\cap\Omega_{n}}\left|\int\frac{d\Lambda_{nq,h}}{dP_{nq,0}}({\bf y}_{nq})dm_{0}(h)-\int\frac{dP_{nq,h}}{dP_{nq,0}}({\bf y}_{nq})dm_{0}(h)\right|dP_{nq,0}({\bf y}_{nq})\nonumber \\
 & \le\int\left\{ \int\mathbb{I}_{A_{n}^{q}}\left|\frac{d\Lambda_{nq,h}}{dP_{nq,0}}({\bf y}_{nq})-\frac{dP_{nq,h}}{dP_{nq,0}}({\bf y}_{nq})\right|dP_{nq,0}({\bf y}_{nq})\right\} m_{0}(h)d\nu(h),\label{eq:pf:lem:posterior approx: 3}
\end{align}
where the second step makes use of Fubini's theorem. The right hand
side of (\ref{eq:pf:lem:posterior approx: 3}) is the same as in (\ref{eq:pf:lem:posterior approx: 1}).
So, by the same arguments as before, $\sup_{q}B_{2n}(q)=o(1)$. The
claim can therefore be considered proved. 
\end{proof}
\begin{lem} \label{Lem: Properties of approimate likelihood}Let
$\mathcal{Y}_{n}$ denote the domain of ${\bf y}_{n}$. Then, $\lim_{n\to\infty}\sup_{\vert h\vert\le\Gamma}\Lambda_{n,h}(\mathcal{Y}_{n})=1$,
and $\Lambda_{n,h}$ is contiguous to $P_{n,\theta_{0}}$. Furthermore,
$\lim_{n\to\infty}\tilde{\bar{P}}_{n}(\mathcal{Y}_{n})=1$, $\tilde{\bar{P}}_{n}$
is contiguous to $P_{n,\theta_{0}}$ and for each $\epsilon>0$ there
exists $M(\epsilon),N(\epsilon)<\infty$ such that $\tilde{\bar{P}}_{n}(A_{n}^{c})<\epsilon$
for all $M\ge M(\epsilon)$ and $n\ge N(\epsilon)$. \end{lem} 
\begin{proof}
Set $P_{n,h}:=P_{n,\theta_{0}+h/\sqrt{n}}$ and $p_{n,h}=dP_{n,h}/d\nu$.
Note that $p_{n,0}({\bf y}_{n})=\prod_{i=1}^{n}p_{0}(Y_{i})$, where
$p_{0}(\cdot)$ is the density function of $P_{\theta_{0}}(Y)$. Then,
by the definition of $\Lambda_{n,h}$ and $\lambda_{n,h}(\cdot)$,
we can write $\Lambda_{n,h}(\mathcal{Y}_{n})\equiv\int\lambda_{n,h}({\bf y}_{n})d\nu({\bf y}_{n})$
as 
\begin{align*}
\Lambda_{n,h}(\mathcal{Y}_{n}) & =(a_{n}(h))^{n}\ \textrm{where}\\
a_{n}(h) & :=\int\exp\left\{ \frac{h}{\sqrt{n}}\psi(Y_{i})-\frac{h^{2}}{2\sigma^{2}n}\right\} p_{0}(Y_{i})d\nu(Y_{i}).
\end{align*}
Denote $g_{n}(h,Y)=\frac{h}{\sqrt{n}}\psi(Y)-\frac{h^{2}}{2\sigma^{2}n}$,
$\delta_{n}(h,Y)=\exp\{g_{n}(h,Y)\}-\{1+g_{n}(h,Y)+g_{n}^{2}(h,Y)/2\}$
and $\mathbb{E}_{p_{0}}[\cdot]$, the expectation corresponding to
$p_{0}(Y)$. Then, 
\begin{align}
a_{n}(h) & =\mathbb{E}_{p_{0}}\left[\exp\left\{ \frac{h}{\sqrt{n}}\psi(Y)-\frac{h^{2}}{2\sigma^{2}n}\right\} \right]\nonumber \\
 & =\mathbb{E}_{p_{0}}\left[1+g_{n}(h,Y)+\frac{1}{2}g_{n}^{2}(h,Y)\right]+\mathbb{E}_{p_{0}}\left[\delta_{n}(h,Y)\right]\nonumber \\
 & :=Q_{n1}(h)+Q_{n2}(h).
\end{align}
Since $\psi(\cdot)$ is the score function at $\theta_{0}$, $\mathbb{E}_{p_{0}}[\psi(Y)]=0$
and $\mathbb{E}_{p_{0}}[\psi^{2}(Y)]=1/\sigma^{2}$. Using these results
and the fact $\vert h\vert\le\Gamma$, straightforward algebra implies
\[
Q_{n1}(h)=1+b_{n},\ \textrm{where }b_{n}\le\Gamma^{4}/8\sigma^{4}n^{2}.
\]
We can expand $Q_{n2}$ as follows:
\begin{equation}
Q_{n2}(h)=\mathbb{E}_{p_{0}}\left[\mathbb{I}_{\psi(Y)\le K}\delta_{n}(h,Y)\right]+\mathbb{E}_{p_{0}}\left[\mathbb{I}_{\psi(Y)>K}\delta_{n}(h,Y)\right].\label{eq:pf:approximate likelihood:decomposition 2-1}
\end{equation}
Since $\vert h\vert\le\Gamma$ and $e^{x}-(1+x+x^{2}/2)=O(\vert x\vert^{3})$,
the first term in (\ref{eq:pf:approximate likelihood:decomposition 2-1})
is bounded by $K^{3}\Gamma^{2}n^{-3/2}$. Furthermore, for large enough
$n$, the second term in (\ref{eq:pf:approximate likelihood:decomposition 2-1})
is bounded by $\mathbb{E}_{p_{\theta_{0}}}[\exp\vert\psi(Y)\vert]/\exp(aK)$
for any $a<1$. Hence, setting $K=(3/2a)\ln n$ gives $\sup_{\vert h\vert\le\Gamma}Q_{n2}(h)=O\left(\ln^{3}n/n^{3/2}\right)$.
In view of the above, 
\[
\sup_{\vert h\vert\le\Gamma}\vert a_{n}(h)-1\vert=O(n^{-c})\ \textrm{for any }c<3/2.
\]
Thus, $\sup_{\vert h\vert\le\Gamma}\vert\Lambda_{n,h}(\mathcal{Y}_{n})-1\vert=\left|\left\{ 1+O(n^{-c})\right\} ^{n}-1\right|=O(n^{-(c-1)})$.
Since it is possible to choose any $c<3/2$, this proves the first
claim. 

Under $P_{n,0},$ the likelihood $d\Lambda_{n,h}/dP_{n,0}$ converges
weakly to some $V$ satisfying $\mathbb{E}_{P_{n,0}}[V]=1$ (the argument
leading to this is standard, see, e.g., \citealp[Example 6.5]{van2000asymptotic}).
Since $\Lambda_{n,h}(\mathcal{Y}_{n})\to1$, an application of Le
Cam's first lemma implies $\Lambda_{n,h}$ is contiguous with respect
to $P_{n,0}$. 

Because $m_{0}(\cdot)$ is supported on $\vert h\vert\le\Gamma$,
$\vert\tilde{\bar{P}}_{n}(\mathcal{Y}_{n})-1\vert\le\int\vert\Lambda_{n,h}(\mathcal{Y}_{n})-1\vert m_{0}(h)d\nu(h)=O(n^{-(c-1)})$.
Thus, $\lim_{n\to\infty}\tilde{\bar{P}}_{n}(\mathcal{Y}_{n})=1$.
Contiguity of $\tilde{\bar{P}}_{n}$ with respect to $P_{n,0}$ follows
from the contiguity of $\Lambda_{n,h}$ with respect to $P_{n,0}$.
The final claim, that $\tilde{\bar{P}}_{n}(A_{n}^{c})<\epsilon$,
follows by similar arguments as in the proof of Lemma \ref{Lem: bound on A_n}. 
\end{proof}
\begin{lem} \label{Lem: Disintegration}The measure, $\tilde{\bar{P}}_{n}$,
can be disintegrated as in equation (\ref{eq:disintegration of tilde P_n}).\end{lem} 
\begin{proof}
Let $\lambda_{nq,h}(\cdot)$, $\tilde{S}_{nq}$ be defined as in the
proof of Lemma \ref{Lem: approximation of posteriors}. Equation (\ref{eq:expansion of lambda_n})
implies
\begin{equation}
\lambda_{n,h}({\bf y}_{n})\cdot m_{0}(h)=\lambda_{n-1,h}({\bf y}_{n-1})\cdot m_{0}(h)\cdot\tilde{p}(Y_{n}\vert h).\label{eq:pf:Disintegration lemma - 1}
\end{equation}
Let $\tilde{S}_{n-1}$ denote the probability measure corresponding
to the density $d\tilde{S}_{n-1}=\lambda_{n-1,h}({\bf y}_{n-1})\cdot m_{0}(h)$.
As argued in the proof of Lemma \ref{Lem: approximation of posteriors},
one can disintegrate this as $d\tilde{S}_{n-1}=p_{n}(h\vert{\bf y}_{n-1})\cdot\tilde{\bar{p}}_{n-1}({\bf y}_{n-1})$,
where $p_{n}(h\vert{\bf y}_{n-1})$ is a conditional probability density
and $\tilde{\bar{p}}_{n-1}({\bf y}_{n-1})=\int\lambda_{n-1,h}({\bf y}_{n-1})m_{0}(h)d\nu(h)$.
Thus, 
\[
\lambda_{n-1,h}({\bf y}_{n-1})\cdot m_{0}(h)=p_{n}(h\vert{\bf y}_{n-1})\cdot\tilde{\bar{p}}_{n-1}({\bf y}_{n-1}).
\]
Combining the above with (\ref{eq:pf:Disintegration lemma - 1}) gives
\[
\lambda_{n,h}({\bf y}_{n})\cdot m_{0}(h)=p_{n}(h\vert{\bf y}_{n-1})\cdot\tilde{\bar{p}}_{n-1}({\bf y}_{n-1})\cdot\tilde{p}(Y_{n}\vert h).
\]
Taking the integral with respect $h$ on both sides, and making use
of the definition of $\tilde{\bar{p}}_{n}(\cdot)$, 
\begin{equation}
\tilde{\bar{p}}_{n}({\bf y}_{n})=\tilde{\bar{p}}_{n-1}({\bf y}_{n-1})\cdot\int\tilde{p}(Y_{n}\vert h)p_{n}(h\vert{\bf y}_{n-1})d\nu(h).\label{eq:pf:Disintegration lemma: iteration}
\end{equation}
There is nothing special about the choice of $n$ here, so iterating
the above expression gives the desired result, (\ref{eq:disintegration of tilde P_n}).
\end{proof}
\begin{lem} \label{Lem: Approximation of tilde P_n}Let $c_{n,i}$
and $\tilde{\mathbb{P}}_{n}$ denote the quantities defined in Step
4 of the proof of Theorem \ref{Thm: General parametric families}.
There exists some non-random $C<\infty$ such that $\sup_{i}\vert c_{n,i}-1\vert\le Cn^{-c}$
for any $c<3/2$. Furthermore, $\lim_{n\to\infty}\left\Vert \tilde{\mathbb{P}}_{n}-\tilde{\bar{P}}_{n}\right\Vert _{\textrm{TV}}=0$.
\end{lem} 
\begin{proof}
Denote 
\[
a_{n}(h):=\int\tilde{p}_{n}(Y_{i}\vert h)d\nu(Y_{i})=\int\exp\left\{ \frac{h}{\sqrt{n}}\psi(Y_{i})-\frac{h^{2}}{2\sigma^{2}n}\right\} p_{0}(Y_{i})d\nu(Y_{i}).
\]
It is shown in the proof of Lemma \ref{Lem: Properties of approimate likelihood}
that $\sup_{\vert h\vert\le\Gamma}\vert a_{n}(h)-1\vert=O(n^{-c})\ \textrm{for any }c<3/2.$
Since $c_{n,i}=\int a_{n}(h)\tilde{p}_{n}(h\vert{\bf y}_{i-1})d\nu(h)$,
and $\tilde{p}_{n}(h\vert{\bf y}_{i-1})$ is a probability density,
this proves the first claim. 

For the second claim, denote $\tilde{p}_{n}(Y_{i}\vert{\bf y}_{i-1}):=\int\tilde{p}_{n}(Y_{i}\vert h)\tilde{p}_{n}(h\vert{\bf y}_{i-1})d\nu(h)$.
We also write $c_{n,i}({\bf y}_{i-1})$ for $c_{n,i}$ to make it
explicit that this quantity depends on ${\bf y}_{i-1}$. The properties
of the total variation metric, along with (\ref{eq:disintegration of tilde P_n})
and (\ref{eq:disintegration result}) imply 
\begin{align*}
\left\Vert \tilde{\mathbb{P}}_{n}-\tilde{\bar{P}}_{n}\right\Vert _{\textrm{TV}} & =\frac{1}{2}\int\left|\frac{d\tilde{\mathbb{P}}_{n}}{d\nu}-\frac{d\tilde{\bar{P}}_{n}}{d\nu}\right|d\nu\\
 & =\frac{1}{2}\int\prod_{i=1}^{n}\tilde{p}_{n}(Y_{i}\vert{\bf y}_{i-1})\left|\prod_{i=1}^{n}\frac{1}{c_{n,i}({\bf y}_{i-1})}-1\right|d\nu({\bf y}_{n})\\
 & \le\frac{1}{2}\sup_{{\bf y}_{n}}\left|\prod_{i=1}^{n}\frac{1}{c_{n,i}({\bf y}_{i-1})}-1\right|\cdot\int\prod_{i=1}^{n}\tilde{p}_{n}(Y_{i}\vert{\bf y}_{i-1})d\nu({\bf y}_{n}).
\end{align*}
Recall from (\ref{eq:disintegration of tilde P_n}) that $\prod_{i=1}^{n}\tilde{p}_{n}(Y_{i}\vert{\bf y}_{i-1})$
is the density (wrt $\nu$) of $\tilde{\bar{P}}_{n}$, so the integral
in the above expression equals $\int d\tilde{\bar{P}}_{n}=\tilde{\bar{P}}_{n}(\mathcal{Y})\to1$
by Lemma \ref{Lem: Properties of approimate likelihood}. Furthermore,
using the first claim of the present lemma, it is straightforward
to show
\[
\sup_{{\bf y}_{n}}\left|\prod_{i=1}^{n}\frac{1}{c_{n,i}({\bf y}_{i-1})}-1\right|=O(n^{-(c-1)}).
\]
Thus, $\left\Vert \tilde{\mathbb{P}}_{n}-\tilde{\bar{P}}_{n}\right\Vert _{\textrm{TV}}=O(n^{-(c-1)})$
and the claim follows.
\end{proof}
\begin{lem} \label{Lem: Auxiliary results for Step 5} For the probability
measure $\tilde{\mathbb{P}}_{n}$ defined in Step 4 of the proof of
Theorem \ref{Thm: General parametric families}, there exists a deterministic
sequence $\xi_{n}\to0$ independent of $s$ and $\pi\in\{0,1\}$ such
that equations (\ref{eq:pf:Thm5:exp under posterior}) - (\ref{eq:pf:Thm5:Step 4: bound on 4th moment})
hold. \end{lem} 
\begin{proof}
Start with (\ref{eq:pf:Thm5:exp under posterior}). We have
\begin{align*}
\mathbb{E}_{\tilde{\mathbb{P}}_{n}}\left[\left.\psi(Y_{nq+1})\right|s\right] & =c_{n,nq+1}^{-1}\int\left\{ \int\psi(Y_{nq+1})\tilde{p}_{n}(Y_{nq+1}\vert h)d\nu(Y_{nq+1})\right\} \tilde{p}(h\vert x,q)d\nu(h)\\
 & =c_{n,nq+1}^{-1}\int\mathbb{E}_{p_{\theta_{0}}}\left[\psi(Y)\exp\left\{ \frac{h}{\sqrt{n}}\psi(Y)-\frac{h^{2}}{2\sigma^{2}n}\right\} \right]\tilde{p}(h\vert x,q)d\nu(h)\\
 & =\left(1+O(n^{-c})\right)\cdot\int\mathbb{E}_{p_{\theta_{0}}}\left[\psi(Y)\exp\left\{ \frac{h}{\sqrt{n}}\psi(Y)-\frac{h^{2}}{2\sigma^{2}n}\right\} \right]\tilde{p}(h\vert x,q)d\nu(h),
\end{align*}
where the second equality follows by the definition of $\tilde{p}(Y_{i}\vert h)$,
and the third equality follows by (\ref{eq:bound on integration factor}),
where it may be recalled we can choose any $c\in(0,3/2)$. Define
$g_{n}(h,Y)=\frac{h}{\sqrt{n}}\psi(Y)-\frac{h^{2}}{2\sigma^{2}n}$
and $\delta_{n}(h,Y)=\exp\{g_{n}(h,Y)\}-\{1+g_{n}(h,Y)\}$. Then,
\begin{align*}
 & \mathbb{E}_{p_{\theta_{0}}}\left[\psi(Y)\exp\left\{ \frac{h}{\sqrt{n}}\psi(Y)-\frac{h^{2}}{2\sigma^{2}n}\right\} \right]\\
 & =\mathbb{E}_{p_{\theta_{0}}}\left[\psi(Y)\left\{ 1+\frac{h}{\sqrt{n}}\psi(Y)-\frac{h^{2}}{2\sigma^{2}n}\right\} \right]+\mathbb{E}_{p_{\theta_{0}}}\left[\psi(Y)\delta_{n}(h,Y)\right].
\end{align*}
Assumption 1(i) implies, see e.g., \citet[Theorem 7.2]{van2000asymptotic},
$\mathbb{E}_{p_{\theta_{0}}}\left[\psi(Y)\right]=0$ and $\mathbb{E}_{p_{\theta_{0}}}\left[\psi^{2}(Y)\right]=1/\sigma^{2}$.
Hence, the first term in the above expression equals $h/(\sqrt{n}\sigma^{2})$.
For the second term, 
\begin{equation}
\mathbb{E}_{p_{\theta_{0}}}\left[\psi(Y)\delta_{n}(h,Y)\right]=\mathbb{E}_{p_{\theta_{0}}}\left[\mathbb{I}_{\psi(Y)\le K}\psi(Y)\delta_{n}(h,Y)\right]+\mathbb{E}_{p_{\theta_{0}}}\left[\mathbb{I}_{\psi(Y)>K}\psi(Y)\delta_{n}(h,Y)\right].\label{eq:pf:Step 4:decomposition 2}
\end{equation}
Since $\vert h\vert\le\Gamma$ and $e^{x}-(1+x)=o(x^{2})$, the first
term in in (\ref{eq:pf:Step 4:decomposition 2}) is bounded by $K^{3}\Gamma^{2}n^{-1}$.
The second term in (\ref{eq:pf:Step 4:decomposition 2}) is bounded
by $\mathbb{E}_{p_{\theta_{0}}}[\exp\vert\psi(Y)\vert]/\exp(aK)$
for any $a<1$. Hence, setting $K=(1/a)\ln n$ gives $\sup_{\vert h\vert\le\Gamma}\vert\mathbb{E}_{p_{\theta_{0}}}\left[\psi(Y)\delta_{n}(h,Y)\right]\vert=O(\ln^{3}n/n)$.
Combining the above results and noting that $\vert h\vert\le\Gamma$,
we obtain
\[
\sqrt{n}\sigma^{2}\mathbb{E}_{\tilde{\mathbb{P}}_{n}}\left[\left.\psi(Y_{nq+1})\right|s\right]=\left(1+O(n^{-c})\right)\cdot\left\{ \int h\tilde{p}(h\vert x,q)d\nu(h)+O(\ln n/\sqrt{n})\right\} =h(s)+\xi_{n},
\]
where $\xi_{n}\asymp\ln n/\sqrt{n}$. This proves (\ref{eq:pf:Thm5:exp under posterior}).
The proofs of (\ref{eq:pf:Thm5:variance under post approximation})
and (\ref{eq:pf:Thm5:Step 4: bound on 4th moment}) are similar.
\end{proof}

\section{Additional details and proof of Theorem \ref{Thm: Non-parametric models}
for non-parametric models\label{subsec:Additional details for non-parametric}}

We start with a formal definition of the parametric sub-models and
priors used in our setup. 

\subsubsection{Parametric sub-models and priors on tangent spaces}

Following \citet{van2000asymptotic}, we define one-dimensional parametric
sub-models,$\{P_{t,\bm{h}}:t\le\eta\}$, to be the class of probability
densities such that 
\begin{equation}
\int\left[\frac{\left(dP_{t,\bm{h}}^{1/2}-dP_{0}^{1/2}\right)}{t}-\frac{1}{2}\bm{h}dP_{0}^{1/2}\right]^{2}d\nu\to0\ \textrm{as}\ t\to0,\label{eq:qmd non-parametrics}
\end{equation}
for some measure function $\bm{h}(\cdot)$. It is well known, see
e.g., \citet{van2000asymptotic}, that (\ref{eq:qmd non-parametrics})
implies $\int\bm{h}dP_{0}=0$ and $\int\bm{h}^{2}dP_{0}<\infty$.
As mentioned in the main text, the set of all such candidate $h$
is termed the tangent space $T(P_{0})$. This is a subset of the Hilbert
space $L^{2}(P_{0})$, endowed with the inner product $\left\langle f,g\right\rangle =\mathbb{E}_{P_{0}}[fg]$
and norm $\left\Vert f\right\Vert =\mathbb{E}_{P_{0}}[f^{2}]^{1/2}$.
As in Section \ref{sec:General-parametric-models}, (\ref{eq:qmd non-parametrics})
implies the SLAN property that for all $\bm{h}\in T(P_{0})$,
\begin{align}
\sum_{i=1}^{\left\lfloor nq\right\rfloor }\ln\frac{dP_{1/\sqrt{n},\bm{h}}}{dP_{0}}(Y_{i}) & =\frac{1}{\sqrt{n}}\sum_{i=1}^{\left\lfloor nq\right\rfloor }\bm{h}(Y_{i})-\frac{q}{2}\left\Vert \bm{h}\right\Vert ^{2}+o_{P_{0}}(1),\ \textrm{ uniformly over }q.\label{eq:SLAN nonparametric setting}
\end{align}

Asymptotic Bayes risk is defined in terms of priors on the tangent
space $T(P_{0})$. To define this formally, we start by selecting
$\{\phi_{1},\phi_{2},\dots\}\in T(P_{0})$ such that $\{\psi/\sigma,\phi_{1},\phi_{2},\dots\}$
form an orthonormal basis for the closure of $T(P_{0})$; the division
of $\psi$ by $\sigma$ is simply to ensure $\left\Vert \psi/\sigma\right\Vert ^{2}=\int x^{2}/\sigma^{2}dP_{0}(x)=1$.
By the Hilbert space isometry, each $\bm{h}\in T(P_{0})$ can then
be associated with an element from the $l_{2}$ space of square integrable
sequences, $(h_{0}/\sigma,h_{1},\dots)$, where $h_{0}=\left\langle \psi,\bm{h}\right\rangle $
and $h_{k}=\left\langle \phi_{k},\bm{h}\right\rangle $ for all $k\neq0$.
A prior on $T(P_{0})$ therefore corresponds to a prior on $l_{2}$.

Let $(\varrho(1),\varrho(2),\dots)$ denote an arbitrary permutation
of $(1,2,\dots)$. As mentioned in the main text, we impose two restriction
on $\rho_{0}$. The first is that $\rho_{0}$ is supported on a finite
dimensional sub-space, 
\[
\mathcal{H}_{I}\equiv\left\{ \bm{h}\in T(P_{0}):\bm{h}=\frac{1}{\sigma}\left\langle \psi,\bm{h}\right\rangle \frac{\psi}{\sigma}+\sum_{k=1}^{I-1}\left\langle \phi_{\varrho(k)},\bm{h}\right\rangle \phi_{\varrho(k)}\right\} 
\]
 of $T(P_{0})$, or equivalently, on a subset of $l_{2}$ of finite
dimension $I$. Crucially, the first component of $\bm{h}\in l_{2}$,
corresponding to $h_{0}/\sigma$, is always included in the support
of the prior. This important as $h_{0}=\left\langle \psi,\bm{h}\right\rangle $
is exactly the mean reward (upto a $\sqrt{n}$ scaling). The second
restriction is that it is possible to decompose $\rho_{0}=m_{0}\times\lambda$,
where $m_{0}$ is a prior on $h_{0}$ and $\lambda$ is a prior on
$(h_{\varrho(1)},h_{\varrho(2)},\dots)$. Recall that $\mu_{n}(\bm{h}):=\mu(P_{1/\sqrt{n},\bm{h}})\approx h_{0}/\sqrt{n}$.
Thus $m_{0}$ is effectively equivalent to a prior on the scaled rewards
$\sqrt{n}\mu_{n}$, just as in Section \ref{sec:Diffusion-asymptotics-and}. 

\subsubsection{Heuristics\label{subsec:Heuristics-non-parametrics}}

We now provide an informal account of why the second component, $\lambda$,
of the product prior $\rho_{0}:=m_{0}\times\lambda$ does not feature
in asymptotics and it is sufficient, asymptotically, to restrict the
state variables to $x_{nq},q,t$.

By construction, the prior $\rho_{0}$ is supported on a finite-dimensional
subset of the tangent space of the form $\left\{ \bm{h}^{\intercal}\bm{\chi}(Y_{i}):\bm{h}\in\mathbb{R}^{I}\right\} $,
where $\bm{\chi}:=(\psi/\sigma,\phi_{\varrho(1)},\dots,\phi_{\varrho(I-1)})$.
In what follows, we drop the permutation $\varrho$ for simplicity.
Consider the posterior density, $p_{n}(\cdot\vert\mathcal{F}_{t})$,
of the vector $\bm{h}$ given $\mathcal{F}_{t}$, where the filtration
$\mathcal{F}_{t}$ is defined as in Section \ref{sec:General-parametric-models}.
By Lemma \ref{Lemma 1}, 

\begin{equation}
p_{n}(\cdot\vert\mathcal{F}_{t})=p_{n}(\cdot\vert{\bf y}_{nq(t)})\propto\left\{ \prod_{i=1}^{\left\lfloor nq(t)\right\rfloor }dP_{1/\sqrt{n},\bm{h}^{\intercal}\bm{\chi}}(Y_{i})\right\} \cdot\rho_{0}(\bm{h}).\label{eq:posterior-non-parametric}
\end{equation}
Here, as before, $q(t)=n^{-1}\sum_{j=1}^{\left\lfloor nt\right\rfloor }\mathbb{I}(A_{j}=1)$.
Now, (\ref{eq:SLAN nonparametric setting}) suggests that the likelihood
term in (\ref{eq:posterior-non-parametric}) can be approximated by
a new likelihood, the density of the `tilted' measure $\Lambda_{nq,\bm{h}}(\cdot)$
defined as
\begin{equation}
d\Lambda_{nq,\bm{h}}({\bf y}_{nq}):=\exp\left\{ \frac{1}{\sqrt{n}}\sum_{i=1}^{\left\lfloor nq\right\rfloor }\bm{h}^{\intercal}\bm{\chi}(Y_{i})-\frac{q}{2}\left\Vert \bm{h}\right\Vert ^{2}\right\} dP_{1/\sqrt{n},0}({\bf y}_{nq}).\label{eq:tilted measure - non-parametrics}
\end{equation}
Le $\bm{\chi}_{nq}:=n^{-1/2}\sum_{i=1}^{\left\lfloor nq\right\rfloor }\bm{\chi}(Y_{i})$.
Then, taking $\tilde{p}_{n}(\cdot\vert{\bf y}_{nq})$ to be the corresponding
approximate posterior density as in Section \ref{sec:General-parametric-models},
we have:
\begin{align}
\tilde{p}_{n}(\bm{h}\vert{\bf y}_{nq}) & \propto d\Lambda_{nq,\bm{h}}({\bf y}_{nq})\cdot\rho_{0}(\bm{h})\nonumber \\
 & \propto\tilde{p}_{q}(\bm{\chi}_{nq}\vert\bm{h})\cdot\rho_{0}(\bm{h});\ \textrm{where }\tilde{p}_{q}(\cdot\vert\bm{h})\equiv\mathcal{N}(\cdot\vert q\bm{h},qI).\label{eq:posterior dist non-parametric - 1st approx}
\end{align}

The approximate posterior of $\bm{h}$ depends on the $I$ dimensional
quantity $\bm{\chi}_{nq}$. However, it is possible to achieve achieve
further dimension reduction for the marginal posterior density, $\tilde{p}_{n}(h_{0}\vert{\bf y}_{nq})$,
of $h_{0}$. Indeed, for any $\bm{h}\in T(P_{0})$,
\begin{align*}
\frac{1}{\sqrt{n}}\sum_{i=1}^{\left\lfloor nq\right\rfloor }\bm{h}(Y_{i})-\frac{q}{2}\left\Vert \bm{h}\right\Vert ^{2} & =\frac{h_{0}}{\sigma\sqrt{n}}\sum_{i=1}^{\left\lfloor nq\right\rfloor }Y_{i}-\frac{q}{2\sigma^{2}}h_{0}^{2}+(\textrm{terms independent of }h_{0})
\end{align*}
where the equality follows from the Hilbert space isometry which implies
$\bm{h}=(h_{0}/\sigma)(\psi/\sigma)+\sum_{k=1}^{I}h_{k}\phi_{k}$,
and $\left\Vert \bm{h}\right\Vert ^{2}=(h_{0}/\sigma)^{2}+\sum_{k=1}^{I}h_{k}^{2}$.
So, defining $x_{nq}=n^{-1/2}\sum_{i=1}^{\left\lfloor nq\right\rfloor }Y_{i},$
we obtain from (\ref{eq:tilted measure - non-parametrics}) and (\ref{eq:posterior dist non-parametric - 1st approx})
that
\begin{align}
\tilde{p}_{n}(h_{0}\vert{\bf y}_{nq}) & \propto\exp\left\{ \frac{h_{0}}{\sigma^{2}}x_{nq}-\frac{q}{2\sigma^{2}}h_{0}^{2}\right\} \cdot m_{0}(h_{0})\nonumber \\
 & \propto\tilde{p}_{q}(x_{nq}\vert h_{0})\cdot m_{0}(h_{0}),\ \ \textrm{where}\quad\tilde{p}_{q}(\cdot\vert h_{0})\equiv\mathcal{N}(\cdot\vert qh_{0},q\sigma^{2}).\label{eq:posterior - non_parametric}
\end{align}
In other words, one can approximate the posterior distribution of
$h_{0}$ under $\mathcal{F}_{t}$ by $\tilde{p}_{n}(h_{0}\vert x_{nq(t)},q(t))\equiv\tilde{p}_{n}(h_{0}\vert{\bf y}_{nq(t)})\propto p_{q(t)}(x_{nq(t)}\vert h_{0})\cdot m_{0}(h_{0})$,
just as in Section \ref{sec:General-parametric-models}. Since the
expected reward depends only on $h_{0}$ due to (\ref{eq:influence function}),
this suggests that it is sufficient, asymptotically, to restrict the
state variables to $x_{nq(t)},q(t),t$.

\subsubsection{Assumptions}

Set $\mathbb{\tilde{E}}[\cdot\vert s]$ to be the expectation under
$\tilde{p}_{n}(h_{0}\vert x,q)$, $\mu^{+}(s):=\tilde{\mathbb{E}}\left[h_{0}\mathbb{I}\{h_{0}>0\}\vert s\right]$
and $\mu(s):=\tilde{\mathbb{E}}[h_{0}\vert s]$. Note that by (\ref{eq:posterior - non_parametric}),
these terms are the same as in Section \ref{subsec:PDE-characterization-of}.
Also, set $h(\bm{\chi}_{nq},q):=\tilde{\mathbb{E}}\left[\bm{h}\vert\bm{\chi}_{nq},q\right]$
where $\tilde{\mathbb{E}}[\cdot\vert\bm{\chi}_{nq},q]$ is the expectation
under $\tilde{p}_{n}(\bm{h}\vert\bm{\chi}_{nq},q)$, defined in (\ref{eq:posterior dist non-parametric - 1st approx}).
We employ the following assumptions for Theorem \ref{Thm: Non-parametric models}: 

\begin{asm2} (i) The sub-models $\{P_{t,h};h\in T(P_{0})\}$ satisfy
(\ref{eq:qmd non-parametrics}). (ii) $\mathbb{E}_{P_{0}}[\vert Y\vert^{3}]<\infty$.
(iii) There exists $\delta_{n}\to0$ such that $\sqrt{n}\mu(P_{1/\sqrt{n},\bm{h}}))=h_{0}+\delta_{n}\left\Vert \bm{h}\right\Vert ^{2}\ \forall\ \bm{h}\in T(P_{0})$.
(iv) $\rho_{0}(\cdot)$ is supported on $\mathcal{H}_{I}(\Gamma)\equiv\left\{ \bm{h}\in\mathcal{H_{I}}:\mathbb{E}_{P_{0}}\left[\exp\vert\bm{h}\vert\right]\le\Gamma\right\} $
for some $\Gamma<\infty$. (v) $\mu(\cdot)$ and $\mu^{+}(\cdot)$
are H{\"o}lder continuous and $\sup_{s}\varpi(s)\le C<\infty$. Furthermore,
$h(\bm{\chi},q)$ is also H{\"o}lder continuous. \end{asm2}

Assumption 2(iii) is a stronger version of (\ref{eq:influence function}),
but is satisfied for all commonly used sub-models. For instance, if
$dP_{1/\sqrt{n},\bm{h}}:=(1+n^{-1/2}\bm{h})dP_{0}$ as in \citet[Example 25.16]{van2000asymptotic},
$\sqrt{n}\mu(P_{1/\sqrt{n},\bm{h}})=\left\langle \psi,\bm{h}\right\rangle =h_{0}$.
Assumption 2(iv) requires the prior to be supported on score functions
with finite exponential moments. As with Assumptions 1(ii) \& 1(iv),
it ensures the tilt $d\Lambda_{nq,\bm{h}}({\bf y}_{nq})/dP_{1/\sqrt{n},0}({\bf y}_{nq})$
in (\ref{eq:tilted measure - non-parametrics}) is uniformly bounded.
It is somewhat restrictive as it implies $\mathbb{E}_{P_{0}}[\exp\vert h_{0}Y\vert]<\infty$
for all $h_{0}\in\textrm{supp}(m_{0})$. However, similar to Assumptions
1(ii) \& 1(iv), we suspect it can be relaxed at the expense of more
intricate proofs. Finally, Assumption 2(v) differs from Assumption
1(v) only in requiring continuity of $h(\bm{\chi},q)$. While $h(\bm{\chi},q)$
is not present in PDE (\ref{eq:PDE optimal bayes risk}), it arises
in the course of various PDE approximations in the proof. The form
of the posterior in (\ref{eq:posterior dist non-parametric - 1st approx})
implies this should be satisfied under mild assumptions on $\rho_{0}$.
It is certainly satisfied for Gaussian $\rho_{0}$.

\subsubsection{Proof of Theorem \ref{Thm: Non-parametric models}}

The proof consists of two steps. First, we show that $V_{n}^{*}(0)$
converges to the solution of a PDE with state variables $(\bm{\chi},q,t)$
where $\bm{\chi}(t):=\bm{\chi}_{nq(t)}$ with $\bm{\chi}_{nq}$ defined
in Section \ref{sec:The-nonparametric-setting}. Recall that the first
component of $\bm{\chi}$ is $x/\sigma$. Next, we show that the PDE
derived in the first step can be reduced to one involving just the
state variables $s=(x,q,t)$. 

The first step follows the proof of Theorem \ref{Thm: General parametric families}
with straightforward modifications. Indeed, the setup is equivalent
to taking $\bm{\chi}(Y_{i})$ to be the vector-valued score function
in the parametric setting (see, Section \ref{subsec:Vector-valued}).
The upshot of these arguments is that $V_{n}^{*}(0)$ converges to
$V^{*}(0)$, where $V^{*}(\cdot)$ solves the PDE 
\begin{align}
\partial_{t}f(\bm{\chi},q,t)+\mu^{+}(x,q)+\min\left\{ -\mu(x,q)+\bar{L}[f](\bm{\chi},q,t),0\right\}  & =0\ \textrm{if }t<1\label{eq:PDE- bayes risk - full version - non-parametrics}\\
f(\bm{\chi},q,t) & =0\ \textrm{if }t=1,\nonumber 
\end{align}
with the infinitesimal generator (here $\triangle$ denotes the Laplace
operator)
\[
\bar{L}[f](\bm{\chi},q,t):=\partial_{q}f+h(\bm{\chi},q)^{\intercal}D_{\bm{\chi}}f+\frac{1}{2}\triangle_{\bm{\chi}}f.
\]
See Section \ref{sec:The-nonparametric-setting} for the definition
of $h(\bm{\chi},q)$. Note that $\mu^{+}(\cdot),\mu(\cdot)$ are functions
only of $(x,q)$. This is because they depend only on the first component,
$h_{0}/\sigma$, of $\bm{h}$ and its posterior distribution can be
approximated by $\tilde{p}_{n}(h_{0}\vert x,q)$, defined in (\ref{eq:posterior - non_parametric}). 

By the arguments leading to (\ref{eq:posterior - non_parametric}),
the first component of the vector $h(\bm{\chi},q)$ is $\sigma^{-1}\tilde{\mathbb{E}}[h_{0}\vert\bm{\chi},q]=\sigma^{-1}\tilde{\mathbb{E}}[h_{0}\vert x,q]=\sigma^{-1}\mu(x,q)$.
Let $\bm{\chi}^{c}$, $h^{c}(\bm{\chi},q)$ denote $\bm{\chi},h(\bm{\chi},q)$
without their first components $\chi_{1}=x/\sigma$ and $h_{1}(\bm{\chi},q)=\sigma^{-1}\mu(x,q)$.
Then, defining 
\[
L[f](x,q,t):=\partial_{q}f+\mu(x,q)\partial_{x}f+\frac{1}{2}\sigma^{2}\partial_{x}^{2}f,
\]
we see that $\bar{L}[f]=L[f]+h^{c}(\bm{\chi},q)^{\intercal}D_{\bm{\chi}^{c}}f+\frac{1}{2}\triangle_{\bm{\chi}^{c}}f$.
Note that in defining $L[f](\cdot)$, we made use of the change of
variables $\partial_{\chi_{1}}f=\sigma\partial_{x}f$ and $\partial_{\chi_{1}}^{2}f=\sigma^{2}\partial_{x}^{2}f$.
We now claim that the solution of PDE (\ref{eq:PDE- bayes risk - full version - non-parametrics})
is the same as that of PDE (\ref{eq:PDE optimal bayes risk}), reproduced
here: 
\begin{align}
\partial_{t}f(x,q,t)+\mu^{+}(x,q)+\min\left\{ -\mu(x,q)+L[f](x,q,t),0\right\}  & =0\ \textrm{if }t<1\label{eq:PDE- bayes risk - non-parametrics-proof}\\
f(x,q,t) & =0\ \textrm{if }t=1.\nonumber 
\end{align}
Intuitively, this is because the state variables in $\bm{\chi}^{c}$
do not affect instantaneous payoffs $\mu^{+}(x,q)-\mu(x,q),\mu^{+}(x,q)$,
nor do they affect the boundary condition, so these state variables
are superfluous. The formal proof makes use of the theory of viscosity
solutions: Under Assumption 2(v), Theorem \ref{theorem 1} implies
there exists a unique viscosity solution to (\ref{eq:PDE- bayes risk - full version - non-parametrics}),
denoted by $V^{*}(\bm{\chi},q,t)$. Then, it is straightforward to
show that $\bar{V}^{*}(x,q,t)=\sup_{\bm{\chi}^{c}}V^{*}(\bm{\chi},q,t)$
is a viscosity sub-solution to (\ref{eq:PDE- bayes risk - non-parametrics-proof}).\footnote{See \citet{crandall1992user} for the definition of viscosity sub-
and super-solutions using test functions. To show $\bar{V}^{*}$ is
a sub-solution one can argue as follows: First, $\bar{V}^{*}(x,q,t)$
is upper-semicontinuous because of the continuity of the solution
$V^{*}(\bm{\chi},q,t)$ to PDE (\ref{eq:PDE- bayes risk - full version - non-parametrics}).
Second, $\bar{V}^{*}$ satisfies the boundary condition in PDE (\ref{eq:PDE- bayes risk - non-parametrics-proof})
by construction. Third, let $\phi\in\mathcal{C}^{\infty}(\mathcal{X},\mathcal{Q},\mathcal{T})$
denote a test function such that $\phi\ge\bar{V}^{*}$ everywhere.
By the definition of $\bar{V}^{*}$ we also have $\phi(x,q,t)\ge V^{*}(\bm{\chi},q,t)$
everywhere. Since $V^{*}(\bm{\chi},q,t)$ is a solution to PDE (\ref{eq:PDE- bayes risk - full version - non-parametrics}),
$\phi$ must satisfy the viscosity requirement for a sub-solution
to PDE (\ref{eq:PDE- bayes risk - full version - non-parametrics}).
But because $\phi$ is constant in $\bm{\chi}^{c}$, this implies
it also satisfies the viscosity requirement for a sub-solution to
PDE (\ref{eq:PDE- bayes risk - non-parametrics-proof}). These three
facts suffice to show $\bar{V}^{*}$ is a sub-solution.} In a similar fashion, $\underline{V}^{*}(x,q,t)=\inf_{\bm{\chi}^{c}}V^{*}(\bm{\chi},q,t)$
is a viscosity super-solution to (\ref{eq:PDE- bayes risk - non-parametrics-proof}).
Under Assumption 2(v), a comparison principle (see, \citealp{crandall1992user})
holds for (\ref{eq:PDE- bayes risk - non-parametrics-proof}) implying
any super-solution is larger than a solution, which is in turn larger
than a sub-solution. But $\bar{V}^{*}(x,q,t)\ge\underline{V}^{*}(x,q,t)$
by definition, so it must be the case $\bar{V}^{*}(x,q,t)=\underline{V}^{*}(x,q,t)=V^{*}(x,q,t)$,
where $V^{*}(x,q,t)$ is the unique viscosity solution to (\ref{eq:PDE- bayes risk - non-parametrics-proof}).
This proves $V^{*}(\bm{\chi},q,t)=V^{*}(x,q,t)$, as claimed.

\section{Theory for MAB and its generalizations \label{sec:Theory-for-extensions}}

\subsection{Multi-armed bandits\label{subsec:Multi-armed-bandits}}

\subsubsection*{Existence of a solution to PDE (\ref{eq:PDE_optimal_bayes_risk_General})}

By \citet[Theorem A.1]{barles2007error}, there exists a unique viscosity
solution to PDE (\ref{eq:PDE_optimal_bayes_risk_General}) if $\mu^{\max}(\cdot)$
and $\mu_{k}(\cdot)$ are H{\"o}lder continuous for all $k$. 

\subsubsection*{Convergence to the PDE}

Let $V_{n}^{*}(\cdot)$ denote the minimal Bayes risk function in
the Gaussian setting. The following analogue of Theorem \ref{Thm: Convergence to minimal PDE}
can then be shown with a straightforward modification to the proof:

\begin{thm}\label{Thm: Convergence to minimal PDE-1} Suppose $\mu(\cdot)$
and $\mu^{\max}(\cdot)$ are H{\"o}lder continuous and the prior
$m_{0}$ is such that $\mathbb{E}[\vert\mu\vert^{3}\vert s]<\infty$
at each $s$. Then, as $n\to\infty$, $V_{n}^{*}(\cdot)$ converges
locally uniformly to $V^{*}(\cdot)$, the unique viscosity solution
of PDE (\ref{eq:PDE_optimal_bayes_risk_General}). \end{thm}

\subsubsection*{Piece-wise constant policies}

The construction of piece-wise constant policies in the multi-armed
setting is analogous to Section \ref{subsec:Optimal-and-approximately}.
Following \citet[Theorem 3.1]{barles2007error}, Theorems \ref{Thm: Approximate control}
and \ref{Thm: Approximate control-2} can be shown to hold under Lipschitz
continuity of $\mu^{\max}(\cdot),\mu_{k}(s)$ and $\sup_{s}\left\{ \mu^{\max}(s)-\max_{k}\mu(s)\right\} <\infty$. 

\subsubsection*{Parametric and non-parametric distributions}

Let $P_{\theta}^{(k)}$ denote the probability distribution over the
rewards from arm $k$. It is without loss of generality to assume
the distributions across arms are independent of each other as we
only ever observe the outcomes from a single arm. The parameter $\theta\in\mathbb{R}^{d}$
may have some components that are shared across all the arms. As in
the one-armed bandit setting, we choose a reference $\theta_{0}$
such that $\mathbb{E}_{P_{\theta_{0}}^{(k)}}\left[Y_{k}\right]=0$,
and focus on local perturbations of the form $\{\theta_{n,h}\equiv\theta_{0}+h/\sqrt{n}:h\in\mathbb{R}^{d}\}$.
We then place a non-negligible prior $M_{0}$ on the local parameter
$h$. 

To simplifty notation, suppose that $\theta$ is scalar. Let $\nu:=\nu_{1}\times\nu_{2}$,
where $\nu_{1}$ is a dominating measure for $\{P_{\theta}^{(k)}:\theta\in\mathbb{R},k=0,\dots,K-1\}$
and $\nu_{2}$ is a dominating measure for the prior $M_{0}$ on $h$.
Define $p_{\theta}^{(k)}=dP_{\theta}^{(k)}/d\nu$, $m_{0}=dM_{0}/d\nu$
(in the sequel, we shorten the Radon-Nikodym derivative $dP/d\nu$
to just $dP$). As in Section \ref{sec:General-parametric-models},
we require the class $\{P_{\theta}^{(k)}\}$ to be quadratic mean
differentiable (q.m.d) around $\theta_{0}$ for each $k$. This in
turn implies the SLAN property that, for each $k$,
\begin{equation}
\sum_{i=1}^{\left\lfloor nq_{k}\right\rfloor }\ln\frac{dp_{\theta_{0}+h/\sqrt{n}}^{(k)}}{dp_{\theta_{0}}^{(k)}}=\frac{1}{\sigma_{k}^{2}}hx_{k,nq_{k}}-\frac{q_{k}}{2\sigma_{k}^{2}}h^{2}+o_{P_{n,\theta_{0}}^{(k)}}(1),\ \textrm{uniformly over }q_{k},\label{eq:SLAN K armed bandits}
\end{equation}
where 
\[
x_{k,nq}:=\sigma_{k}^{2}\frac{1}{\sqrt{n}}\sum_{i=1}^{\left\lfloor nq\right\rfloor }\psi_{k}(Y_{i}^{(k)}),
\]
$\psi_{k}(\cdot)$ is the score function corresponding to $P_{\theta_{0}}^{(k)}$,
and $\sigma_{k}^{2}$ is the corresponding inverse information matrix,
i.e., $\sigma_{k}^{2}=\left(\mathbb{E}_{P_{\theta_{0}}^{(k)}}\left[\psi_{k}^{2}\right]\right)^{-1}$. 

Recall that ${\bf y}_{n}^{(k)}:=(Y_{1}^{(k)},\dots,Y_{n}^{(k)})$
denotes the vector of stacked outcomes for each arm $k$. Then, in
the fixed $n$ setting, the posterior distribution of $h$ is (compare
the equation below with (\ref{eq:True posterior}))
\begin{align*}
p_{n}(h\vert\mathcal{F}_{t})=p_{n}\left(h\vert\left\{ {\bf y}_{nq_{k}(t)}^{(k)}\right\} _{k}\right) & \propto\left[\prod_{k=0}^{K-1}\prod_{i=1}^{\left\lfloor nq_{k}(t)\right\rfloor }p_{\theta_{0}+h/\sqrt{n}}^{(k)}(Y_{i}^{(k)})\right]\cdot m_{0}(h).
\end{align*}
As in Section \ref{sec:General-parametric-models}, we approximate
the likelihood (the bracketed term in the above expression) with an
approximation implied by (\ref{eq:SLAN K armed bandits}). So, the
approximate posterior is 
\begin{equation}
\tilde{p}_{n}(h\vert s)\propto\left[\prod_{k=0}^{K-1}\tilde{p}_{q_{k}}\left(x_{k}\vert h\right)\right]\cdot m_{0}(h);\ \textrm{where }\ \tilde{p}_{q_{k}}(\cdot\vert h)\equiv\mathcal{N}(\cdot\vert q_{k}h,q_{k}\sigma_{k}^{2}).\label{eq:approximate posterior MAB}
\end{equation}
The above suggests Theorem \ref{Thm: General parametric families}
can be extended to the $K$ armed case. This is done under the following
assumptions: Define $\mu_{n}^{(k)}(h)=\mathbb{E}_{P_{\theta_{0}+h/\sqrt{n}}^{(k)}}\left[Y_{i}^{(k)}\right]$. 

\begin{asm3} (i) The class $\{P_{\theta}^{(k)}\}$ is q.m.d around
$\theta_{0}$ for each $k$. (ii) $\mathbb{E}_{P_{\theta_{0}}^{(k)}}[\exp\vert\psi_{k}(Y)\vert]<\infty$
for each $k$. (iii) For each $k$, there exists $\dot{\mu}_{0}^{(k)}<\infty$
such that $\sqrt{n}\mu_{n}^{(k)}(h)=\dot{\mu}_{0}^{(k)}h+o(\vert h\vert^{2})$.
(iv) The support of $m_{0}(\cdot)$ is a compact set $\{h:\vert h\vert\le\Gamma\}$
for some $\Gamma<\infty$. (v) $\mu(\cdot)$ and $\mu^{\max}(\cdot)$
are H{\"o}lder continuous. Additionally, $\sup_{s}\left\{ \mu^{\max}(s)-\max_{k}\mu(s)\right\} \le C<\infty$.
\end{asm3}

Let $V_{\pi,n}(\cdot)$ denote the Bayes risk of policy $\pi$ and
$V_{n}^{*}(\cdot)$ the minimal Bayes risk, both under fixed $n$.
Define $\Pi^{\mathcal{S}}$ as the class of all sequentially measurable
policies that are functions only of $s=\{\{x_{k},q_{k}\}_{k},t\}$,
and $V_{n}^{\mathcal{S}*}(0)$ the fixed $n$ minimal Bayes risk when
the policies are restricted to $\Pi^{\mathcal{S}}$. Also, take $\pi_{\Delta t}^{*}$
to be the optimal piece-wise constant policy with $\Delta t$ increments.
Finally, denote by $L_{k}[\cdot]$ the infinitesimal generator 
\begin{equation}
L_{k}[f]:=\partial_{q_{k}}f+h(s)\partial_{x_{k}}f+\frac{1}{2}\sigma_{k}^{2}\partial_{x_{k}}^{2}f,\label{eq:infinitesimal generator MAB}
\end{equation}
where $h(s):=\tilde{\mathbb{E}}[h\vert s]$ and $\tilde{\mathbb{E}}[\cdot\vert s]$
is the expectation under $\tilde{p}_{n}(\cdot\vert s)$, defined in
(\ref{eq:approximate posterior MAB}). 

\begin{thm} \label{Thm: General parametric families-1}Suppose that
Assumption 3 holds. Then: (i) $\lim_{n\to\infty}\left|V_{n}^{*}(0)-V_{n}^{\mathcal{S}*}(0)\right|=0$.
(ii) $\textrm{\ensuremath{\lim}}_{n\to\infty}V_{n}^{*}(0)=V^{*}(0)$,
where $V^{*}(\cdot)$ solves PDE (\ref{eq:PDE_optimal_bayes_risk_General})
with the infinitesimal generators given by (\ref{eq:infinitesimal generator MAB}).
(iii) If, further, $\mu(\cdot)$, $\mu^{\max}(\cdot)$ are Lipschitz
continuous, $\lim_{n\to\infty}\vert V_{\pi_{\Delta t}^{*},n}(0)-V^{*}(0)\vert\lesssim\Delta t{}^{1/4}$
for any fixed $\Delta t$.\end{thm} 

The proof is analogous to that of Theorem \ref{Thm: General parametric families},
with the key difference being that the relevant likelihood is 
\[
\prod_{k=0}^{K-1}\prod_{i=1}^{\left\lfloor nq_{k}(t)\right\rfloor }p_{\theta_{0}+h/\sqrt{n}}^{(k)}(Y_{i}^{(k)})
\]
instead of $\prod_{i=1}^{\left\lfloor nq(t)\right\rfloor }p_{\theta_{0}+h/\sqrt{n}}(Y_{i})$.
The independence of the reward distributions across arms is convenient
here, and helps simplify the proof.\footnote{For instance, it implies that the joint probability $\prod_{k=0}^{K-1}P_{nq_{nk},h}^{(k)}$
is contiguous to $\prod_{k=0}^{K-1}P_{nq_{k},0}^{(k)}$ for any $(q_{n0},\dots,q_{n(K-1)})\to(q_{0},\dots,q_{K})$
as $n\to\infty$, as long as $P_{nq_{nk},h}^{(k)}$ is contiguous
to $P_{nq_{k},0}^{(k)}$ for each $k$. This enables us to prove an
analogue to Lemma \ref{Lem: bound on A_n}, which is a key step in
the proof.} See \citet{adusumilli2022Wald} for an example of the formal argument. 

Similar adaptations can be made for the results in Section \ref{sec:The-nonparametric-setting}. 

\subsection{Best arm identification\label{subsec:Best-arm-identification}}

Best arm identification describes a class of sequential experiments
in which the DM is allowed to experiment among $K$ arms of a bandit
until a set time $t=1$ (corresponding to $n$ time periods). At the
end of the experimentation phase, an arm is selected for final implementation.
Statistical loss is determined by expected payoffs during the implementation
phase, but not on payoffs generated during experimentation, i.e.,
there is no exploitation motive. In the Gaussian setting, it is sufficient
to use the same state variables $s=\{\{x_{k},q_{k}\}_{k},t\}$ as
in $K$ armed bandits. 

Let $\bm{\mu}:=(\mu_{0},\dots,\mu_{K-1})$ denote the mean rewards
of each arm, and $\pi^{(I)}\in\{0,\dots,K-1\}$ the action of the
DM in the implementation phase. Following the best arm identification
literature, see, e.g., \citet{kasy2019adaptive}, we take the loss
function to be expected regret in the implementation phase (also known
as ``simple regret'')
\[
L(\pi^{(I)},\bm{\mu})=\max_{k}\mu_{k}-\sum_{k}\mu_{k}\mathbb{I}(\pi^{(I)}=k).
\]
Suppose that the state variable at the end of experimentation is $s$.
The Bayes risk of policy $\pi^{(I)}$ given the terminal state $s$
is
\[
V_{\pi^{(I)}}(s)=\mathbb{E}\left[L(\pi^{(I)},\bm{\mu})\vert s\right]=\mu^{\max}(s)-\sum_{k}\mu_{k}(s)\mathbb{I}(\pi^{(I)}=k).
\]
Hence, the optimal Bayes policy is $\pi^{(I)}=\argmax_{k}\mu_{k}(s)$
and the minimal Bayes risk at the end of experimentation, i.e., when
$t=1$, is $V^{*}(s)=\mu^{\max}(s)-\max_{k}\mu_{k}(s)$. This determines
the boundary condition at $t=1$. 

We can obtain a PDE characterization of $V^{*}(\cdot)$ through similar
heuristics as in Section \ref{subsec:Bayes-risk}. By (\ref{eq:diffusion limit}),
the change to $q_{k}$ and $x_{k}$ in a short time period $\Delta t$
following state $s$ is approximately
\begin{align*}
\Delta q_{k} & \approx\pi_{k}\Delta t;\quad\Delta x_{k}\approx\pi_{k}\mu_{k}\Delta t+\sigma_{k}\sqrt{\pi_{k}}\Delta W(t).
\end{align*}
Now, for `interior states' with $t<1$, the recursion 
\[
V^{*}(s)=\inf_{\pi\in[0,1]^{K}}\mathbb{E}\left[\left.V^{*}\left(\left\{ x_{k}+\Delta x_{k},q_{k}+\Delta q_{k}\right\} _{k},t+\Delta t\right)\right|s\right]
\]
must hold for any small time increment $\Delta t$. Thus, by similar
(heuristic) arguments as in Section \ref{subsec:Bayes-risk}, $V^{*}(\cdot)$
satisfies 
\begin{align}
\partial_{t}V^{*}+\min_{k}L_{k}[V^{*}](s) & =0\quad\quad\!\textrm{if }t<1;\label{eq:PDE- bayes risk- BAI}\\
\qquad V^{*}(s) & =\varpi(s)\ \textrm{if }t=1,\nonumber 
\end{align}
where $\varpi(s):=\mu^{\textrm{max}}(s)-\max_{k}\mu_{k}(s).$

As we show below, all previous theoretical results (including for
parametric and non-parametric models) continue to apply with minor
modifications to the statements and the proofs. See also \citet{adusumilli2022minimax}
for the derivation of the minimax optimal policy in the two arm case.
The assumptions required are the same as that for multi-armed bandits.

\subsubsection*{Existence of a solution to PDE (\ref{eq:PDE- bayes risk- BAI})}

This is again a direct consequence of \citet[Theorem A.1]{barles2007error}. 

\subsubsection*{Convergence to the PDE }

Recall that the relevant state variables are $s=\{\{x_{k},q_{k}\}_{k},t\}$.
In analogy with (\ref{eq:discrete approximation}), the Bayes risk
in the fixed $n$ setting is given by 
\begin{align}
 & V_{n}^{*}\left(x_{1},q_{1},\dots,x_{K},q_{K},t\right)=\mathbb{I}_{n}^{c}\cdot\varpi(s)+\dots\nonumber \\
 & \dots+\min_{\pi_{1},\dots,\pi_{K}\in[0,1]}\mathbb{E}\left[\left.\mathbb{I}_{n}\cdot V_{n}^{*}\left(\left\{ x_{k}+\frac{\pi_{k}Y_{nq_{k}+1}^{(k)}}{\sqrt{n}},q_{k}+\frac{\pi_{k}}{n}\right\} _{k},t+\frac{1}{n}\right)\right|s\right]\label{eq:discrete approximation-BAI}
\end{align}
where $\mathbb{I}_{n}:=\mathbb{I}\{t\ge1/n$\}. The solution, $V_{n}^{*}(\cdot)$,
of the above converges locally uniformly to the viscosity solution,
$V^{*}(\cdot)$, of PDE (\ref{eq:PDE- bayes risk- BAI}). We can show
this by modifying the proof of Theorem \ref{Thm: Convergence to minimal PDE}
to account for the non-zero boundary condition. As in that proof,
after a change of variables $\tau=1-t$, we can characterize $V_{n}^{*}(\cdot)$
as the solution to $S_{n}(s,\phi(s),[\phi])=0$, where for any $u\in\mathbb{R}$
and $\phi:\mathcal{S}\to\mathbb{R}$, and $\mathbb{I}_{n}:=\mathbb{I}\{\tau>1/n\}$,
\begin{align*}
 & S_{n}(s,u,[\phi]):=-\mathbb{I}_{n}^{c}\cdot\frac{\left(\varpi(s)-u\right)}{n}-\cdots\\
 & \dots-\mathbb{I}_{n}\cdot\min_{\pi_{1},\dots,\pi_{K}\in[0,1]}\mathbb{E}\left[\left.\phi\left(\left\{ x_{k}+\frac{\pi_{k}Y_{nq_{k}+1}^{(k)}}{\sqrt{n}},q_{k}+\frac{\pi_{k}}{n}\right\} _{k},\tau-\frac{1}{n}\right)-u\right|s\right].
\end{align*}
Define $F(D^{2}\phi,D\phi,s)=\partial_{\tau}\phi-\min_{k}L_{k}[\phi](s).$ 

We need to verify monotonicity, stability and consistency of $S_{n}(\cdot)$.
Monotonicity of $S_{n}(s,u,[\phi])$ is clearly satisfied. Stability
is also straightforward under the assumption $\sup_{s}\varpi(s)<\infty$.
The consistency requirement is more subtle. For interior values, i.e.,
when $s:=(x,q,\tau)$ is such that $\tau>0$, the usual conditions
(\ref{eq:upper consistency}) and (\ref{eq:lower consistency}) are
required to hold with the definitions of $S_{n}(\cdot),F(\cdot)$
above. These can be shown using the same Taylor expansion arguments
as in the proof of Theorem \ref{Thm: Convergence to minimal PDE}.
For boundary values, $s\in\partial\mathcal{S}\equiv\{(x,q,0):x\in\mathcal{X},q\in[0,1]\}$,
the consistency requirements are (see, \citealp{barles1991convergence})
\begin{align}
\limsup_{\substack{n\to\infty\\
\rho\to0\\
z\to s\in\partial\mathcal{S}
}
}nS_{n}(z,\phi(z)+\rho,[\phi+\rho]) & \le\max\left\{ F(D^{2}\phi(s),D\phi(s),s),\phi(s)-\varpi(s)\right\} ,\label{eq:upper consistency-BAI}\\
\liminf_{\substack{n\to\infty\\
\rho\to0\\
z\to s\in\partial\mathcal{S}
}
}nS_{n}(z,\phi(z)+\rho,[\phi+\rho]) & \ge\min\left\{ F(D^{2}\phi(s),D\phi(s),s),\phi(s)-\varpi(s)\right\} .\label{eq:lower consistency-BAI}
\end{align}
We can show (\ref{eq:upper consistency-BAI}) as follows (the proof
of (\ref{eq:lower consistency-BAI}) is similar): By the definition
of $S_{n}(\cdot)$, for every sequence $(n\to\infty,\rho\to0,z\to s\in\partial\mathcal{S})$,
there exists a sub-sequence such that either $nS_{n}(z,\phi(z)+\rho,[\phi+\rho])=\phi+\rho-\varpi(z)$
or 
\[
nS_{n}(z,\phi(z)+\rho,[\phi+\rho])=-\min_{\pi_{1},\dots,\pi_{K}\in[0,1]}\mathbb{E}\left[\left.\phi\left(\left\{ x_{k}+\frac{\pi_{k}Y_{nq_{k}+1}^{(k)}}{\sqrt{n}},q_{k}+\frac{\pi_{k}}{n}\right\} _{k},\tau-\frac{1}{n}\right)-u\right|s\right].
\]
In the first instance, $nS_{n}(z,\phi(z)+\rho,[\phi+\rho])\to\phi(s)-\varpi(s)$
by the continuity of $\varpi(\cdot)$, while the second instance gives
rise to the same expression for $S_{n}(\cdot)$ as being in the interior,
so that $nS_{n}(z,\phi(z)+\rho,[\phi+\rho])\to F(D^{2}\phi(s),D\phi(s),s)$
by similar arguments as in the proof of Theorem \ref{Thm: Convergence to minimal PDE}.
Thus, in all cases, the limit along subsequences is smaller than the
right hand side of (\ref{eq:upper consistency-BAI}). 

\subsubsection*{Piecewise-constant policies}

The results on piece-wise constant policies continue to apply since
\citet[Theorem 3.1]{barles2007error} holds under any continuous boundary
condition.

\subsubsection*{Parametric and non-parametric distributions}

The analogues of Theorems \ref{Thm: General parametric families}
and \ref{Thm: Non-parametric models} follow by the same reasoning
as that employed for multi-armed bandits in Appendix \ref{subsec:Multi-armed-bandits}.
In fact, the proofs are even simpler since the loss function is just
the regret payoff at $t=1$. 

\subsection{Discounting\label{subsec:Discounting}}

Our methods also apply to bandit problems without a definite end point.
Suppose the rewards in successive periods are discounted by $e^{-\beta/n}$
for some $\beta>0$. Here, $n$ is to be interpreted as a scaling
of the discount factor; it is the number of periods of experimentation
in unit time when the DM experiments in regular time increments and
intends to discount rewards by the fraction $e^{-\beta}$ after $\Delta t=1$.
Discounting ensures the cumulative regret is finite. It also changes
the considerations of the DM, who will now be impatient to start `exploitation'
sooner as future rewards are discounted. Popular bandit algorithms
such as Thompson sampling do not admit discounting and will therefore
be substantially sub-optimal. 

In the Gaussian setting with one arm, the relevant state variables
under discounting are $s:=(x,q)$, where $x,q$ are defined in the
same manner as before, but $q$ can now take values above $1$ (it
is the number of times the arm is pulled divided by $n$). The counterpart
of PDE (\ref{eq:PDE optimal bayes risk}) for discounted rewards is
\begin{equation}
\beta V^{*}-\mu^{+}(s)-\min\left\{ -\mu(s)+L[V^{*}](s),0\right\} =0.\label{eq:discounted PDE}
\end{equation}
Note that PDE (\ref{eq:discounted PDE}) does not require a boundary
condition. 

All the previous theoretical results continue to apply to discounted
bandits, as we demonstrate below. The assumptions required are the
same as in Theorems \ref{theorem 1}-\ref{Thm: Non-parametric models}
in the main text, along with $\beta>0$.

\subsubsection*{Existence of a solution to PDE (\ref{eq:discounted PDE})}

By \citet[p. 29]{barles2007error}, there exists a unique viscosity
solution to PDE (\ref{eq:discounted PDE}). 

\subsubsection*{Convergence to the PDE}

The analogue to (\ref{eq:discrete approximation}) under discounting
is 
\begin{align}
V_{n}^{*}\left(x,q\right) & =\min_{\pi\in[0,1]}\mathbb{E}\left[\left.\frac{\mu^{+}(s)-\pi\mu(s)}{n}+e^{-\beta/n}V_{n}^{*}\left(x+\frac{A_{\pi}Y_{nq+1}}{\sqrt{n}},q+\frac{A_{\pi}}{n}\right)\right|s\right].\label{eq:discrete approximation- Discounting}
\end{align}
A straightforward modification of the proof of Theorem \ref{Thm: Convergence to minimal PDE}
then shows $V_{n}^{*}(\cdot)$ converges locally uniformly to $V^{*}(\cdot)$,
the viscosity solution of PDE (\ref{eq:discounted PDE}). There is
no analogue to piece-wise constant policies in the discounted setting. 

\subsubsection*{Parametric and non-parametric distributions}

The proofs of Theorems \ref{Thm: General parametric families} and
\ref{Thm: Non-parametric models} are slightly complicated by the
fact $q$ is now unbounded. While the SLAN property (\ref{eq:LAN})
applies even if $q>1$, it does require $q<\infty$. We can circumvent
this issue by exploiting the fact that the infinite horizon problem
is equivalent to a finite horizon problem with a very large time limit.
In other words, we prove the relevant results for the PDE 
\begin{align}
\partial_{t}V^{*}-\beta V^{*}+\mu^{+}(s)+\min\left\{ -\mu(s)+L[V^{*}](s),0\right\}  & =0\ \textrm{if }t<1,\nonumber \\
V^{*}(s) & =0\ \textrm{if }t=T,\label{eq:PDE optimal bayes - discounting with fixed T}
\end{align}
with the boundary condition set at $t=T$, and then let $T\to\infty$. 

Let $V^{*}(0),V^{*}(0;T)$ denote the viscosity solutions to PDEs
(\ref{eq:discounted PDE}) and (\ref{eq:PDE optimal bayes - discounting with fixed T}),
evaluated at $s_{0}$. Following the first step in Appendix (\ref{subsec:Proof-sketch-of-Theorem 5}),
the Bayes risk under a policy $\pi$ in the fixed $n$ setting with
discounting can be shown to be
\begin{align}
V_{\pi,n}(0) & =\mathbb{E}_{({\bf y}_{n},h)}\left[\frac{1}{n}\sum_{j=1}^{\infty}e^{-\beta j/n}R_{n}(h,\pi_{j})\right].\label{eq:discrete approximation-policy-parametric-discounting}
\end{align}
Analogously, if we terminate the experiment at a suitably large $T$,
we have 
\[
V_{\pi,n}(0;T)=\mathbb{E}_{({\bf y}_{n},h)}\left[\frac{1}{n}\sum_{j=1}^{nT}e^{-\beta j/n}R_{n}(h,\pi_{j})\right].
\]
Under Assumption 1, $R_{n}(h,\pi)\le C<\infty$ (due to the compactness
of the prior $m_{0}$), so $\sup_{\pi\in\Pi}\vert V_{\pi,n}(0)-V_{\pi,n}(0;T)\vert\apprle e^{-\beta T}$.
Now, a straightforward modification of the proof of Theorem \ref{Thm: General parametric families}
implies $\lim_{n\to\infty}\inf_{\pi\in\Pi}V_{\pi,n}(0;T)=V^{*}(0;T)$,
where $V^{*}(0;T)$ is the viscosity solution to PDE (\ref{eq:PDE optimal bayes - discounting with fixed T})
evaluated at $s_{0}$. Finally, it can be shown, e.g., by approximating
the PDEs with dynamic programming problems as in Theorem \ref{Thm: Convergence to minimal PDE},
that $\vert V^{*}(0;T)-V^{*}(0)\vert\apprle e^{-\beta T}$. Since
we can choose $T$ as large as we want, it follows $\lim_{n\to\infty}\inf_{\pi\in\Pi}V_{\pi,n}(0)=V_{n}^{*}(0)$.
The proof of Theorem \ref{Thm: Non-parametric models} can be modified
in a similar manner. 

\section{Computation using finite-difference methods\label{sec:Details-on-computation}}

As mentioned in the main text, PDE (\ref{eq:PDE optimal bayes risk})
also be solved using `upwind' finite-difference methods. The method
is more accurate than the Monte-Carlo algorithm (Algorithm 1) but
scales less favorably with increasing number of arms. To implement
this method we first discretize both the spatial (i.e., $\mathcal{X}$
and $\mathcal{Q}$) and time domains. Let $i,j$ index the grid points
for $x,q$ respectively, with the grid lengths being $\Delta x,\Delta q$.
PDEs of the form (\ref{eq:PDE optimal bayes risk}) are always solved
backward in time, so, for this section, we switch the direction of
time (i.e., $t=1$ earlier is now $t=0$) and discretize it as $0,\Delta t,\dots,m\Delta t,\dots,1$.
Denote $V_{i,j}^{m}$ as the approximation to the PDE solution $V^{*}$
at grid points $i,j$ and time period $m\Delta t$.

We approximate the second derivative $\partial_{x}^{2}V^{*}$ using
\[
\partial_{x}^{2}V^{*}\approx\frac{V_{i+1,j}^{m}+V_{i-1,j}^{m}-2V_{i,j}^{m}}{(\Delta x)^{2}}.
\]
As for the first order derivatives, we approximate by either $\frac{V_{i+1,j}^{m}-V_{i,j}^{m}}{\Delta x}$
or $\frac{V_{i,j}^{m}-V_{i-1,j}^{m}}{\Delta x}$ depending on whether
the associated drift, i.e., the coefficient multiplying $\partial_{x}V^{*}$
is positive or negative. This is known as up-winding and is crucial
for ensuring the resulting approximation procedure is `monotone' (see
Appendix \ref{subsec:Proof-of-Theorem-2}, and also \citet{achdou2017income}
for a discussion of monotonicity, and its necessity for showing convergence
of the approximation procedures). In our setting, this implies 
\begin{align*}
\partial_{x}V^{*} & \approx\frac{V_{i+1,j}^{m}-V_{i,j}^{m}}{\Delta x}\mathbb{I}(\mu(s)\ge0)+\frac{V_{i,j}^{m}-V_{i-1,j}^{m}}{\Delta x}\mathbb{I}(\mu(s)<0)\\
 & :=\left(\frac{V_{i+1,j}^{m}-V_{i,j}^{m}}{\Delta x}\right)_{+},
\end{align*}
while $\partial_{q}V^{*}$, which is associated with the coefficient
1, is approximated as 
\[
\partial_{q}V^{*}\approx\frac{V_{i,j+1}^{m}-V_{i,j}^{m}}{\Delta q}.
\]
Finally, let $\mu_{i,j}^{+},\mu_{i,j}$ denote the values of $\mu^{+}(\cdot),\mu(\cdot)$
evaluated at the grid points $i,j$. 

Following the derivative approximations, the PDE can be solved using
explicit, implicit or hybrid schemes. The previous version of this
manuscript discussed these different approaches and their convergence
properties.\footnote{This version can be accessed at \href{https://arxiv.org/abs/2112.06363}{arXiv:2112.06363v14}.}
Our recommendation is to use the hybrid scheme. It is faster than
the standard implicit scheme as it does not require policy iteration.
At the same time, it is more numerically stable than the explicit
scheme as it does not require the CFL condition that $\Delta t\le0.5\min\left\{ (\Delta x)^{2},(\Delta q)^{2}\right\} $;
instead, we only need $\Delta t\to0$. 

The algorithm is based on a recursion whereby $V_{i,j}^{0}=0$, and
an estimate of the action-value function, $\tilde{V}_{i,j}^{m+1,1}$,
corresponding to the case where the arm was pulled in step $m+1$,
is computed in terms of $V_{i,j}^{m}:=\min\left\{ \tilde{V}_{i,j}^{m,1},\tilde{V}_{i,j}^{m,0}\right\} $
as the solution to
\begin{align}
\tilde{V}_{i,j}^{m+1,1} & =V_{i,j}^{m}+\mu_{i,j}^{+}-\mu_{i,j}+\frac{\tilde{V}_{i,j+1}^{m+1,1}-\tilde{V}_{i,j}^{m+1,1}}{\Delta q}\nonumber \\
 & \hfill+\mu_{i,j}\left(\frac{\tilde{V}_{i+1,j}^{m+1,1}-\tilde{V}_{i,j}^{m+1,1}}{\Delta x}\right)_{+}+\frac{1}{2}\sigma^{2}\frac{\tilde{V}_{i+1,j}^{m+1,1}+\tilde{V}_{i-1,j}^{m+1,1}-2\tilde{V}_{i,j}^{m+1,1}}{(\Delta x)^{2}}=0.\label{eq:hybrid_scheme}
\end{align}
As for the action-value function corresponding to the case where the
arm was not pulled, we have
\[
\tilde{V}_{i,j}^{m+1,0}:=V_{i,j}^{m}+\mu_{i,j}^{+}.
\]
We then set $V_{i,j}^{m+1}:=\min\left\{ \tilde{V}_{i,j}^{m+1,1},\tilde{V}_{i,j}^{m+1,0}\right\} $
and continue the iterations until $m=M-1$. The pseudo-code for the
hybrid FD scheme is described in Algorithm \ref{alg:Algorithm 2}. 

\begin{algorithm}
\begin{raggedright}
\caption{Hybrid FD}
\label{alg:Algorithm 2}
\par\end{raggedright}
\raggedright{}{\small\begin{algorithmic}[1]
\Require{M (number of time periods)}
\State \textbf{initialize} $V_{i,j}^{0} = 0$
\For{$m=0,\dots,M-1$:}
\State{Write (\ref{eq:hybrid_scheme}) as $A{\bf \tilde{V}}_{m+1}^{1}-{\bf V}_{m}+{\bf X}=0$ where ${\bf \tilde{V}}_{m}^{(1)}=\textrm{vec}(\tilde{V}_{i,j}^{m,1};i,j)$ }
\State{${\bf \tilde{V}}_{m+1}^{1} = A^{-1}\left({\bf V}_{m}-{\bf X}\right)$ }
\State{${\bf \tilde{V}}_{m+1}^{0}={\bf V}_{m}+\bm{\mu^{+}}$ where $\bm{\mu^{+}}=\textrm{vec}(\mu_{i,j}^{+};i,j)$}
\State{${\bf V}_{m+1} = \min\left\{ {\bf \tilde{V}}_{m+1}^{1},{\bf \tilde{V}}_{m+1}^{0}\right\} $ where the minimum is computed element-wise}
\EndFor
\end{algorithmic}}{\small\par}
\end{algorithm}

\subsubsection{Implementation details for Section \ref{subsec:A-one-armed-bandit}}

For the empirical illustration in Section \ref{subsec:A-one-armed-bandit},
we used $\Delta x=1/1500$, $\Delta q=1/600$ and $\Delta t=1/1000$.
Since $x$ is unbounded, for the purposes of computation we set its
upper and lower bounds to $l-3\sigma$ and $u+3\sigma$, where $l$
and $u$ are the support points of the least favorable prior. 
\end{document}